%% file: pipeline.tex
\documentclass[acmsmall,screen]{acmart}

\setcopyright{rightsretained}
\acmPrice{}
\acmDOI{10.1145/3341702}
\acmYear{2019}
\copyrightyear{2019}
\acmJournal{PACMPL}
\acmVolume{3}
\acmNumber{ICFP}
\acmArticle{98}
\acmMonth{8}
\startPage{1}

\usepackage{booktabs}   

\usepackage{tikz}
\usepackage{pgfplots}
\pgfplotsset{compat=1.13}
\usetikzlibrary{fit,matrix,positioning,shadows,shapes.geometric,shapes.symbols,shapes.misc}
\tikzset{doc/.style={
    draw,
    chamfered rectangle,
    chamfered rectangle corners=north east,
    append after command={edge
        [transform canvas={xshift=-0.2pt, yshift=-0.2pt},
         draw, -,
         line cap=butt,
         to path=(\tikztostart.before north east) -|
                 (\tikztostart.after north east)]
        (\tikzlastnode)}}}

\usepackage{hyperref}
\usepackage[nameinlink,noabbrev,capitalize,nosort]{cleveref}
\crefname{equation}{equation}{equations}
\crefname{lstlisting}{Listing}{Listings}

\usepackage{prooftree1}
\usepackage{mathbbol}
\usepackage{tabularx}
\usepackage{mdwtab}
\usepackage{hyphenat}
\usepackage{bm}
\usepackage{xspace}
\usepackage{comment}
\usepackage{listings}
\usepackage{fancyvrb}
\usepackage{pgfplots}
\usepackage{natbib1}

\newcommand{\quant}[2]{\mathopen{\text{\divide\thinmuskip2 \divide\medmuskip2 \divide\thickmuskip2 $#1#2.$}\,}}
\newcommand{\fun}{\quant\lambda}
\newcommand{\altern}{\mathrel{\big|}}
\newcommand\nomathbreak{\relpenalty=10000 \binoppenalty=10000 }
\newcommand\mathbox[1]{{\nomathbreak\ensuremath{#1}}}

\DeclareSymbolFont{AMSb}{U}{msb}{m}{n}%
\DeclareSymbolFontAlphabet{\mathbbl}{AMSb}%
\newcommand{\Real}{\ensuremath{\mathbbl{R}}}
\newcommand{\Integer}{\mathbbl{Z}}
\newcommand{\Bool}{\mathbbl{B}}
\newcommand{\Unit}{\ensuremath\Eins}
\newcommand{\rplus}{\ensuremath{\Real^+}}
\newcommand{\iplus}{\mathbbl{N}}
\newcommand{\Measure}{\mathinner{\mathbbl{M}}}
\newcommand{\Array}{\mathinner{\mathbbl{A}}}
\newcommand{\subst}[3]{#3\{#2\mapsto #1\}}

\newcommand\anintegrand{\ensuremath h}
\newcommand\aspace{\ensuremath X}
\newcommand\anotherspace{\ensuremath Y}
\let\xh\hbar

\newcommand\vecdots[1]{\bm{\vec{#1}}}
\newcommand\vecdotsvec[1]{\bm{\vec{#1}}}

\newcommand\DefineConstructor[2][]{\expandafter\newcommand\csname#2#1\endcsname{\textsf{#2}}}
\DefineConstructor{Uniform}
\DefineConstructor{Gaussian}
\DefineConstructor{Cauchy}
\DefineConstructor{StudentT}
\DefineConstructor[D]{Beta}
\DefineConstructor[D]{Gamma}
\DefineConstructor{Categorical}
\DefineConstructor{Ret}
\DefineConstructor{Bind}
\DefineConstructor{Weight}
\DefineConstructor{Msum}
\DefineConstructor[ThenElse]{If}
\DefineConstructor{ary}
\DefineConstructor{Plate}
\DefineConstructor{LO}
\DefineConstructor{integrate}
\DefineConstructor{unintegrate}
\DefineConstructor{unproduct}
\DefineConstructor{Fanout}
\newcommand{\Index}{\Idx}
\DefineConstructor{Idx}
\DefineConstructor{Split}
\DefineConstructor{Nop}
\DefineConstructor{Add}
\DefineConstructor{newArray}
\DefineConstructor{clearArray}
\newcommand{\size}{\#}
\newcommand{\Bucket}{\mathop{\textsf{Hist}}\nolimits}
\newcommand{\summary}{\mathit{hist}}
\newcommand{\summarize}{\ensuremath{\mathop\textsf{histogram}\nolimits}}

\newcommand{\literalArray}[1]{\mathopen{\textsf[}#1\mathclose{\textsf]}}
\newcommand{\idx}[1]{\penalty\binoppenalty [#1]}
\newcommand{\Exp}{\mathrm{e}^}
\newcommand{\Frac}[3][1]{\frac{#1}{#3}#2}
\newcommand{\sumvec}[1]{\sum_{i=0}^{n-1}\vec{#1}\idx{i}}
\newcommand{\Loop}[3]{\!\begin{smallmatrix}#3\\#1=#2\end{smallmatrix}\!}

\newcommand{\pw}[4]{\left\{\!\begin{smallmatrix}#1&#2\\#3&#4\end{smallmatrix}\!\right.}

\newcommand{\pwow}[3]{\pw{#1}{#2}{#3}{\text{otherwise}}}

\newcommand{\guard}{\mathbin?}
\newcommand{\XI}{\mathit{xi}}
\newcommand{\Let}[3]{\textsf{let\ }#1=#2\textsf{\ in\ }#3}
\newcommand{\Lets}[3][=]{%
    \begin{array}[t]{@{}l@{}}
        \textsf{let\ }\begin{array}[t]{@{}l@{${}#1{}$}>{\displaystyle}l@{}}#2\end{array}
     \\ #3
    \end{array}}
\newcommand{\For}[4]{%
    \textsf{for\ }#1 = #2\textsf{\ to\ }#3\textsf{:} \\
    \begin{array}{@{\quad}l@{}}#4\end{array} }

\newcommand{\X}{\ensuremath{\times}}
\newcommand{\mulequal}{\mathrel{\mathord\times\mathord=}}
\newcommand{\plusequal}{\mathrel{\mathord+\mathord=}}

\newcommand{\hakaru}{Hakaru\xspace}

\makeatletter
\patchcmd{\@addmarginpar}{\ifodd\c@page}{\ifodd\c@page\@tempcnta\m@ne}{}{}%
\makeatother
\reversemarginpar
\newif\ifcomments
\ifcomments
    \advance\marginparwidth \marginparsep
    \advance\marginparwidth -4pt
    \advance\marginparwidth 210mm
    \advance\marginparwidth -6.75in
    \long\def\remark#1{
        \let\SAVEDspacefactor\relax
        \ifhmode\edef\SAVEDspacefactor{\the\spacefactor}\fi
        \raisebox{-3.5pt}[0pt][0pt]{\makebox[0pt][c]{$\scriptstyle\diamond$}}%
        \marginpar{\raggedright\hbadness=10000
        \parindent=8pt \parskip=2pt
        \def\baselinestretch{0.8}\footnotesize
        \itshape\noindent #1\par}%
        \ifx\SAVEDspacefactor\relax\else\spacefactor\SAVEDspacefactor\fi\relax}
\else
    \long\def\remark#1{\relax}%
\fi
\newcommand{\diff}[2]{\remark{\color{red}#1: #2}}
\newcommand{\diffopen}[2]{\remark{\color{red}[#1: #2}}
\newcommand{\diffclose}[1]{\remark{\color{red}#1]}}

\begin{document}
\lstset{
  basicstyle=\ttfamily\footnotesize,
  stringstyle=\ttfamily,
  resetmargins=true,
  columns=fixed
}

\title{From High-Level Inference Algorithms to Efficient Code}


\author{Rajan Walia}
\email{rawalia@indiana.edu}
\author{Praveen Narayanan}
\email{pravnar@umail.iu.edu}
\affiliation{
  \department{Department of Computer Science}
  \institution{Indiana University}
  \city{Bloomington}
  \country{USA}
}

\author{Jacques Carette}
\email{carette@mcmaster.ca}
\affiliation{
  \department{Department of Computing and Software}
  \institution{McMaster University}
  \country{Canada}
}

\author{Sam Tobin-Hochstadt}
\email{samth@cs.indiana.edu}
\author{Chung-chieh Shan}
\email{ccshan@indiana.edu}
\affiliation{
  \department{Department of Computer Science}
  \institution{Indiana University}
  \city{Bloomington}
  \country{USA}
}

\begin{abstract}
Probabilistic programming languages are valuable because they allow
domain experts to express probabilistic models and inference algorithms
without worrying about irrelevant details.  However, for decades there
remained an important and popular class of probabilistic inference
algorithms whose efficient implementation required manual low-level
coding that is tedious and error-prone.  They are algorithms whose idiomatic expression requires
random array variables that are \emph{latent} or whose likelihood is
\emph{conjugate}.  Although that is how practitioners
communicate and compose these algorithms on paper, executing such
expressions requires \emph{eliminating} the latent variables and
\emph{recognizing} the conjugacy by symbolic mathematics.  Moreover,
matching the performance of handwritten code requires speeding up loops
by more than a constant factor.

We show how probabilistic programs that directly and concisely express
these desired inference algorithms can be compiled
while maintaining efficiency.  We introduce new transformations that
turn high-level probabilistic programs with arrays into pure loop code.
We then make great use of domain-specific invariants and norms to
optimize the code, and to specialize and JIT-compile the code per execution.
The resulting performance is competitive with manual implementations.
\end{abstract}



\maketitle


\section{Simplifying and Optimizing Probabilistic Programming}

\ifcomments
\begin{tikzpicture}[overlay, remember picture]
    \path (current page.north west)++(46pt,-29pt)
     node [inner sep=0, anchor=north west, text=red, text width=\textwidth]
     {All differences between our original submission and the
      second-stage submission, except minor edits, are described in
      the right margin and marked using diamonds~$\diamond$ in this
      document.  Each change is marked with the letter(s) of the
      review(s) that motivated~it.  The extent of larger changes is
      marked at the beginning and at the end by [square brackets].};
\end{tikzpicture}%
\fi
Many users of an algorithm would rather not worry about the details
of its efficient implementation or correctness proof.  Whether the
algorithm is copied from a textbook by a programmer or generated from a
domain-specific language by a compiler, the vocabulary used to express
the algorithm needs to be mapped to executable code before the algorithm
can be run.  For example, if the algorithm invokes sorting, then it is
easier to turn into executable code using a language or library that
features a sorting routine.  To take a more recent example, if the
algorithm refers to the gradient of a function, then it is easier to
turn into executable code using
automatic differentiation.

In the realm of probabilistic programming, while a wide variety of
languages~\citep{de-raedt-problog,pfeffer-design,milch-blog,goodman-church,figaro,tristan-augur,huang-compiling,wu-swift,de-salvo-braz-lifted,lunn-winbugs,wood-new,goodman-design,patil-pymc,carpenter-stan,kiselyov-embedded-dsl,kiselyov-probabilistic,narayanan-probabilistic,mansinghka-venture,fischer-autobayes,tran-deep,nori-r2}
have made many algorithms easier to express, many
practically-important inference methods continue to require manual
transformation and implementation.
In this paper, we extend the range of probabilistic inference
algorithms that can be turned automatically into executable code, to
include \emph{arrays} whose distributions need to be \emph{simplified}
and whose loops need to be \emph{optimized}.
\begin{itemize}
    \item Simplification includes \emph{eliminating latent variables}
          and \emph{recognizing conjugate likelihoods}.
          \begin{itemize}
              \item Briefly,\diff{A}{Explain latent variables} a~latent variable is a random variable whose
                    value may be used in the program but is not returned.
                    Elimination is widely applied to discrete and
                    continuous variables
                    \citep{zhang-simple,dechter-bucket,zhang-exploiting,poole-exploiting,sanner-symbolic,de-salvo-braz-lifted}
                    and is known in various contexts as
                    \emph{Rao-Blackwellization} \citep{rao-information,blackwell-conditional,kolmogorov-unbiased,gelfand-sampling-based,casella1996rao,murray-delayed},
                    \emph{collapse} \citep{koller-pgm,liu-covariance,liu-collapsed,venugopal-dynamic},
                    \emph{marginalization} \citep{meng-seeking,obermeyer-automated}, and
                    \emph{integrating out} \citep{resnik-gibbs,griffiths-finding}.
              \item Briefly,\diff{A}{Explain conjugate likelihood} a~conjugate likelihood is a weight on samples
                    that can be made constant while preserving semantics by
                    changing how the samples are generated in the first place.
                    Conjugacy is a preferred starting point and basic
                    building block of Bayesian data modeling
                    \citep[page 36]{gelman-bayesian} and underlies such
                    popular applications as Naive Bayes classification
                    \citep{bayes-essay} and Bayesian linear regression
                    \citep{borgstrom-fabular}.
          \end{itemize}
    \item Loop optimization includes reordering sums to achieve
          superlinear speedups, and fusing and specializing loops to
          obtain one more order of magnitude in performance.
\end{itemize}
As the description above suggests, the importance
of this class of algorithms has been established in applied statistics
for decades.  However,
turning the vocabulary used to express them into executable
code had required manual calculation and coding that is tedious and
error-prone \citep{cook-validation,geweke-getting}.  Our work thus paves
the way for programmers and compilers alike to target a higher-level
probabilistic language with arrays and to worry less about the details
of the correctness of distribution simplifications and the efficiency of
loop optimizations.

One major reason that turning high-level algorithms into efficient code
is difficult---whether by hand or by machine---is that it requires
sophisticated symbolic mathematics.  Recent research has
started to automate such reasoning on probabilistic programs
\citep{carette-simplifying-padl,gehr-psi,tran-deep,hoffman-autoconj}.
However, even systems that support arrays
either fail to perform popular transformations
such as latent-variable elimination (as in Augur \citep{tristan-augur},
AugurV2 \citep{huang-compiling}, and
Edward \citep{tran-deep}) or unroll random choices\diff{E}{Corrected: it is not arrays but
random choices that PSI needs to unroll in order to simplify.} entirely at prohibitive
performance cost (as in PSI \citep{gehr-psi}).  Given that
arrays are key in almost any inference algorithm,
unrolling is a non-starter for efficient execution.

We remove these obstacles to automation. Probabilistic
programmers can express high-level algorithms and expect sophisticated
transformations to automate efficient execution on large arrays of data.
We present a domain-specific compilation pipeline that
meets all these goals.  
Specifically:
\begin{enumerate}
\item We extend probabilistic programs and their simplification to
  those with arrays of large or arbitrary size, such as arrays of size
  $n$ or of size $n_1$-by-$n_2$-by-$n_3$, where each $n$ is large and/or
  unknown (\cref{s:simplify}).
  Our array simplification transformation is modular in that it reuses
  existing technology underlying scalar simplification and, like
  that technology, eschews brittle pattern matching of specific
  distributions and extends easily to new primitive distributions.
        \begin{itemize}
            \item We extend symbolic integration in computer algebra to
                  high- and arbitrary-dimensional integrals, such as
                  integrals over $\Real^n$ or over
                  \smash{${{\Real^{n_3}}^{n_2}}^{n_1}$}, where each $n$
                  is large and/or unknown.
            \item We introduce the symbolic \emph{unproduct} operation
                  to uncover independence underlying a program so as to
                  apply our simplification transformation.
                  This process traverses
                  an input term systematically and recursively to
                  uncover its equivalence to a sequence of products
                  $\prod_i\prod_j\prod_k$ of any given length.
        \end{itemize}
\item We introduce the \emph{histogram} optimization
  (\cref{s:histogram}), which asymptotically speeds up loops by
  rewriting them as map-reduce expressions in a modular and general way.
        \begin{itemize}
            \item This optimization \emph{unnests}
                  loops, by locating conditionals buried deep inside
                  any level of nested loop bodies.  It is particularly
                  effective on simplified array probabilistic programs.
        \end{itemize}
\item We optimize the resulting array\hyp manipulating code aggressively
  yet safely, by taking advantage of the domain-specific features of
  probabilistic programs (\cref{s:codegen}).
        \begin{itemize}
            \item We carefully engineer loop-invariant code motion
                  (LICM) and loop fusion, so that they apply soundly,
                  widely, and profitably.
            \item We further use just-in-time (JIT) compilation to
                  propagate static information.
        \end{itemize}
\item We show that while each of our techniques is valuable, their
  composition---our \emph{pipeline}---is dramatically more effective.
  In other words, each bullet item above is a significant and essential
  contribution.
\end{enumerate}
\Cref{s:overview} lays out our compilation pipeline and
sets the stage for these technical contributions.

We emphasize that our aim is not to improve the compilation of models
already handled by existing systems, but rather to enable the
compilation of algorithms not handled by existing systems and not
expressed by previous probabilistic programmers.  We compile
probabilistic programs that directly and concisely express a new and
open class of algorithms of lasting and current significance that
previously required manual, tedious, and error-prone mathematics and
coding.  Of course, we can only measure our system against other systems
on tasks that they can also~do.
The quantitative
evaluation in \cref{s:eval} demonstrates that our proof-of-concept system
achieves the competitive performance expected of the newly expressed
algorithms, relative to handwritten code for the same
algorithms and other state-of-the-art systems carrying out different algorithms.
\diffopen{D}{Emphasize that we contribute more than a system}%
Whereas our system automates exact inference and collapsed
Metropolis-Hastings (MH) sampling, our modular tools and techniques
automate tasks often performed manually by practitioners of many other
inference methods.  Hence, for example, it is promising to
incorporate our contributions in a future system that supports
Hamiltonian Monte Carlo (HMC) or variational inference.\diffclose{D}

\section{Compilation Pipeline Overview}
\label{s:overview}

The heart of many popular inference algorithms is to calculate
a conditional distribution exactly and possibly sample from it.  This
pattern is clearest and most challenging in Gibbs sampling, which
repeatedly updates a sample by conditioning on some of its dimensions.
But the same pattern recurs in MH\@,
HMC\@, and importance sampling, because they
require computing a density, which is the total of a conditional
distribution.  And in important cases such as Bayesian linear
regression, an exact solution is available, because
conditioning the model on the observed data results in
a distribution that can be represented in a closed form.
\diffopen{CE}{Explain exact computation on inference distributions for approximate inference}%
Due to this pattern, practitioners communicate and compose not only
probabilistic models (such as hidden Markov models
\citep{rabiner-speech}) but also inference algorithms (such as Markov
Chain Monte Carlo (MCMC) \citep{mackay-monte}) using the same
probabilistic programming constructs: sampling, sequencing, looping,
conditioning, and so~on.  Therefore, it~makes sense to express both
modeling and inference distributions in a single declarative language of
measures, as pioneered by the proof-of-concept probabilistic programming
system \hakaru \citep{narayanan-probabilistic,zinkov-composing}.\diffclose{CE}

The compilation pipeline in this paper is designed to express such
inference algorithms concisely and execute them efficiently.  The
starting point is a probabilistic program that expresses the desired
inference algorithm by denoting the conditional distribution to
calculate and possibly sample.  That is, we represent the inference
distribution as a \emph{generative} process, which is a step-by-step
procedure for drawing random variables and computing a final outcome.
Some procedures score their outcome so its \emph{importance weight}
varies from run to run; other procedures make no random choice so the
computation is deterministic.

\begin{figure}
\advance\baselineskip-1pt
\tikzset{>=stealth,
    maple/.style={draw, rounded corners, inner sep=1.125ex, text depth=0, align=center},
    ll/.style={draw, tape, tape bend top=none, inner xsep=.875ex, inner ysep=1.25ex},
    transform/.style={font=\sffamily}}
\begin{tikzpicture}
    \matrix (m) [row sep=1.5pc, column sep=8pc, inner sep=0, nodes={inner sep=.3333em}] {
        \node (conditional) [doc, text depth=0, anchor=south west] {conditional distribution};
        & \node (patently) [maple, anchor=south] {patently linear\\expression};
    \\
        \node (simplified) [doc, text depth=0, anchor=north west] {simplified distribution};
        & \node (simplified-patently) [maple, anchor=north] {simplified\\patently linear\\expression};
    \\[-2em]
        \node (summarized) [doc, text depth=0, anchor=west] {map-reduce expressions};
    \\
        \node (licm) [doc, text depth=0, anchor=west] {combined let bindings};
    \\
        \node (sham1) [ll, anchor=west] {Sham IR};
    \\[-1ex]
        \node (sham2) [ll, anchor=west] {Sham IR};
    \\[-1ex]
        \node (sham3) [ll, anchor=west] {Sham IR};
        & \node (code) [ll] {x86 code};
    \\
    };
    \path (conditional.north) ++(0,4pc)
     node (model) [doc, text depth=0, anchor=south] {model};
    \path (sham3.east) -- node (llvm) [midway, xshift=1em, ll] {LLVM IR} (code.west);
    \draw [<-, out=90, in=-135] (conditional.north) ++(-3pc,0)
        to node [transform, left, near end] {disintegrate} (model);
    \draw [<-, dotted, out=90, in=-45] (conditional.north) ++(+3pc,0)
        to node (misc) [transform, pos=.7, xshift=.75em, fill=white, draw, inner sep=0,
                        cloud, cloud ignores aspect] {MH, \dots} (model);
    \draw [->, out=90, in=90] (conditional.north) ++(-2pc,0)
        to node [transform, above] {Gibbs} ++(+4pc,0);
    \draw [->, dotted, out=90, in=180] (conditional.north) ++(-2.5pc,0) to (misc);
    \draw [->] (conditional) -- node [transform, above] {integrate (\cref{s:high-dimensional})} (conditional -| patently.west);
    \draw [->] (patently) -- node [transform, right] {reduce (\cref{s:unproducts})} (simplified-patently);
    \draw [->] (simplified-patently.west |- simplified) -- node [transform, above] {recognize (\cref{s:unproducts})} (simplified);
    \node [fit=(m.west |- conditional.north) (simplified.south), inner sep=0, left delimiter=\{, label={[xshift=-1em, transform, align=right, inner sep=0]left:simplify\\(\cref{s:simplify})}] {};
    \draw [->] (simplified.south -| sham1) -- node [transform, right] {histogram (\cref{s:histogram})} (summarized.north -| sham1);
    \node [fit=(m.west |- summarized.south) (simplified.south), inner sep=0, left delimiter=\{, label={[xshift=-1em, transform, align=right, inner sep=0]left:histogram\\(\cref{s:histogram})}] {};
    \draw [->] (summarized.south -| sham1) -- node [transform, right] {A-normalization; loop-invariant code motion (\cref{s:licm})} (licm.north -| sham1);
    \draw [->] (licm.south -| sham1) -- node [transform, right] {loop fusion; lowering (\cref{s:licm})} (sham1);
    \draw [->] (sham1) -- node [transform, right, align=left] {common indexing-expression elimination (\cref{s:licm})} (sham2);
    \draw [->] (sham2) -- node [transform, right, align=left] {constant propagation; array pre-allocation (\cref{s:sham})} (sham3);
    \draw [->] (sham3) -- node [transform, below] {lowering} (llvm);
    \draw [->] (llvm) -- node [transform, below] {--O3} (code);
    \node [fit=(m.west |- llvm.south east) (summarized.south), inner sep=0, left delimiter=\{, label={[xshift=-1em, transform, align=right, inner sep=0]left:code gen\\(\cref{s:codegen})}] {};
    \node (static) at (sham2.east) [xshift=3pc, anchor=west, draw, align=center] {\strut static data such as array sizes};
    \draw [->] (static) to (sham2);
\end{tikzpicture}%
\DeclareRobustCommand\legendP{\begin{tikzpicture}[baseline] \node [doc, anchor=base, text depth=0] {\!probabilistic\!};\end{tikzpicture}}%
\DeclareRobustCommand\legendI{\begin{tikzpicture}[baseline] \node [ll, anchor=base, text depth=0] {imperative};\end{tikzpicture}}%
\DeclareRobustCommand\legendM{\begin{tikzpicture}[baseline] \node [maple, anchor=base] {\!math\vphantom{y}\!};\end{tikzpicture}}%
\DeclareRobustCommand\legendD{\begin{tikzpicture}[baseline] \node [draw, anchor=base] {data};\end{tikzpicture}}%
\caption{Our pipeline, compiling \legendP\ programs via \legendM\ into \legendI\ code to process \legendD}
\label{fig:pipeline}
\end{figure}

\Cref{fig:pipeline} lays out our compilation pipeline from model to
code.  Because this paper starts with the conditional distribution near
the top, it leaves open the issue of how to find the desired inference
algorithm.  After all, there is no single inference method
that works well for all models, and knowing what works well
takes domain expertise not available to a compiler.  In~\hakaru,
which our implementation builds on,
the inference distribution is typically produced
by metaprogramming constructs that form a directed graph of choices
\citep{shan-exact,narayanan-symbolic,zinkov-composing}, depicted
schematically at the very top of the figure.  In another context, the
inference distribution may be produced by hand.  Either way, producing
the inference distribution\diff{CE}{Explain exact computation on inference distributions for approximate inference} is outside the scope of this paper. Rather, the
contributions of this paper start as soon as the inference algorithm is
expressed as a conditional distribution, as it is naturally in the
literature.

The probabilistic IR in which we express and simplify inference distributions is the
\hakaru language.  Because \hakaru eschews general recursion and is typed
and terminating, all abstractions can be beta-reduced away near the
start of the pipeline, leaving a first-order core whose constructs
express mathematical operations and the measure monad
(\cref{fig:syntax}).  Other probabilistic languages that allow general
recursion may well profit from selectively applying our pipeline,
but that is outside the scope of this paper.

\paragraph{An elementary tour}

\diffopen{AC}{Changed the introductory example to match and tie
together \cref{s:simplify,s:histogram,s:codegen,s:eval}. Added more
explanation using concrete and elementary math.}%
We sketch a simple application to give an impression of the parts of our
pipeline.  The rest of the paper elaborates on this \emph{Gaussian
mixture} example.

Suppose we observe $n$ data points.  Each data point lies at some
location along the real line and belongs to one of $m$ classes, so we
store our observations in two arrays of size~$n$: the locations in
$\vec{s}\in \Real^n$ and the class labels in $\vec{y}\in
\{0,\dotsc,m-1\}^n$.  A~simple model of how these data points came to be
might say that each class has an underlying location, not directly
observed, and the location of each data point is a noisy measurement of
the underlying location of the class of the data point.  Whether we are
interested in estimating the locations of the classes or predicting the
locations of upcoming data points, the \emph{disintegration}
transformation can produce a probabilistic program that denotes the
distribution of our quantity of interest, conditioned on our
observations.  In this program, each possible underlying location of
a class is weighted, or scored, by the likelihood of the noisy
measurements that we observed from the class.

Starting with this probabilistic program for a conditional distribution,
our \emph{simplification} transformation produces a closed-form formula
that (a)~estimates the underlying location of each class
or predicts the location of each upcoming data point,
and (b)~evaluates how well the observed data fits the model.
As one might expect, part~(a) is based on the mean of the
data points in that class, and part~(b) is based on the
variance of the data points in that class.  Thus, the formula generated
by simplification contains sums such as
\begin{align}
\label{e:histo-sources}
    \sum\nolimits_{j=0}^{n-1} \pwow{\vec{s}\idx{j}^2}{i=\vec{y}\idx{j}}{0}
&&
    \sum\nolimits_{j=0}^{n-1} \pwow{\vec{s}\idx{j}  }{i=\vec{y}\idx{j}}{0}
&&
    \sum\nolimits_{j=0}^{n-1} \pwow{1               }{i=\vec{y}\idx{j}}{0}
\end{align}
where $i\in\{0,\dotsc,m-1\}$ is a class label.  Each of the three sums
take time $O(n)$ to compute, so looping over all $m$ classes takes
$O(mn)$ time.  Improving upon this situation, our \emph{histogram}
transformation discovers that each of the three sums can be computed for
all $m$ classes in a single pass through the data that creates an array
of size~$m$:
\begin{align}
\label{e:sketch-histograms}
    \arrayextrasep=0pt
    \Lets[:=]{
        \summary_2 & \newArray(m) \\
    }{\For{j}{0}{n-1}{
        \summary_2\idx{\vec{y}\idx{j}} \plusequal \vec{s}\idx{j}^2}}
&&
    \arrayextrasep=0pt
    \Lets[:=]{
        \summary_1 & \newArray(m) \\
    }{\For{j}{0}{n-1}{
        \summary_1\idx{\vec{y}\idx{j}} \plusequal \vec{s}\idx{j}}}
&&
    \arrayextrasep=0pt
    \Lets[:=]{
        \summary_0 & \newArray(m) \\
    }{\For{j}{0}{n-1}{
        \summary_0\idx{\vec{y}\idx{j}} \plusequal 1}}
\end{align}
Now the computation takes only $O(n)$ time for all $m$ classes.  Even
better, the code in~\labelcref{e:sketch-histograms} never materializes
because our \emph{domain-specific code generator} aggressively optimizes these
loops: it fuses the three loops into one, eliminates the common indexing
expressions $\vec{y}\idx{j}$ and $\vec{s}\idx{j}$, pre-allocates the
three output arrays, and JIT-specializes the machine code not only to
the given size~$m$ but also to the addresses of the pre-allocated
output arrays.

These aggressive optimizations might be overkill if we only need to
perform the computation once for a given size~$m$.  But the generated
code may well be the inner loop of an approximate inference algorithm
that solves a harder problem.  For example, suppose that the class
labels~$\vec{y}$ are not observed.  Then, a closed-form solution is no
longer available.  A~popular solution approach, called MCMC\@, is to
design a random walk among the $m^n$ possible values of~$\vec{y}$ that
approximates the target distribution.  The transition probabilities of
this random walk are calculated by repeating part~(b) above at
every step.  Hence these optimizations are worthwhile, and we
automate them.%
\diffclose{AC}

\section{Simplifying Array Programs}
\label{s:simplify}

To \emph{simplify} a probabilistic program is to produce a more efficient
(or readable) program while still representing the same
distribution.
\Citet{carette-simplifying-padl} introduced a simplifier that applies
computer algebra strategically to the linear
operator denoted by a probabilistic program: their simplifier eliminates
latent variables and recognizes conjugate likelihoods by exploiting
domain constraints.
We extend that simplifier to
handle probabilistic programs with arrays,
which naturally represent high- and arbitrary-dimensional distributions
that arise in inference algorithms.

Our extended
simplifier handles latent variables and conjugacy
by exploiting constraints on array indices.
A key part, the \emph{unproduct} operation~(\cref{s:unproducts}),
uncovers independence in the mathematical denotations of array
programs; this operation is derived from first principles and subsumes
AugurV2's rewrite rule for indirect indexing~\citep{huang-compiling}.
Without unrolling an array
or even knowing its concrete size, our
simplifier computes exact distribution parameters that recover
sufficient statistics such as sample mean, sample variance, and word
counts by document class.  These informative symbolic parameters let us
compile inference algorithms such as MCMC on
Dirichlet\hyp multinomial mixtures.

Simplification depends heavily on computer algebra.
Our extended simplifier is implemented in Maple, but
we do not rely on features specific to Maple, and
we have experimented with SymPy and obtained promising results.

The rest of this section uses
a progression of examples to explain what our
extended simplifier does, why it's useful, and how it works.
To pump intuition about
Bayesian inference, these examples use simplification as a form of exact
inference, even though simplification is also essential for efficient
approximate inference, as discussed in \cref{s:overview}.

\subsection{Background}
\label{s:background}

We tour \citearound{'s simplifier}\citet{carette-simplifying-padl} with
an example.
Consider the distribution over~$\Real^2$ generated~by
\begin{enumerate}
    \item drawing $x\in\Real$ from the normal distribution with some fixed
          mean~$\mu$ and standard deviation~$1$;
    \item drawing $y,z\in\Real$
          from the normal distribution with mean~$x$
          and standard deviation~$1$; and
    \item returning the pair $\literalArray{y,z}$.
\end{enumerate}
These steps model two noisy measurements $y,z$ of the unknown location~$x$ of a
particle along the real line.  To model that we do not directly observe the
location~$x$, the returned outcome $\literalArray{y,z}$ omits~$x$, and we
say that the random variable~$x$ is \emph{latent}.
We represent this distribution by the term
\begin{equation}
\label{e:starting-pre}
        \Bind(\Gaussian(\mu,1),x,
        \Bind(\Gaussian(x,1),y,
        \Bind(\Gaussian(x,1),z,
        \Ret(\literalArray{y,z}))))
        \text,
\end{equation}
in which $\mu$ is a free variable, and $x,y,z$ are bound and take scope
to their right.  To create generative processes, we use two
\diffopen{C}{Clarified relation to \citealp{giry-categorical,ramsey-stochastic,staton-commutative}}constructs of \citearound{'s monad of probability
distributions}\citet{giry-categorical} (which was popularized by
\citet{ramsey-stochastic}):
\begin{itemize}
    \item $\Ret(e)$ produces the outcome $e$ deterministically.
    \item $\Bind(m,x,m')$ carries out the process~$m$ (such as the
          primitive distribution $\Gaussian(\mu,1)$) and binds the
          outcome to the variable~$x$ then carries out~$m'$ to get the
          final outcome.
\end{itemize}
\Cref{fig:syntax} shows the essential part of the language. We write the
informal type $\Measure T$ for distributions (measures) over the type~$T$.
Whereas \citeauthor{giry-categorical}'s and
\citeauthor{ramsey-stochastic}'s works concerned probability
distributions, our language includes the $\Weight(e,m)$ construct for
weighting samples or scaling distributions.  Because the weight~$e$ is
not bounded, our language can express not just probability or
sub-probability distributions but the more general class of
\emph{s-finite} measures \citep{staton-commutative}.\diffclose{C}

\input{syntax-arrays}

One way to interpret the term~\labelcref{e:starting-pre} is as a monadic program
that samples three random numbers each time it is run.
But before running the program,
we can first use \citearound{'s simplifier}\citet{carette-simplifying-padl} to
turn it into
\begin{equation}
\label{e:starting-post}
    \textstyle
        \Bind(\smash{\Gaussian(\mu,\sqrt{2})},y,
        \Bind(\smash{\Gaussian(\frac{1}{2}(\mu+y),\frac{\sqrt{6}}{2})},z,
        \Ret(\literalArray{y,z})))
        \text.
\end{equation}
The latent variable~$x$ was eliminated, and
the distributions of $y$ and~$z$ adjusted accordingly.
Compared to the program~\labelcref{e:starting-pre},
the program~\labelcref{e:starting-post} makes fewer random choices yet
produces the same distribution.  That is, the two programs are
equivalent if we interpret $\Measure$ as the distribution monad,
but \labelcref{e:starting-post} uses randomness more efficiently if we
interpret $\Measure$ as the sampling monad \citep{ramsey-stochastic}.
Moreover, we can read off from the form
of~\labelcref{e:starting-post} exactly how to perform a kind of Bayesian
\emph{inference}:\diff{D}{Rephrased ``enable inference''}  If~we have measured $y$ but not~$z$, we can
predict~$z$ using
\begin{equation}
\label{e:starting-conditional}
    \textstyle
    \Gaussian(\frac{1}{2}(\mu+y),\frac{\sqrt{6}}{2})
    \text,
\end{equation}
a subterm of~\labelcref{e:starting-post}.  (In particular, we can
estimate~$z$ using the mean $\frac{1}{2}(\mu+y)$.)  That is, the simplifier
has computed \labelcref{e:starting-conditional} to be the
\emph{conditional} distribution of~$z$ given~$y$ in our model.

To pump intuition about Bayesian inference, we ordered the random
variables $x,y,z$ in \labelcref{e:starting-pre} so that simplification
produces a conditional distribution~\labelcref{e:starting-conditional}.
If we had commuted the bindings of $y$
and~$z$, then simplification would instead produce the
conditional distribution of~$y$ given~$z$.  This illustrates
that simplification, like a typical optimization pass, is sensitive to
syntactic choices in semantically equivalent inputs, even though it
preserves semantics.

We now zoom into how simplification works.
\Cref{fig:pipeline} illustrates the structure of
\citearound{'s simplifier}\citet{carette-simplifying-padl}, whose parts we
extend with arrays.  It turns \labelcref{e:starting-pre}
into~\labelcref{e:starting-post} by three steps.

First, the simplifier converts the program~\labelcref{e:starting-pre} into
\begin{gather}
\label{e:starting-lo}
\let\cdot\relax
    \int_\Real
    \int_\Real
    \int_\Real
    \frac{\Exp{-\Frac{(x-\mu)^2}{2}}}{\sqrt{2\cdot\pi}} \cdot
    \frac{\Exp{-\Frac{(y-x)^2}{2}}}{\sqrt{2\cdot\pi}} \cdot
    \frac{\Exp{-\Frac{(z-x)^2}{2}}}{\sqrt{2\cdot\pi}} \cdot
    \mathinner{\anintegrand(\literalArray{y,z})}
    dz\,dy\,dx
    \text.
\raisetag{2ex}
\end{gather}
This quantity is the \emph{expectation} of an arbitrary measurable\diff{D}{Added ``measurable'' restriction} function
$\anintegrand : \Real^2\to\rplus$ with respect to the distribution.
In other words, the simplifier interprets~$\Measure$ as the expectation
monad \citep{ramsey-stochastic}.
The expectation~\labelcref{e:starting-lo} is linear in~$\anintegrand$.
To understand this integral, consider when
$\anintegrand(\literalArray{y,z}) = \bigl\{\begin{smallmatrix}1&\literalArray{y,z}\in
S\\0&\text{otherwise}\end{smallmatrix}$ for some $S\subseteq\Real^2$; the integral is then just the
probability of~$S$.
Each factor in~\labelcref{e:starting-lo}, such as
$\frac{\Exp{-\Frac{(x-\mu)^2}{2}}}{\sqrt{2\pi}}$, is
the density of a primitive distribution, here $\Gaussian(\mu,1)$ at~$x$.

Second, noticing that the variable~$x$ is latent (that is, no argument
to~\anintegrand\ contains~$x$ free), the simplifier
symbolically integrates over~$x$ to get
\begin{equation}
\label{e:starting-loI}
\let\cdot\relax
    \int_\Real
    \int_\Real
    \frac{\Exp{-\Frac{\mu^2}{3}} \cdot
          \Exp{-\Frac{y^2}{3}} \cdot \Exp{\Frac{\mu\cdot y}{3}} \cdot
          \Exp{-\Frac{z^2}{3}} \cdot \Exp{\Frac{\mu\cdot z}{3}} \cdot \Exp{\Frac{y\cdot z}{3}}
    }{2\cdot\sqrt{3}\cdot\pi} \cdot
    \mathinner{\anintegrand(\literalArray{y,z})}
    dz\,dy
\text.
\end{equation}

Third, inverting the first step, the simplifier converts
\labelcref{e:starting-loI} back to a program,
namely~\labelcref{e:starting-post}.  \diffopen{CD}{Explain holonomic representation more}This conversion requires the
simplifier to recognize that a factor, such as the fraction
in~\labelcref{e:starting-loI}, is~the density of a primitive
distribution, here~\labelcref{e:starting-conditional} at~$z$.
Recognizing a factor as the density of a distribution subsumes
recognizing the conjugacy of a likelihood with respect to
a distribution.  To recognize a factor by matching it against syntactic
patterns would be brittle and ad-hoc.  Instead, the simplifier
characterizes the factor~$f(z)$ by its \emph{holonomic}
representation \citep{WilfZeil1992,ChSa98}, a first-order linear differential equation
(here $3f'(z)=(\mu+y-2z)f(z)$) whose coefficients (here $3$ and $\mu+y-2z$)
are polynomials in~$z$.

Fortunately, functions with holonomic representations constitute a large
class with useful closure properties, such as closure under integration,
differentiation, and composition with algebraic functions
\citep{Kauers2013}.  Taking advantage of these closure properties, the
simplifier computes the holonomic representation from the factor
expression compositionally and not by pattern matching.
Moreover, because the coefficients are polynomials,
their ratios can be matched efficiently and
robustly using existing algorithms such as Euclid's algorithm.\diffclose{CD}
Therefore, this third and final step of the simplifier is
robust against syntactic perturbations,
general across primitive distributions, and
modular so that implementing each primitive distribution separately
suffices for conjugacy relationships among them to be recognized
\citep{carette-simplifying-padl}.



The first of the three steps,
$\integrate(m,\anintegrand)$, produces
an expression patently linear in~$\anintegrand$ by
structural induction on the program~$m$.
The expression produced by $\integrate(m,\anintegrand)$ denotes the
\emph{expectation}, or \emph{Lebesgue integral}, of the function~$\anintegrand$ with
respect to the distribution~$m$; for example,
$\integrate(\labelcref{e:starting-pre},\anintegrand)$
produces~\labelcref{e:starting-lo}, and
$\integrate(\labelcref{e:starting-post},\anintegrand)$ produces something that expands
to~\labelcref{e:starting-loI}.  Because distributions~$m$ and the linear
operators $\fun\anintegrand\integrate(m,\anintegrand)$ are in
one-to-one correspondence \citep[Section 2.3]{pollard-measure}, any
simplification of $\integrate(m,\anintegrand)$ that preserves its
meaning also preserves the distribution denoted by~$m$.
But feeding~\labelcref{e:starting-lo} willy-nilly to a computer algebra
system will not out-of-the-box improve it to~\labelcref{e:starting-loI}
and may even make it worse.  Instead,
\citearound{'s simplifier}\citet{carette-simplifying-padl} operates strategically on parts
of a patently linear expression, guided by the grammar in
\input{patently-linear}\cref{fig:patently-linear}.

\subsection{Scalar Simplification Is Not Enough}
\label{s:not-enough}

Given that \citearound{'s simplifier}\citet{carette-simplifying-padl}
works on scalar probabilistic programs, one might hope that array
probabilistic programs can be simplified by applying the same simplifier
to scalars in loop bodies.  Unfortunately, the array programs that
express desired inference algorithms require extending the simplifier at
the level of mathematical denotations, not just applying it
strategically at the level of source programs.  Before describing our
extended simplifier, we motivate it with four increasingly tricky
examples.  Along the way, we introduce the array constructs of our
language.

We begin with an example of an array program that is trivial to handle
using the scalar simplifier.  The distribution over $\Real^2$ in
\cref{s:background} generalizes to one over $\Real^{2n}$, generated~by
repeating the following for $i=0,\dotsc,n-1$:
\begin{enumerate}
    \item drawing $x\in\Real$ from the normal distribution with some fixed
          mean~$\mu$ and standard deviation~$1$;
    \item drawing $y,z\in\Real$
          from the normal distribution with mean~$x$
          and standard deviation~$1$; and
    \item returning the pair $\literalArray{y,z}$.
\end{enumerate}
This distribution models $2n$ noisy measurements of
the unknown locations of $n$ particles along the real line.
Because the loop body returns a pair of reals, the loop returns an array
of $n$ pairs of reals.  We represent this distribution by
\begin{equation}
\label{e:fused}
    \Plate(n,i,
        \Bind(\Gaussian(\mu,1),x,
        \Bind(\Gaussian(x,1),y,
        \Bind(\Gaussian(x,1),z,
        \Ret(\literalArray{y,z}))))
    )
    \text.
\end{equation}
The new construct $\Plate$ forms a monadic loop: the variable~$n$ above
is free like~$\mu$ and denotes an arbitrary iteration count, and the
variable~$i$ is an index that takes scope over the monadic action to its right.
In general, $\Plate(n,i,m)$ is a monadic action whose outcome is
an array of $n$ elements,
independently drawn from the distributions $\subst{0}{i}{m}, \penalty0 \dotsc, \penalty0
\subst{n-1}{i}{m}$.  (Indices begin at~$0$.)  The informal typing rule
for $\Plate$ in \cref{fig:syntax} says accordingly that if $m$ has type
$\Measure T$ then $\Plate(n,i,m)$ has type $\Measure(\Array T)$, where
$\Array T$ means arrays of~$T$.  This $\Plate$ construct is named after
\emph{plate notation} for repetition in Bayes nets
\citep{buntine-operations,koller-pgm}.
It is like \texttt{Data.Vector.generateM} in Haskell, but
since each array element is drawn
independently, a $\Plate$ is a \emph{parallel}
comprehension~\citep{huang-compiling}.

Of course, we can apply the scalar simplifier to the subexpression
\labelcref{e:starting-pre} in \labelcref{e:fused}, and the result is an
improvement for the same reasons as for~\labelcref{e:starting-pre}: it
makes fewer random choices ($2n$ instead of $3n$) and enables
probabilistic inference (from measuring each~$y$ to predicting
each~$z$).

\paragraph{But pointwise simplification is not enough}

It is just as natural to express essentially the same
distribution by multiple loops: we can generate a pair of arrays of $n$
reals~by
\begin{enumerate}
    \item drawing $\vec{x}\idx{i}\in\Real$ from
          the normal distribution with mean~$\mu$ and
          standard deviation~$1$, for $i=0,\dotsc,n-1$;
    \item drawing $\vec{y}\idx{i},
          \vec{z}\idx{i}\in\Real$ from
          the normal distribution with mean~$\vec{x}\idx{i}$ and
          standard deviation~$1$, for $i=0,\dotsc,n-1$; and
    \item returning the pair $\literalArray{\vec{y},\vec{z}}$.
\end{enumerate}
We use accents on the three variables $\vec{x},\vec{y},\vec{z}$
to remind ourselves that they denote arrays, so
their type is $\Array\Real$, and
element~$i$ of $\vec{x}$ is $\vec{x}\idx{i}$, not $x\idx{i}$.
Again using $\Plate$, we represent this distribution by
\begin{equation}
\label{e:plate2-pre}
    \begin{array}{@{}l@{}}
        \Bind(\Plate(n,i,\Gaussian(\mu,1)),\vec{x}, \\
        \Bind(\Plate(n,i,\Gaussian(\vec{x}\idx{i},1)),\vec{y}, \\
        \Bind(\Plate(n,i,\Gaussian(\vec{x}\idx{i},1)),\vec{z},
        \Ret(\literalArray{\vec{y},\vec{z}}))))
    \end{array}
\end{equation}
and we want to simplify this probabilistic program to
\begin{equation}
\label{e:plate2-post}
  \begin{array}[t]{@{}l@{}}
    \Bind(\smash{\Plate(n,i,\Gaussian(\mu,\sqrt{2}))},\vec{y}, \\
    \Bind(\smash{\Plate(n,i,\Gaussian(\frac{1}{2}(\mu+\vec{y}\idx{i}),\frac{\sqrt{6}}{2}))},\vec{z},
    \Ret(\literalArray{\vec{y},\vec{z}})))
    \text.
  \end{array}
\end{equation}
Before we can apply the scalar simplifier, we seem to have to first fuse
the three $\Plate$ loops in~\labelcref{e:plate2-pre}, to form a single
loop body to simplify.

\paragraph{Loop fusion is still not enough}

Fusing loops may seem promising, but the following richer classic
example illustrates the broader variety of array programs that
simplification ought to improve.
Suppose we would like to model \diffopen{E}{Better introduce GMM}$n+1$ data points drawn from
a \emph{mixture} of $m$ normal distributions.  Each component~$i$ of the
mixture might represent a different subpopulation, such as researchers
of different specialties.
A~\emph{Gaussian mixture} distribution~\citep{Pearson71} can be
generated by
\begin{enumerate}
    \item \label{loop:theta}drawing the \emph{mixture weights} \smash{$\vec{\theta}$}, an array of $m$ non-negative reals
          that sum to~$1$, from some \emph{Dirichlet} distribution;
    \item \label{loop:x}drawing $m$ \emph{component means} $\vec{x}\idx{i}$ from
          $\Gaussian(\mu,\sigma)$, for $i=0,\dotsc,m-1$;
    \item \label{loop:y}drawing $n$ class labels
          $\vec{y}\idx{j}\in\{0,\dotsc,m-1\}$ from the discrete
          distribution~\smash{$\vec{\theta}$},
          for $j=0,\dotsc,n-1$;
    \item \label{loop:s}drawing $n$ data points
          $\vec{s}\idx{j}$ from
          $\Gaussian(\vec{x}\idx{\vec{y}\idx{j}}, 1)$,
          for $j=0,\dotsc,n-1$;
    \item \label{loop:z}drawing one more class label $z\in\{0,\dotsc,m-1\}$ from the discrete
          distribution~\smash{$\vec{\theta}$};
    \item \label{loop:t}drawing one more data point $t$ from
          $\Gaussian(\vec{x}\idx{z},1)$;\diffclose{E} and
    \item returning the tuple $\literalArray{\vec{y},\vec{s},z,t}$.
\end{enumerate}
By first drawing the random indices $\vec{y},z$ then using those
class labels to decide which means in~$\vec{x}$ to draw
$\vec{s},t$ around, this process models how different subpopulations
share different characteristics.  Again, we want to automate the process
by which human experts simplify this program to make fewer choices
(eliminating \smash{$\vec\theta,\vec{x}$}) and enable inference
(predicting $z,t$ from $\vec{y},\vec{s}$).

We first examine how to eliminate~$\vec{x}$, then turn to
eliminating~\smash{$\vec{\theta}$}.  At first glance, it is not obvious
how to eliminate the latent array variable~$\vec{x}$, because the loop
where $\vec{x}$ is drawn and the loop where $\vec{x}$ is used (to
draw~$\vec{s}$) range over different domains ($i=0,\dotsc,m-1$ and
$j=0,\dotsc,n-1$) and cannot be fused.  However, we can group the
iterations of the latter loop by which element of~$\vec{x}$ they use:
each $\vec{x}\idx{i}$ is used to draw exactly those $\vec{s}\idx{j}$ for
which $i=\vec{y}\idx{j}$.  In other words,\diff{D}{Added grouping explanation} we can group the elements
of~$\vec{s}$ by how they are classified in~$\vec{y}$.
Hence we can transform steps
\labelcref{loop:x,loop:s,loop:t} above into a single loop that,
informally speaking, repeats the following for $i=0,\dotsc,m-1$:
\begin{enumerate}
    \item[($\ref{loop:x}'$)] drawing $\vec{x}\idx{i}$ from
          $\Gaussian(\mu,\sigma)$;
    \item[($\ref{loop:s}'$)] drawing $\vec{s}\idx{j}$ from
          $\Gaussian(\vec{x}\idx{i}, 1)$,
          for each $j=0,\dotsc,n-1$ such that $i=\vec{y}\idx{j}$; and
    \item[($\ref{loop:t}'$)] drawing $t$ from
          $\Gaussian(\vec{x}\idx{i},1)$ if $i=z$.
\end{enumerate}
Because each $\vec{x}\idx{i}$ drawn in the new step~$\ref{loop:x}'$ is
used only in steps $\ref{loop:s}'$ and~$\ref{loop:t}'$ in the same
iteration over~$i$ and not beyond, scalar simplification can
eliminate~$\vec{x}\idx{i}$.  More formally, eliminating
each~$\vec{x}\idx{i}$
requires performing the integral $\int_\Real
\Exp{f(\vec{x}\idx{i})}\,d\vec{x}\idx{i}$,
whose integrand $\Exp{f(\vec{x}\idx{i})}$ multiplies together the
densities of the same $\Gaussian(\vec{x}\idx{i},1)$ at all the elements
of~$\vec{s}$ whose classification in~$\vec{y}$ is~$i$.\diff{D}{Added grouping explanation}
Because just one $\Gaussian(\vec{x}\idx{i},1)$ is involved,
the exponent
\begin{equation}
f(x) =
        \sum_{j=0}^{n-1}
        \pwow{-\Frac{(\vec{s}\idx{j}-x)^2}{2}\!}
             {i = \vec{y}\idx{j}}
             {0}
=
        -\Frac{\biggl(
               \sum_{j=0}^{n-1}
               \pwow{\vec{s}\idx{j}^2\!}{i = \vec{y}\idx{j}}{0}
               \biggr)}{2}
        + x
        \biggl(
        \sum_{j=0}^{n-1}
        \pwow{\vec{s}\idx{j}\!}{i = \vec{y}\idx{j}}{0}
        \biggr)
        -\Frac{x^2
               \biggl(
               \sum_{j=0}^{n-1}
               \pwow{1}{i = \vec{y}\idx{j}}{0}
               \biggr)}{2}
\label{e:gmm-unproduct-post}
\end{equation}
depends on just one element~$x$ of~$\vec{x}$ at a time.  The result of
the integration is expressed in terms of the three summations in the
right-hand side of~\labelcref{e:gmm-unproduct-post}
(same as~\labelcref{e:histo-sources}).  They are the
square-sum, sum, and count of just those elements of~$\vec{s}$ labeled
by~$\vec{y}$ to belong to class~$i$; these summations recover the
sufficient statistics of the input data.  Thus, simplifying array
programs requires extracting per-element formulas such as
\labelcref{e:gmm-unproduct-post} and conjuring the conditionals therein
to preserve semantics.

\paragraph{Even iteration reordering is not enough}

Grouping loop iterations in the source program is enough to eliminate
the latent variable~$\vec{x}$ but not~\smash{$\vec\theta$}.  To explain
why, we first need to explain what Dirichlet distributions are.
Dirichlet distributions are a family of distributions over arrays of
non-negative numbers that sum to~1\diff{C}{Improved weird wording} (the informal type is
$\Measure(\Array\rplus)$).  For simplification, we expand
step~\labelcref{loop:theta} above, ``draw \smash{$\vec\theta$} from some
Dirichlet distribution'', as a macro to the following:
\begin{enumerate}
    \item[(\labelcref{loop:theta}a)]
        drawing $\vec{p}\idx{i} \in [0,1]$ from some \emph{Beta}
        distribution, for $i=0,\dotsc,m-2$; and
    \item[(\labelcref{loop:theta}b)]
        returning the array $\smash{\vec\theta}=\literalArray{
            \begin{tabular}[t]{@{}Ml@{}}
            1-\vec{p}\idx{0},\\
            \vec{p}\idx{0} \cdot (1-\vec{p}\idx{1}),\\
            \vec{p}\idx{0} \cdot \vec{p}\idx{1} \cdot (1-\vec{p}\idx{2}),\\
            \vec{p}\idx{0} \cdot \vec{p}\idx{1} \cdot \vec{p}\idx{2} \cdot (1-\vec{p}\idx{3}),\\
            \dotsc,\\
            \vec{p}\idx{0} \cdots \vec{p}\idx{m-3} \cdot (1-\vec{p}\idx{m-2}),\\
            \vec{p}\idx{0} \cdots \vec{p}\idx{m-3} \cdot \vec{p}\idx{m-2}
        }\text.\end{tabular}$\\
	(Here we notate an array by a bracketed list of elements.)
\end{enumerate}
This expansion is a well-known, finite\hyp dimensional variant of the
\emph{stick-breaking process} \citep[page 583]{gelman-bayesian}.  The
intuition behind the name is to start with a stick of length~$1$ and
break off a piece of proportion $\vec{p}\idx{0}$, then from that piece
break off a piece of proportion $\vec{p}\idx{1}$ (that is, of length
$\vec{p}\idx{0}\cdot\vec{p}\idx{1}$), then from \emph{that}
piece break off a piece of proportion $\vec{p}\idx{2}$ (that is, of
length $\vec{p}\idx{0}\cdot\vec{p}\idx{1}\cdot\vec{p}\idx{2}$), and so
on.\footnote{We use $m-1$ Beta distributions, not
$m$ Gamma distributions, even though normalizing an array of
$m$ independent Gamma variables is another well-known way to obtain the
Dirichlet distribution.  The reason is that every element of the
normalized array depends on every element of the unnormalized array, so
this more symmetric way to obtain the Dirichlet distribution actually
makes it harder to eliminate \smash{$\vec{\theta}$} and to recognize the
conjugacy of $\vec{y}$ and~$z$.}\diff{D}{Explain why not use $m$ Gamma distributions}
We represent this process by the term\diffopen{E}{Name elided exponents}
\begin{equation}
\label{e:dirichlet}
    \textstyle
    \Bind\bigl(\Plate\bigl(m-1,i,\BetaD(\alpha(i)+1,\beta(i)+1)\bigr),\vec{p},
    \Ret\bigl(\ary\bigl(m,i,
        \bigl(\prod_{k=0}^{i-1}\vec{p}\idx{k}\bigr)
        \cdot
        \pw{1-\smash[t]{\vec{p}\idx{i}}}{i<m-1}{1}{i=m-1}
    \bigr)\bigr)\bigr)
    \text,
\end{equation}
where $\alpha(i)+1$ and $\beta(i)+1$ are parameters of the $\BetaD$
distribution that may depend on~$i$.\diffclose{E}
Here $\ary$ is an array comprehension construct: the term $\ary(m,i,e)$
denotes an array of size~$m$ whose elements are $\subst{0}{i}{e},
\dotsc, \subst{n-1}{i}{e}$.  As the informal typing rules in
\cref{fig:syntax} show, the difference between $\ary$ and $\Plate$ is
that $\ary$ is non-probabilistic: it neither requires nor produces
a distribution (like \texttt{Data.Vector.generate} in Haskell).
Hence $\ary(m,i,e)\idx{e'}$ reduces to
$\subst{e'}{i}{e}$.\footnote{We leave the meaning of indexing out of
bounds undefined.}

Eliminating the latent array variable \smash{$\vec{\theta}$} is trickier
than eliminating~$\vec{x}$.  Because most elements
of~\smash{$\vec\theta$} use multiple elements of~$\vec{p}$, we cannot
eliminate~\smash{$\vec\theta$} just by reordering the iterations of the
loop where \smash{$\vec\theta$} is used (step~\labelcref{loop:y} above).
Rather, we need to work with mathematical denotations underlying the
source program.  Eliminating~\smash{$\vec\theta$} amounts to performing
the $(m-1)$-dimensional integral
\begin{equation}
\label{e:dirichlet-integral}
    \int_{\Real^{m-1}}
    \smash[b]{
        \underbrace{
            \biggl(\smash{\prod_{i=0}^{m-2}}
                \vec{p}\idx{i}^{\alpha(i)}
                (1
                 \mskip.5\thinmuskip\mathord-\mskip.5\thinmuskip
                 \vec{p}\idx{i})^{\beta(i)}
            \biggr)
        }_{\text{step \labelcref{loop:theta}}}
        \underbrace{
            \biggl(\smash{\prod_{j=0}^{n-1}}\textstyle
                \bigl(\prod_{k=0}^{\vec{y}\idx{j}-1}\vec{p}\idx{k}\bigr)
                \!
                \pw{1-\smash[t]{\vec{p}\idx{\vec{y}\idx{j}}}}{\smash[t]{\vec{y}\idx{j}}<m-1}{1}{\smash[b]{\vec{y}\idx{j}}=m-1}
            \biggr)
        }_{\text{step \labelcref{loop:y}}}
        \underbrace{\textstyle
            \vphantom{\biggl(}
            \bigl(\prod_{k=0}^{z-1}\vec{p}\idx{k}\bigr)
            \!
            \pw{1-\smash[t]{\vec{p}\idx{z}}}{z<m-1}{1}{z=m-1}
        }_{\text{step \labelcref{loop:z}}}
    }
    \,d\vec{p}
    \vphantom{\underbrace{\biggl(}}
    \text.
\end{equation}
The factors in the integrand arise from steps
\labelcref{loop:theta,loop:y,loop:z} above and the macro
expansion~\labelcref{e:dirichlet}.  The way to calculate this integral
is to group the factors $\vec{p}\idx{k}$,
$1-\vec{p}\idx{\vec{y}\idx{j}}$, and $1-\vec{p}\idx{z}$ by which
elements of~$\vec{p}$ they use.  For instance, the factors
\smash[t]{$\prod_{k=0}^{\smash[b]{\vec{y}}\idx{j}-1}\vec{p}\idx{k}$}
include $\vec{p}\idx{i}$ at exactly those $j$ for which
$i<\vec{y}\idx{j}$.  Thus, the integral \labelcref{e:dirichlet-integral}
can be rewritten to the form\diffopen{E}{Name elided exponents}
\begin{equation}
\label{e:dirichlet-integral-unproduct}
    \int_{\Real^{m-1}}
        \biggl(\prod_{i=0}^{m-2}
            \vec{p}\idx{i}^{\alpha'(i)}
            (1-\vec{p}\idx{i})^{\beta'(i)}
        \biggr)
    d\vec{p}
    =
    \prod_{i=0}^{m-2} \int_\Real
            \vec{p}\idx{i}^{\alpha'(i)}
            (1-\vec{p}\idx{i})^{\beta'(i)}
    d\vec{p}\idx{i}
    \text,
\end{equation}
whose exponents
\begin{flalign}
    \;
    \alpha'(i)
    &=
        \!\! \underbrace{
            \vphantom{\biggl(}
            \alpha(i)
        }_{\text{step \labelcref{loop:theta}}} \!\!
    +
        \underbrace{
            \biggl(\smash{\sum_{j=0}^{n-1}\pwow{1}{i<\smash[t]{\vec{y}\idx{j}}}{0}}\biggr)
        }_{\text{step \labelcref{loop:y}}}
    +
        \underbrace{
            \mathopen{\vphantom{\biggl(}}
            \pwow{1}{i<z\vphantom{y}}{0}
        }_{\text{step \labelcref{loop:z}}}
&
    \beta'(i)
    &=
        \!\! \underbrace{
            \vphantom{\biggl(}
            \beta(i)
        }_{\text{step \labelcref{loop:theta}}} \!\!
    +
        \underbrace{
            \biggl(\smash{\sum_{j=0}^{n-1}\pwow{1}{i=\smash[t]{\vec{y}\idx{j}}<m-1}{0}}\biggr)
        }_{\text{step \labelcref{loop:y}}}
    +
        \underbrace{
            \mathopen{\vphantom{\biggl(}}
            \pwow{1}{i=z\vphantom{y}<m-1}{0}
        }_{\text{step \labelcref{loop:z}}}
\end{flalign}
have absorbed
terms from steps \labelcref{loop:y,loop:z}.\diffclose{E}\footnote{This
absorption can also be viewed as the conjugacy of binomial likelihoods with
respect to Beta distributions.}
The right-hand side of~\labelcref{e:dirichlet-integral-unproduct} is
a product of \emph{independent} one-dimensional integrals that existing
computer algebra can finally calculate.\footnote{Expressing
a multivariate distribution by transforming an array of independent
random variables is a general strategy that may apply beyond Dirichlet
distributions.  For example, it~is promising to express a multivariate
Gaussian distribution by transforming an array of independent
one-dimensional Gaussian random variables, but we have only tried it
(successfully) with known (non-diagonal) covariance
matrices.}\diff{D}{Discuss generalizing beyond Dirichlet distributions}

In sum, simplifying an adequate variety of array programs
requires representing high- and arbitrary-dimensional integrals
and uncovering independence among the dimensions that is not necessarily expressible
at the source level.  We flesh out this approach below.
It succeeds on all the examples above.

\subsection{High- and Arbitrary-Dimensional Integrals}
\label{s:high-dimensional}

Our simplifier handles arrays by converting them to high- and arbitrary-dimensional integrals.
It \nolinebreak[3] takes the same three steps as \citearound{'s scalar
simplifier}\citet{carette-simplifying-padl}.  We illustrate these steps using
the relatively simple example~\labelcref{e:plate2-pre} above.  First, our
simplifier converts~\labelcref{e:plate2-pre} into the expression
\begin{equation}
\label{e:plate2-lo}
    \int_{\Real^n}
    \biggl(\prod_{i=0}^{n-1}\frac{\Exp{-\Frac{(\vec{x}\idx{i}-\mu)^2}{2}}}{\sqrt{2\pi}}\biggr)
    \int_{\Real^n}
    \biggl(\prod_{i=0}^{n-1}\frac{\Exp{-\Frac{(\vec{y}\idx{i}-\vec{x}\idx{i})^2}{2}}}{\sqrt{2\pi}}\biggr)
    \int_{\Real^n}
    \biggl(\prod_{i=0}^{n-1}\frac{\Exp{-\Frac{(\vec{z}\idx{i}-\vec{x}\idx{i})^2}{2}}}{\sqrt{2\pi}}\biggr)
    \mathinner{\anintegrand(\literalArray{\vec{y},\vec{z}})}
    d\vec{z}\,d\vec{y}\,d\vec{x}
    \text.
\end{equation}
Second, it integrates over the latent variable~$\vec{x}$ to get
\begin{multline}
\label{e:plate2-loI}
    \smash[b]{\int_{\Real^n} \int_{\Real^n}}
    2^{-n} 3^{-\Frac{n}{2}} \pi^{-n}
    \Exp{-\Frac{n\mu^2}{3}}
    \Exp{-\Frac{\sum_{i=0}^{n-1}\vec{y}\idx{i}^2}{3}}
    \Exp{\Frac{\mu\sumvec{y}}{3}}
    \Exp{-\Frac{\sum_{i=0}^{n-1}\vec{z}\idx{i}^2}{3}}
    \Exp{\Frac{\mu\sumvec{z}}{3}}
    \Exp{\Frac{\sum_{i=0}^{n-1}\vec{y}\idx{i}\vec{z}\idx{i}}{3}}
\\[-1ex]
    \mathinner{\anintegrand(\literalArray{\vec{y},\vec{z}})}
    d\vec{z}\,d\vec{y}
    \text.
\end{multline}
Third, it converts this expression back to the
program~\labelcref{e:plate2-post}.

Although conceptually straightforward, extending these three steps to
handle arrays is challenging because computer algebra systems today only
support integrals whose dimensionality is low and known, not high and
arbitrary.  Even just to represent the integrals---let alone compute
with them---we had to extend the language of expressions.

Our representation for high- and arbitrary-dimensional
integrals appears at the end of \cref{fig:patently-linear}:
\begin{align}
    g & \Coloneqq \dotsb \altern \textstyle \int_\aspace g\,d\vecdotsvec{x}
    & \aspace & \Coloneqq (a,b) \altern \textstyle \prod_{i=c}^d \aspace
\end{align}
Whereas in $\int_a^b g\,dx$ the variable~$x$ ranges over reals,
in $\int_\aspace
g\,d\vecdotsvec{x}$ the variable~$\vecdotsvec{x}$ ranges
over arrays of (arrays of~\dots) reals.  The space~$\aspace$
is either a real interval $(a,b)$ or a Cartesian
product $\prod_{i=c}^d \anotherspace(i)$
indexed by integers~$i$ between $c$ and~$d$.  For example,
$\smash[b]{\int_a^b f(x) \,dx}$ is equivalent to
$\smash[b]{\int_{(a,b)} f(x) \,dx}$, and
\(
    \int_{a_0}^{b_0}
    \int_{a_1}^{b_1}
    \int_{a_2}^{b_2}
    f(\literalArray{x,y,z})
    \,dz\,dy\,dx
\)
is equivalent to
\(
    \int_{\prod_{i=0}^2 (a_i,b_i)} f(\vec{x}) \,d\vec{x}
\).

\diffopen{E}{Replaced ``a huge jungle of ellipses'' by an explicit metavariable $B$}In the integral $\smash{\int_\aspace g\,d\vecdotsvec{x}}$ over the space
$X = \smash[b]{\prod_{i_1=c_1}^{d_1} \dotsm \prod_{i_r=c_r}^{d_r}(a,b)}$,
the set of valid indices into the array~$\vecdotsvec{x}$ is determined by
the sequence of index-variable bindings
$[\Loop{i_1}{c_1}{\smash[b]{d_1}},\dotsc,\Loop{i_r}{c_r}{\smash[b]{d_r}}]$.
We~notate this sequence of name-bounds pairs by the metavariable~$B$,
then define the syntactic sugar
\begin{align}
    \textstyle \prod_B X
    &= \textstyle \prod_{i_1=c_1}^{d_1} \dotsm \prod_{i_r=c_r}^{d_r} X\text, &
    \ary(B,e)
    &= \ary(d_1-c_1+1, i_1, \dotsc \ary(d_r-c_r+1, i_r, e) \dotsc)\text,
\\
    \textstyle \prod_B e
    &= \textstyle \prod_{i_1=c_1}^{d_1} \dotsm \prod_{i_r=c_r}^{d_r} e\text, &
    e\idx{B}
    &= e\idx{i_1+c_1}\dotso\idx{i_r+c_r}\text.
\end{align}

\input{forth-arrays}

Our new first step is defined using this new notation.
\Cref{fig:forth-arrays} shows the key cases.  In the call
$\integrate(m,B,h)$, the second argument~$B$ is a new accumulator that
tracks the
$\Plate$ levels nested around~$m$.  This list starts empty, and
grows when $\integrate$ encounters $\Plate$.  When $\integrate$
arrives at a primitive distribution such as $\Gaussian$, it
generates an integral whose body nests as many
definite products as the list is long.

Our second step seeks to eliminate latent array variables by integrating
over them.
In~\labelcref{e:plate2-lo} for example, we seek to integrate
$\int_{\Real^n} (\prod_i\dotsm) (\prod_i\dotsm) (\prod_i\dotsm) \,d\vec{x}$
symbolically.
We perform such an integral by factoring it into
a product of \emph{independent} one-dimensional integrals.
Formally, suppose we want to perform an integral
$\smash{\int_X f(\vecdotsvec{t}) \,d\vecdotsvec{t}}$
over the space $\smash{X =
\prod_B (a,b)}$.
We try to re-express its body \smash{$f(\vecdotsvec{t})$}~as
\begin{equation}
\label{e:unproducts}
\textstyle
    e_0 \cdot \prod_B g(\vecdotsvec{t}\idx{B})
    \text,
\end{equation}
where~$g$ depends on just one element of~$\smash{\vecdotsvec{t}}$
at a time.  If this rewrite succeeds,
then the integral factors into a product of
one-dimensional integrals over a scalar variable~$t$:
\begin{equation}
\label{e:ints-prod}
\textstyle
    \int_X f(\vecdotsvec{t}) \,d\vecdotsvec{t} =
    \int_X e_0 \cdot \prod_B g(\vecdotsvec{t}\idx{B}) \,d\vecdotsvec{t} =
    e_0 \cdot
    \prod_B
    \int_a^b g(t)\,dt
\end{equation}
\diffclose{E}In our running example,
the array case reduces to the scalar case of integrating over~$x$
in~\labelcref{e:starting-lo}:
\begin{equation}
\smash[b]{
    \int_{\Real^n}
    \prod_{i=0}^{n-1}
    \frac{\Exp{-\Frac{(\vec{x}\idx{i}-\mu)^2}{2}}}{\sqrt{2\pi}}
    \frac{\Exp{-\Frac{(\vec{y}\idx{i}-\vec{x}\idx{i})^2}{2}}}{\sqrt{2\pi}}
    \frac{\Exp{-\Frac{(\vec{z}\idx{i}-\vec{x}\idx{i})^2}{2}}}{\sqrt{2\pi}}
    d\vec{x}
    = \prod_{i=0}^{n-1}
    \int_\Real
    \frac{\Exp{-\Frac{(t-\mu)^2}{2}}}{\sqrt{2\pi}}
    \frac{\Exp{-\Frac{(\vec{y}\idx{i}-t)^2}{2}}}{\sqrt{2\pi}}
    \frac{\Exp{-\Frac{(\vec{z}\idx{i}-t)^2}{2}}}{\sqrt{2\pi}}
    dt
}\vphantom{\Big|}
\label{e:plate2-elim}
\end{equation}
Existing routines for
integrals and definite products then directly apply
to eliminate the latent~$\vec{x}$,
even if $n$ were unknown.
(If~rewriting to \labelcref{e:unproducts} fails, then the latent
variable would not be eliminated.)

To recognize array distributions, the third step tries to rewrite
a density~\smash[t]{$f(\vecdotsvec{t})$} to a product~\labelcref{e:unproducts}.  If this
succeeds and the resulting factor~$g$ is the density
of some one-dimensional distribution~$m$, then $f$ is the density
of $r$ levels of $\Plate$ nested around~$m$.
Continuing the example, the right-hand-side
of~\labelcref{e:plate2-elim} is already a product whose body depends on
just one element of~$\vec{z}$ at a time,
so again the array case reduces to the
scalar case~\labelcref{e:starting-conditional}, and our
simplifier recognizes \labelcref{e:plate2-elim} to be the density of
\smash{$\Plate(n,i,\Gaussian(\frac{1}{2}(\mu+\vec{y}\idx{i}),\frac{\sqrt{6}}{2}))$}
at~$\vec{z}$.

\subsection{Rewriting an Expression as a Product}
\label{s:unproducts}

The purpose of the \emph{unproduct} operation is to rewrite an expression as
a product~\labelcref{e:unproducts}.  As just described, this rewrite is
key to eliminating array variables and recognizing array distributions
in the second and third steps of our extended simplifier.  Because the
running example~\labelcref{e:plate2-pre} above is simple, the unproduct
rewrite to~\labelcref{e:plate2-elim} is trivial.  It turns out that we
can handle a much broader variety of array programs that express desired
algorithms---including all the examples in~\cref{s:not-enough}---by
making the unproduct operation succeed more often.

The unproduct operation enables the automation of many common
simplifications, by uncovering independence among random variables and
likelihood factors that is prevalent yet often hidden in the source
program.  It generalizes the \emph{normalization} rewrite rule for
indirect indexing in AugurV2~\citep{huang-compiling}, as illustrated by
the Gaussian mixture model in~\cref{s:not-enough}.  It also generalizes
\emph{inversion} in the lifted inference
literature~\citep{de-salvo-braz-lifted-modulo} from discrete
distributions to continuous ones.  At the very least, because the
unproduct operation is the only way for our extended simplifier to
produce $\Plate$, it must succeed in order for a program containing
$\Plate$ to even just simplify to itself unscathed.  (Our test suite has
many such \emph{round-trip} tests.)  Hence, unproduct needs to succeed
even though factors tend to have their parts shuffled by computer
algebra.  In particular, because our simplifier rewrites
$\prod\Exp{\cdots}$ to $\Exp{\sum\cdots}$ so as to expose holonomy, the
two forms need to be treated equivalently by the unproduct operation.

More formally,
given a term~$e$ and an array variable~$\vec{x}$,
the goal of the operation $\unproduct(e,\vec{x})$ is to
produce a pair of expressions $(e',g)$ such that $g$ does not contain
$\vec{x}$ free yet $e = e'
\cdot \prod_i g(i,\vec{x}\idx{i})$.
Because the produced factor~$g$ does not contain $\vec{x}$ free but
rather takes an index~$i$ and an element~$\vec{x}\idx{i}$ as inputs,
it~only gets to use one element of~$\vec{x}$ at a time.
We call this operation $\unproduct$ because its specification is that
putting $\prod_i$ on its output should be equal to its input.

The unproduct operation proceeds by structural recursion over a
term, remembering the path to the subterm currently in focus.
We represent the path as a \emph{heap}. It is a
context---an expression with a single hole~$[~]$ where a subexpression can be plugged
in.  The result of plugging
an expression~$e$ into a heap~$H$ is notated $H[e]$.
\diffopen{DE}{Clarified description of heaps}We distinguish between heaps of two \emph{modes} by what they \emph{factor over}:
$H^\times$~of mode~$\times$
factors over multiplication, whereas~$H^+$~of
mode~$+$ factors over addition.
For example, $H^\times$ could be $[~]^2$ because $(e_1\cdot e_2)^2
= e_1^2 \cdot e_2^2$, whereas $H^+$ could be $\Exp{[~]}$ because
$\Exp{e_1+e_2} = \Exp{e_1}\cdot\Exp{e_2}$.  More generally, we maintain
the factoring invariants
\begin{align}
\label{e:factoring-mul-prod}
    H^\times[1] &= 1
&
    H^\times[e_1 \cdot e_2] &= H^\times[e_1] \cdot H^\times[e_2]
&
    \textstyle H^\times\bigl[\prod_{i=a}^b e\bigr] &= \textstyle \prod_{i=a}^b H^\times[e]
\\
\label{e:factoring-add-sum}
    H^+     [0] &= 1
&
    H^+     [e_1 +     e_2] &= H^+     [e_1] \cdot H^+     [e_2]
&
    \textstyle H^+     \bigl[\sum _{i=a}^b e\bigr] &= \textstyle \prod_{i=a}^b H^\times[e]
\end{align}
by defining a restricted grammar of heaps
\begin{alignat}{4}
    H^\times & \Coloneqq [~]
    && \altern H^\times\bigl[[~]^c\bigr]
    && \altern \textstyle H^\times\bigl[\prod_{i=a}^b [~]\bigr]
    && \altern H^\times\bigl[\pwow{[~]}{e}{1}\bigr]
\\
    H^+ & \Coloneqq H^\times\bigl[c^{[~]}\bigr]
    && \altern H^+\bigl[c\cdot [~]\bigr]
    && \altern \textstyle H^+\bigl[\sum_{i=a}^b [~]\bigr]
    && \altern H^+\bigl[\pwow{[~]}{e}{0}\bigr]
\end{alignat}
where the expressions~$a,b,c$ are constants in the sense that they do not
contain $\vec{x}$ free.
An occurrence of $\prod_{i=a}^b$ or $\sum_{i=a}^b$ in a heap binds the
index variable~$i$.\diffclose{DE}

The goal of $\unproduct(e,\vec{x},H)$, where the accumulator
argument~$H$ is initially the empty heap~$[~]$, is to produce a pair of
expressions $(e',g)$ such that $g$ does not contain $\vec{x}$ free yet
\mathbox{H[e] = e' \cdot \prod_i g(i,\vec{x}\idx{i})}.
Again, because $g$ does not contain $\vec{x}$
free but rather takes $i$ and $\vec{x}\idx{i}$ as
inputs, it~only gets to use one element of~$\vec{x}$ at a time.

\input{unproduct}

The definition of $\unproduct$ appears in \cref{fig:unproduct}.
The notation\diffopen{CDE}{Explain unproduct workhorse better} $e'\guard e$ means the conditional $\pwow{e}{e'}{I}$ where $I$ is
the identity of the mode of the surrounding heap. That is, we define
\begin{equation}
    H^\times[e'\guard e] = H^\times\bigl[\pwow{e}{e'}{1}\bigr] \text,\qquad
    H^+     [e'\guard e] = H^+     \bigl[\pwow{e}{e'}{0}\bigr] \text.
\end{equation}

The second case in \cref{fig:unproduct} is the workhorse;
it~is the source of any $g$ returned that is not just constantly~$1$.
It~applies when there is a unique index~$a$ where the term~$e$ uses
the array~$\vec{x}$.  It~would then be correct to return
\begin{equation}
\label{e:workhorse}
    \bigl(1, \fun{(i,\XI)} H[(i=a)\guard e(\XI)] \bigr) \text.
\end{equation}
For example, $\unproduct$ can rewrite $f(\vec{x}\idx{k})$ to
$\delimiterfactor=850 \prod_{i=0}^{n-1}\smash[b]{\pwow{f(\vec{x}\idx{i})}{i=k}{1}}$.
However, to~prevent subsequent computer algebra from stumbling over the
conditional, we further reduce the result~\labelcref{e:workhorse}
algebraically in four steps.
\begin{enumerate}
    \item Given $H[e'\guard e]$, hoist the test~$e'$ as far out of~$H$
          as possible:  First, decompose $H$ into $H=H_1[C]$, where the
          context $C$ is the maximal inner portion of~$H$ that does not
          bind any free variable in~$e'$.  (In particular, if $H$ does
          not bind any free variable in~$e'$ at all, then $H_1=[~]$ and
          $C=H$.  Otherwise, $H_1$ has the form
          \smash{$H_1^\times\bigl[\prod_j[~]\bigr]$} or
          \smash{$H_1^+\bigl[\sum_j[~]\bigr]$}, where $j$ is the
          innermost-scoped index variable bound by~$H$ that $e'$
          contains free.)  Then, rewrite $H_1\bigl[C[e'\guard e]\bigr]$
          to $H_1\bigl[e'\guard C[e]\bigr]$.  (The identity in $e'\guard
          e$ may differ from the identity in $e'\guard C[e]$, because
          the mode of~$H_1[C]$ may differ from the mode of~$H_1$.)
    \item Try to turn a loop into a let:
          Given the loop $\prod_j\pwow{e(j)}{e'(j)}{1}$
                      or $\sum _j\pwow{e(j)}{e'(j)}{0}$,
          where the test~$e'(j)$ contains~$j$ free, try to solve for $j$
          in~$e'(j)$ to yield an equivalent equation $j=b$.  If the
          solving succeeds, then rewrite the loop to $e(b)$.  For
          example, $\unproduct$ can rewrite
          $\prod_{j=1}^{n}f(j,\vec{x}\idx{j-1})$ to
          $\prod_{i=0}^{n-1}f(i+1,\vec{x}\idx{i})$, by solving the test
          $i=j-1$ symbolically to yield the equivalent equation $j=i+1$.
          A~more substantial example is that $\unproduct$ enables
          eliminating a Dirichlet distribution by rewriting
          \labelcref{e:dirichlet-integral} to
          \labelcref{e:dirichlet-integral-unproduct}.
          However, if the test is $i=\vec{y}\idx{j}$, as in the
          mixture-model example~\labelcref{e:gmm-unproduct-post}, then
          the solving fails and this step does nothing.
    \item Try to turn $\prod$ into~$\sum$ by pushing it inward:
          Given the loop $\prod_j\pwow{e(j)}{e'(j)}{1}$ (or just $\prod_j e(j)$),
          if the body $e(j)$ has the form $e_0^{e_1(j)}$ (or just~$e_0$) where
          $e_0$ does not contain $j$ free, then rewrite the loop to
          $e_0^{\sum_j\pwow{e_1(j)}{e'(j)}{0}}$.
          If the body $e(j)$ consists of several factors multiplied together,
          then deal with each factor separately.  Similarly, rewrite any
          factor \smash[b]{$\pwow{e_0^{e_1}}{e'}{1}$} at the top level, where $e_0$
          is closed, to $e_0^{e_1\cdot\pwow{1}{e'}{0}}$.
    \item Try to push $\sum$ inward:
          Given the loop $\sum_j\pwow{e(j)}{e'(j)}{0}$ (or just $\sum_j e(j)$),
          if the body $e(j)$ has the form $e_0\cdot e_1(j)$ (or just~$e_0$)
          where $e_0$ does not contain $j$ free, then rewrite the loop to
          $e_0\cdot \sum_j\pwow{e_1(j)}{e'(j)}{0}$.  If the body $e(j)$
          expands to several terms added together, then deal with each
          term separately.
\end{enumerate}
Roughly, these steps work together to reduce the
conditional expressions produced by $\unproduct$, so that subsequent
computer algebra successfully
eliminates latent variables and recognizes primitive
distributions in \cref{s:not-enough} and our classification benchmarks.  These benchmarks use
indexing heavily to express clusters, topics, and Dirichlet
distributions.
For example, $\unproduct$ produces the expression on the
\emph{right}-hand-side of~\labelcref{e:gmm-unproduct-post}, which is
expanded so that the sums and conditionals do not contain the
integration variable~$x$ free; existing computer algebra can thus
perform the integral $\int_\Real \Exp{f(x)}\,dx$ automatically by
treating those sums and conditionals atomically.

The last case in \cref{fig:unproduct} is the fallback for when $e$ uses
$\vec{x}$ at multiple indices but cannot be decomposed recursively.
This fallback is the source of any $e'$ returned
that uses $\vec{x}$ as a whole.  Because it trivially satisfies the
equational specification $H[e] = e'\cdot\prod_i g(i,\vec{x}\idx{i})$,
simplification still preserves semantics, but does not eliminate
a variable or recognize a conjugacy.  In some cases, this is because
there is really nothing to do; in other cases, our optimizing compiler
lags behind the mathematical prowess of applied statisticians.\diffclose{CDE}


\section{The Histogram Transformation}
\label{s:histogram}

We introduce the \emph{histogram} transformation, which improves the
asymptotic running time of loops that arise from simplifying mixture
models, by rewriting loops into map-reduce
expressions.

Recall that the goal of our compilation pipeline is the efficient
execution of array inference algorithms expressed as probabilistic
programs denoting conditional distributions.
\diffopen{AD}{Improved histogram discussion}Simplifying these programs produces loops, such as
\begin{equation}\label{e:histo-source}
        \sum\nolimits_{j=0}^{n-1}
        \pwow{\vec{s}\idx{j}}{i = \vec{y}\idx{j}}{0}
\end{equation}
and other summations in~\labelcref{e:histo-sources}
(same as in the right-hand-side of~\cref{e:gmm-unproduct-post}).  When
the program performs indirect indexing, the
resulting loops are nested: the outer loop iterates over classes~$i$
and the inner loop iterates over all
individuals~$j$ but only considers
those that belong to the current class ($i = \vec{y}\idx{j}$).  By
generalizing loops from scalar summation to other map-reduce expressions, we
can dramatically speed up such nested loops to run in time independent
of the number of classes.  For example, by looking up the class of
every individual, a~single pass over the population can
produce the sum of every class; a~summation such
as~\labelcref{e:histo-source} can be computed for
all~$i$ in $O(n)$ rather than $O(mn)$ time.

Loop nests like~\labelcref{e:histo-source} often arise for inference
when the model divides array elements into subpopulations, as~mixture
models do.  Eliminating latent variables proliferates such loop nests,
because intuitively, after eliminating a variable~$x$, the information
that used to be required to infer~$x$ becomes required to infer the
variables that depend on~$x$.  For example, in the Gaussian mixture
model in \cref{s:not-enough}, each point~$\vec{s}\idx{j}$ only requires
one quantity---the mean $\vec{x}\idx{\vec{y}\idx{j}}$ underlying the
class~$\vec{y}\idx{j}$---but eliminating $\vec{x}$ makes the
point~$\vec{s}\idx{j}$ require the other points in the same class.

Wherever a nested formula arises,
an~applied statistician would translate it manually
to unnested code as a matter of course; we automate
this asymptotic improvement here.\diffclose{AD}  As \cref{fig:pipeline} suggests,
this histogram optimization of ours composes with simplification and
applies to both exact and approximate inference procedures.
In fact, it applies to probabilistic and non-probabilistic programs
alike, even though probabilistic programming is the context where we
needed it and invented~it.
This
modularity and generality sets our work apart from other systems that
incorporate this optimization only for MCMC inference on
mixture models \citep{tristan-augur,huang-compiling}.

As the name implies, the histogram transformation recognizes nested
loops that are usually visualized as (generalized) histograms.
These histogram computations manifest as sums such
as~\labelcref{e:histo-source}.
We thus introduce a term construct $\Bucket$ to represent
such computations. The transformation rewrites such sums to an
equivalent
let-expression that binds a $\Bucket$ term to a $\summary$ variable.
For example, in the scope of~$i\in\{0,\dotsc,m-1\}$, the histogram
transformation rewrites~\labelcref{e:histo-source} to
\begin{equation}
\label{e:histogram-post}
    \Let{\summary}
        {\Bucket_{j=0}^{n-1}\bigl(\Index_i^m(\vec{y}\idx{j}, \Add(\vec{s}\idx{j}))\bigr)}
        {\summary\idx{i} \text,}
\end{equation}
where the capitalized keywords are new (in \cref{fig:reducer}).  The
$\summary$ variable is bound to an array whose size is~$m$ and whose
element at each index~$i$ is the sum of those $\vec{s}$ whose
corresponding $\vec{y}$ matches~$i$.
The sequential code we generate for computing $\summary$ initializes
it to an all-zero mutable array then adds $\vec{s}\idx{j}$ to
$\summary\idx{\vec{y}\idx{j}}$ for each $j$ from $0$ to~$n-1$.
Roughly,\diff{E}{Explain how the semantics of \labelcref{e:histogram-post} arises from its components} $\Add$ means to add,
$\Index_i^m$ means to index into an array of size~$m$,
and $\Bucket_{j=0}^{n-1}$ means to loop for $j$ from $0$ to~$n-1$.
We leave further speedups of such map-reduce computations using
parallelization, vectorization, and GPUs to future work.

Out of context, the let-expression~\labelcref{e:histogram-post} seems
like a waste because it computes $\summary$ then uses only one element
of~it.  But because the class variable~$i$ does not occur free in the
$\Bucket$ expression (the subscript $i$ is a binder), LICM
(\cref{s:licm}) will later lift the binding of $\summary$ out
of the scope of~$i$, thus reusing it across all $m$ classes.  To~pave
the way, a $\Bucket$ term should depend on as few inner-scoped
variables as possible, and the index variable~$i$ in $\summary\idx{i}$
should be loop-bound.

\subsection{Syntax and Semantics of Reducers}

\Cref{fig:reducer} formalizes the sublanguage of reducers, which
constitute the body of a $\Bucket$ expression.  The judgment $r
\triangleright_j T$ means that $r$ is a reducer of type~$T$ over
index~$j$.
\diffopen{DE}{Explain reducers better}A~reducer~$r$ constitutes the body of a histogram expression
$\Bucket_{j=a}^b(r)$, whose typing rule is shown at the bottom of the
figure.  The scope of the variable~$j$ is special, because the histogram
expression \smash[t]{$\Bucket_{j=a}^b(r)$} interprets the reducer~$r$ in
two ways: first to initialize a mutable histogram to zero independently
of~$j$, and then to update the histogram iteratively by looping over the
index~$j$.  Thus, $j$ can only appear free in certain parts of~$r$,
marked intentionally by ``$[j:\iplus]\dotso$'' in \cref{fig:reducer}.

\input{reducer}

Mathematically, a reducer~$r$ of type~$T$ denotes a monoid whose carrier
is~$T$ (that is, an associative binary operation~$+_r$ on~$T$ that has
an identity~$r^0$), along with a map~$r^1$ from indices~$j$ to elements
of~$T$.  Intuitively, the histogram expression
\smash{$\Bucket_{j=a}^b(r)$} first initializes a mutable histogram using
the identity~$r^0$, then updates the histogram iteratively using the
map~$r^1$ and the monoid operation~$+_r$.  That is, $\Bucket_{j=a}^b(r)$
denotes the monoidal sum $r^1(a) +_r \dotsb +_r r^1(b)$ (which equals
$r^0$ in case $a=b+1$).  Thus, to describe the operational semantics of
a histogram expression on sequential hardware, we associate with each
reducer~$r$ two methods: initializing a mutable~$T$, and updating it at
a given index~$j$.  The expression $\Bucket_{j=a}^b(r)$ uses $r$ to
initialize a mutable histogram~$T$ then updates it at each index
$j=a,\dotsc,b$.  We now describe the denotation and operation of each
reducer construct in turn.
\begin{itemize}
    \item $\Add(e)$ denotes addition on~$\Real$ along with the map
          $\fun{j} e$.
          Accordingly, $\Add(e)$ initializes a real to~$0$ and updates it by
          adding~$e$.
    \item $\Index_i^b(e,r(i))$ denotes the product of the monoids
          denoted by $r(0),\dotsc,r(b-1)$, along with the map
          \begin{equation}
            \Index_i^b(e,r(i))^1 =
            \fun{j}\ary\smash[t]{\Bigl(b,i,\pwow{r(e)^1(j)}{i=e}{r(i)^0}\Bigr)}
            \text.
          \end{equation}
          Accordingly, $\Index_i^b(e,r(i))$ initializes an array of size~$b$ by
          initializing its elements using $r(0),\dotsc,r(b-1)$, and
          updates the array by updating just the element at~$e$ using
          $r(e)$.

          Note that the denoted monoid and the initialization method are
          independent of~$e$ and thus independent of~$j$.  In
          particular, the size expression~$b$ is evaluated during
          initialization without using~$j$.  However, $\Index_i^b$ binds
          $i$ in~$r(i)$, so the monoid and initialization of each
          histogram element $\summary\idx{i}$ can depend on~$i$.  In
          particular, $r(i)$ may contain an inner
          \smash[t]{$\Idx_{i'}^{b'}$} whose size expression~$b'$ does
          depend on~$i$ (not~$j$).  Such a nested $\Idx$ reducer
          produces a histogram that is a ragged array of arrays.  We
          cannot forego the bound variable~$i$ by substituting $e$
          for~$i$ in~$b'$, because $e$ may contain~$j$ free.
    \item $\Split(e,r_1,r_2)$ and $\Fanout(r_1,r_2)$ both denote the
          product of the monoids denoted by $r_1$ and~$r_2$.  But
          \begin{align}
            \Split(e,r_1,r_2)^1 &=
            \smash[t]{\fun{j}\pwow{(r_1^1(j),r_2^0)}{e}{(r_1^0,r_2^1(j))}}
            \text, &
            \Fanout(r_1,r_2)^1 &=
            \fun{j}\bigl(r_1^1(j),r_2^1(j)\bigr)
            \text.
          \end{align}
          Accordingly, $\Split(e,r_1,r_2)$ and $\Fanout(r_1,r_2)$ both initialize
          a pair by initializing its parts using $r_1$ and~$r_2$.  But
          $\Split$ uses $r_1$ to update the first part when $e$ is true
          and uses $r_2$ to update the second part when $e$ is false,
          whereas $\Fanout$ always updates both parts.
    \item $\Nop$ denotes the trivial monoid and the constant map.
          Accordingly, $\Nop$ initializes a unit value and does nothing to~it.\diffclose{DE}
\end{itemize}

\subsection{Histogram Transformation Implementation}

We recognize when a $\sum_{j=0}^{n-1} e$ can be
rewritten in terms of an equivalent $\Bucket$ computation that can then be
hoisted by LICM for reuse. Formally,
we describe a program transformation \summarize\ such that
if $\summarize(e, j) = (r, f)$ then
$\sum_{j=0}^{n-1} e = f\bigl(\Bucket_{j=0}^{n-1}( r )\bigr)$.
To~facilitate LICM\@, $r$~should depend on as few inner-scoped variables as
possible.

\input{summarize}

The entire definition of \summarize\ appears in \cref{fig:summarize}.
Whenever we encounter a summation $\sum_{j=0}^{n-1} e$,
we apply the rules in~\cref{fig:summarize}
to evaluate $\summarize(e,j)$ to $(r,f)$, then replace
$\sum_{j=0}^{n-1} e$ by $f\bigl(\Bucket_{j=0}^{n-1}( r )\bigr)$ if
$r$ looks profitable (that is, contains $\Index$ or $\Fanout$).

The \summarize\ transformation is profitable when the summand
chooses among alternatives, typically depending on some contextual
information (such as $i$ in~\labelcref{e:histogram-post}).
The first rule takes all expressions defined by cases
which do not depend on the summation variable $j$, and
translates them to a $\Fanout$. Further case expressions
are translated to either a \Split\ or an \Index, by
pulling out conditions while prioritizing outermost bound variables.
Once all case expressions are gone, the
remainder is emitted either as \Nop\ (if zero) or \Add.

Continuing with the example \labelcref{e:histo-source}, we try
$\summarize\smash[t]{\left(\pwow{\vec{s}\idx{j}}{i = \vec{y}\idx{j}}{0},j\right)}$.
The first rule does not apply, as the condition $i = \vec{y}\idx{j}$
depends on~$j$.  The next two rules are both applicable: the \Split\
rule incurs the free variables $\{i,\vec{y}\}$ whereas the \Index\ rule
only incurs~$\{\vec{y}\}$.  The \Index\ rule wins, as the
input~$\vec{y}$ is bound outside~$i$. We end up with
\mathbox{\summarize(\vec{s}\idx{j},j)}, which only matches the last
rule.  Assembling the results gives
\(
    \bigl(\Index_i^m(\vec{y}\idx{j}, \Add(\vec{s}\idx{j})),
     \fun\summary \summary\idx{i}\bigr)
\)
as desired.


\section{Code Generation}
\label{s:codegen}

Our code generator uses the domain specific properties of \hakaru
programs to generate optimized x86 code at runtime. This
generator is designed to fit into the pipeline of
\cref{fig:pipeline}---after the programs have undergone the
simplification and histogram transformations---although it
applies to any \hakaru program.  In fact, the optimizations performed by
the generator make sense for a general-purpose language (GPL) and are
not new, but thanks to the invariants present in \hakaru programs, we
can implement them much more easily, perform them much more
aggressively, and reap much more performance benefit.  And we need to:
as we demonstrate in the ablation study in \cref{s:ablation},
simplifying array programs that express desired inference algorithms
produces residual code---such as repeated traversals of arrays---that
would be prohibitively slow without optimization.

The time-consuming computations of probabilistic programs come from
pure numerical expressions involving tuples and arrays.  It is
straightforward to translate these programs into any GPL\@.
However, the domain-specific nature of
\hakaru provides several advantages for generating efficient code,
advantages not typically available to GPLs:
\begin{enumerate}
\item All arrays in Hakaru programs are immutable and unaliased, and
  loops operate over arrays.
\item The histogram transformation produces loops that are nested yet
      independent.
\item Hakaru programs not only contain loops but typically \emph{are} the loop body of an inference
  method, so they are both short and called repeatedly on a
  particular data set.
\end{enumerate}
Using these insights, the second half of our pipeline (the bottom half of
\cref{fig:pipeline}) optimizes programs in two ways that
are novel in the context of probabilistic programming languages:
\begin{enumerate}
\item We perform LICM \citep{aho-dragon} to hoist inner loops out of
      outer loops.  We then fuse loops of the same bounds together while
      lowering the program into \emph{Sham IR}\@, an IR with
      \textsf{for} loops and mutation that compiles to x86 via LLVM\@.
      We carry out these simple optimizations freely and aggressively,
      without worrying about side effects (\cref{s:licm}). These
      optimizations yield a $1289\times$ speedup (\cref{fig:opt-cmp}).
\item We JIT-compile Hakaru programs at run time, allowing for
  extensive specialization (\cref{s:sham}) yielding a $9.5\times$
  speedup (\cref{fig:opt-cmp}).
\end{enumerate}

\diffopen{E}{Clarified our arithmetic}%
Our code generator uses exact arithmetic but generates code that uses
floating-point arithmetic.  It~is well known that floating-point
probabilities should be computed in log-space in order to avoid
underflow.  We use this log-representation for all numbers of type~$\rplus$.\diffclose{E}



\subsection{Loop Optimizations}
\label{s:licm}

LICM and loop fusion are the two most significant optimizations
performed by our code generator.  As depicted in \cref{fig:pipeline},
LICM operates on A-normal forms~\citep{flanagan-anf} in our pure (probabilistic)
language, before loop fusion lowers them into Sham's imperative IR\@.
This design makes the optimizations easier to implement and more
effective, as we now describe.

\begin{figure}
\begin{tabularx}{\textwidth}{@{}X@{\qquad}X@{}}
\hfil\begin{array}[t]{@{}l@{}}
    \fun{\vec\alpha : \Array\rplus}
    \fun{\vec y : \Array\iplus}
    \fun{\vec s : \Array\Real}
    \fun{u : \iplus}
\\
    \Lets{
        \summary_1 & \smash[t]{
                     \Bucket_{k=0}^{\size\vec s-1}\bigl(
                         \Index_\_^{\size\vec\alpha}(\vec y \idx k,
                         \Add(1))\bigr)} \\
        \summary_2 & \Bucket_{k=0}^{\size\vec s-1}\bigl(
                         \Index_\_^{\size\vec\alpha}(\vec y \idx k,
                         \Add(\vec s \idx k))\bigr) \\
    }{\Lets{\mathit{array}_1 &\ary(\size\vec\alpha,i,}{} \\[2ex]
      \qquad\Lets{\mathit{prod}_1
                   & \smash[t]{
                     \prod_{j=0}^{\size\vec\alpha-1}\Bigl(\summary_1 \idx j + \pwow{\vec y \idx u}{j=i}{0}\Bigr)} \\
                   \mathit{sum}_1
                   & \sum _{j=0}^{\size\vec\alpha-1}\Bigl(\summary_2 \idx j + \pwow{\vec s \idx u}{j=i}{0}\Bigr) \\
             }{\mathit{prod}_1 + \mathit{sum}_1)} \\
      \Categorical(\mathit{array}_1)}
\end{array}
\caption{An excerpt from one of our examples after performing LICM\@. Here $\summary_1$ and $\summary_2$ were moved out of the $\mathit{prod}_1$ and $\mathit{sum}_1$ loops respectively, and out of the $\mathit{array}_1$ loop together.}
\label{lst:after-anf}           %
&
\divide\arrayextrasep 3
\hfil\begin{array}[t]{@{}l@{}}
    \fun{\vec\alpha : \Array\rplus}
    \fun{\vec y : \Array\iplus}
    \fun{\vec s : \Array\Real}
    \fun{u : \iplus}
\\
    \Lets[:=]{
        \summary_1 & \newArray(\size\vec\alpha) \\
        \summary_2 & \newArray(\size\vec\alpha) \\
    }{\For{k}{0}{\size\vec s-1}{
        \summary_1\idx{\vec y \idx k} \plusequal 1 \textsf{;\ }
        \summary_2\idx{\vec y \idx k} \plusequal \vec s \idx k\\
      }\\
      \Lets[:=]{
          \mathit{array}_1 & \newArray(\size\vec\alpha) \\
      }{\For{i}{0}{\size\vec\alpha-1}{
            \Lets[:=]{
                \mathit{prod}_1 & 1 \textsf{;\ } \mathit{sum}_1 := 0
            }{
                \delimiterfactor=850
                \For{j}{0}{\size\vec\alpha-1}{
                    \begin{array}{@{}l@{}>{{}}c<{{}}@{}l@{}}
                        \mathit{prod}_1 &\mulequal  &\summary_1 \idx j + \smash[t]{\pwow{\smash{\vec y} \idx u}{j=i}{0}} \\[1.9pt]
                        \mathit{sum}_1  &\plusequal &\summary_2 \idx j + \pwow{\smash{\vec s} \idx u}{j=i}{0} \\
                    \end{array}
                }\\
                \mathit{array}_1\idx{i} := \mathit{prod}_1 + \mathit{sum}_1
            }
        }\\
        \Categorical(\mathit{array}_1)
      }
    }
\end{array}
\caption{The result of loop fusion and lowering on the example in \cref{lst:after-anf}}
\label{lst:fusion}
\end{tabularx}
\end{figure}

The input language to our LICM pass makes it easy to identify loops
and compute their dependencies.  That is important as we want to find
where we can convert a nest of loops into a sequence of
loops---that is, when an inner loop does not depend on an outer loop's
index variable.  Such code motion yields our biggest performance gain,
in part due to the preceding histogram transformation. Identifying
loops is simple, because \hakaru has only four specialized loop
constructs ($\sum$,~$\prod$, $\ary$, $\Bucket$) and no general recursion.
Computing dependencies using A-normalization in a pure
language ensures that code motion preserves semantics: we hoist
\textsf{let}-bindings as far out as the scope of their free variables
allows.  \Cref{lst:after-anf} shows how a typical program looks like
after LICM and before loop fusion and lowering; the two $\Bucket$
expressions, which were originally nested inside two loops, did not
depend on them and have been safely hoisted.

Next, multiple independent loops with identical bounds can be fused.
In our domain, aggressive loop fusion improves performance because most
loops iterate over arrays and fusion reduces the number
of indexing operations.  In contrast, loop fusion in a GPL may worsen
performance by disturbing locality of reference.

Although \hakaru makes loop
fusion straightforward, it is inappropriate as the output language,
because a single fused loop may need to maintain many accumulators
without tupling them.  Instead, our loop-fusion pass produces Sham
IR\@, which has \textsf{for}-loops and mutation.  A single pass fuses
loops and lowers them to Sham IR\@, to avoid the harder task of
identifying independent loops in Sham IR\@.
\Cref{lst:fusion} shows
the result of loop fusion on the example from \Cref{lst:after-anf}.

Applying LICM and loop fusion to histogram operations introduces
multiple array indexing operations that were previously implicit.  If
two histograms over the same array were fused, the resulting loop body
would contain repeated indexing operations, such as $\vec y \idx k$ in
\cref{lst:fusion}.  To avoid this repeated indexing, we follow loop
fusion by a hoisting pass in Sham IR that applies only to indexing
operations into input arrays, which are known to be constant.
This helps reduce memory lookup and improve cache locality
should these loops be unrolled later.

\subsection{Run-Time Specialization and Code Generation}
\label{s:sham}

Our programs are small and typically run as the
body of an outer loop over fixed-size data.
To use this fact, we perform several
optimizations that can only be performed in a JIT compiler.
Inside the outer loop, some information stays the same across
iterations; in particular, arrays whose values change may well stay a
constant size nevertheless. Thus we allow the programmer to mark
arguments with such binding-time information.

When array sizes are known, exact loop bounds tend to become known for
most loops. LLVM can then optimize those loops more aggressively. From
input array sizes we can even infer intermediate array sizes.  When we
know the constant size of an intermediate array, we
pre-allocate it only once and reuse it across iterations, removing
per-iteration allocation overhead. Thus for array arguments,
two different specialization directives can be given: known
size, and known size and values.

By waiting until we know array sizes before generating code, we can
prepone allocation even further: we can allocate intermediate arrays
before we even emit the code!  In other words, upon execution of a
program, we can use the size of input data to allocate arrays of the
appropriate size for intermediate data.  The machine code we emit
then embeds the intermediate arrays' sizes as well as \emph{addresses}
as constants\diff{D}{Clarified why pre-allocating memory is related to JITing}, which no longer need to be kept in registers.
We end up with extra registers that can be used for other variables,
reducing the need to store and load things on stack.

To perform the run-time specializations as described, we build LLVM IR
in memory and JIT-compile it using LLVM's C-API\@. The outcome, as shown
in \cref{s:eval}, is highly optimized code compared to traditional
implementations of domain-specific languages.



\section{Evaluation}
\label{s:eval}


The main claim of this paper is that array inference algorithms,
expressed as probabilistic programs denoting conditional distributions,
can be compiled automatically to efficient code.
It is impossible to evaluate how existing systems compile the same
programs to implement the same algorithms, because they don't.
Instead, we justify our claim by ballpark quantitative comparisons on
flagship applications of the decades of work in applied statistics
that established the importance of this class of algorithms.
We make two overall findings:
\begin{itemize}
    \item Compared against handwritten code for the same algorithms, we
          find that \hakaru's generated code achieves competitive speed
          (and of course the same accuracy).
    \item Compared against existing systems that use different inference
          algorithms for the same models, we find that \hakaru delivers
          the expected increase in accuracy and/or speed.
\end{itemize}

We measure the performance of both
approximate and exact inference algorithms.
For approximate inference using Gibbs sampling, we are
\begin{itemize}
    \item more accurate and 2--12\X\ as fast as JAGS \cite{Plummer2003}\@, a popular
          probabilistic\hyp programming system specialized for Gibbs
          sampling that cannot eliminate latent variables;
    \item more accurate or faster than STAN \citep{carpenter-stan}\@,
          a popular probabilistic\hyp programming system that carries
          out other inference algorithms and cannot eliminate latent
          variables;
    \item 9\X\ as fast as MALLET \cite{mallet}\@, a popular document\hyp
          classification package whose handwritten code performs the same computation as
          our inference procedure; and
    \item more accurate than AugurV2 \cite{huang-compiling}\@,
          a~recent research system that like \hakaru can compile models
          with arrays into fast MCMC samplers, but cannot
          eliminate variables.
\end{itemize}
For exact inference, we are
\begin{itemize}
    \item over 5000000\X\diff{E}{Updated PSI benchmark results using \texttt{build-release.sh} and the \texttt{--nocheck} flag}\ faster while handling 10\X\ more data than PSI
          \cite{gehr-psi}, another system that can perform exact
          inference on models containing arrays; and
    \item 3--11\X\ as fast as handwritten-quality Haskell code
          emitted by an earlier backend.
\end{itemize}
All benchmarks were executed on a 6-core AMD-Ryzen 5 with 16 GB of RAM\@,
running Linux 4.15\@. We used Racket 6.12\@, LLVM 5.0.1\@, Maple 2017.2\@,
and GHC 8.0.2\@.

Our benchmarks span inference tasks that are unsupervised and
supervised, with observed and inferred variables that are
continuous and discrete.
We do not compare against Figaro \citep{figaro} and Anglican
\citep{wood-new} because those shallowly
embedded languages do not use conjugacy to handle unlikely
continuous observations gracefully: Figaro produces no Gibbs samples
whereas Anglican produces very inaccurate samples.  (Our preliminary
testing also found Figaro an order of magnitude slower than JAGS on
models with just a few discrete variables.)

\subsection{Approximate Inference}
\label{s:apprxi}

We report three benchmarks of approximate inference using Gibbs sampling:
\begin{enumerate}
    \item clustering\diff{D}{Changed ``unsupervised classification'' to ``clustering''} of data points using a Gaussian
          mixture model (\cref{s:not-enough})
    \item supervised document classification using a Naive Bayes
          model \citep{mccallum1998comparison}
    \item unsupervised topic modeling using Latent
          Dirichlet Allocation (LDA) \citep{blei-latent}
\end{enumerate}
Gibbs sampling works by repeatedly \emph{sweeping} through all
unobserved random variables and \emph{updating} their currently inferred
values randomly.  Thus a sweep consists of as many updates as there are
random variables that are unobserved and uneliminated (such as unclassified data points or
documents).

On each benchmark, we compare with
\begin{itemize}
    \item AugurV2, a~probabilistic\hyp programming research system
          focused on composable and performant MCMC\@.  (Applying
          AugurV2 required a small patch to make its algebraic rewriting
          more robust.)
\end{itemize}
On the first two benchmarks, we further compare with
\begin{itemize}
    \item JAGS\@, a widely used probabilistic\hyp programming
          system specialized for Gibbs sampling.
          (JAGS does not scale to the third benchmark.)
\end{itemize}
Both JAGS and AugurV2 perform different computations than \hakaru,
because those systems do not eliminate latent variables
\citep{casella1996rao} as our simplification transformation does.
In the first benchmark, Gaussian mixture classification, we further
compare with
\begin{itemize}
    \item STAN\@, another widely used probabilistic\hyp programming
          system that cannot perform Gibbs sampling but defaults to
          a very different MCMC inference algorithm, namely HMC
          \citep{neal-hamiltonian,betancourt-conceptual}
          with No-U-Turn Sampling \citep{hoffman-nuts}.  (Applying STAN
          required the manual elimination of latent discrete array variables,
          a transformation automated by our simplification
          transformation.)
\end{itemize}
In the second benchmark, Naive Bayes document classification, we further
compare with
\begin{itemize}
    \item MALLET\@, a popular Java-based package for statistical
          natural\hyp language processing that can be configured to
          perform the same computation as \hakaru.
\end{itemize}

To summarize the results across benchmarks, our generated
code turns out to be faster than JAGS and
MALLET\@, and more accurate for a given time budget than AugurV2 and STAN\@. As
noted above, our system executes a different algorithm than JAGS,
AugurV2, and STAN\@, which we credit for the higher eventual
accuracy we achieve. We reiterate that the purpose of these benchmarks is to show that
\hakaru compiles a new class of inference algorithms while maintaining competitive performance,
not to rehash or analyze the superiority of a particular inference algorithm.

\paragraph{Gaussian mixture model}

The first benchmark uses synthetic data, and we show two variations.
Following the Gaussian mixture model in \cref{s:not-enough}, we draw $n=10000$ ($5000$) data points
from a mixture of $m=50$ ($25$) normal distributions, whose standard
deviations are all~$1$ and whose means are independently generated
with standard deviation $\sigma=14$ and mean~$\mu=0$.  We then hold
out all the labels~$\vec{y}$ and use Gibbs sampling to infer them.

\begin{figure}
  \includegraphics[width=.5\linewidth]{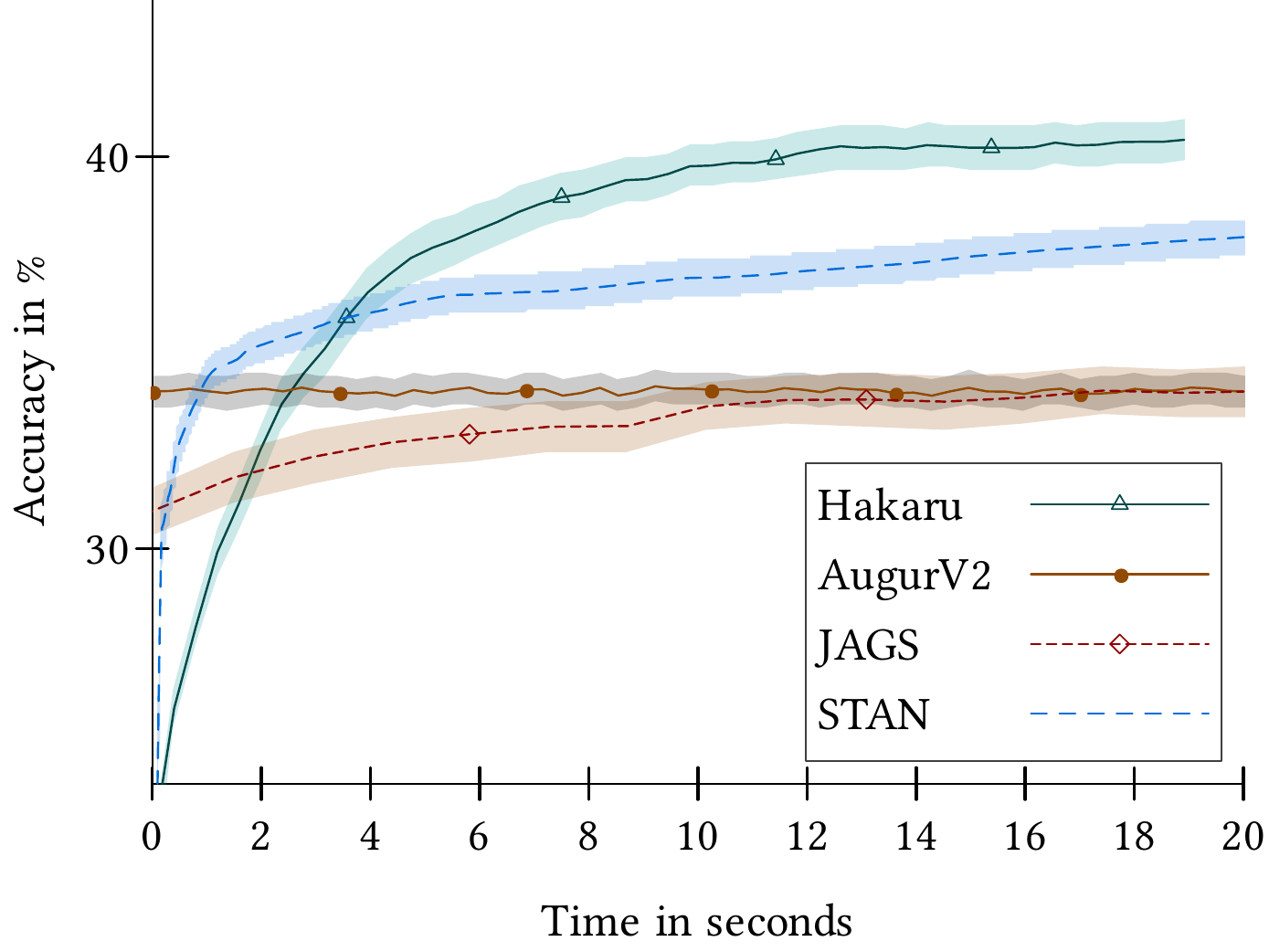}%
  \includegraphics[width=.5\linewidth]{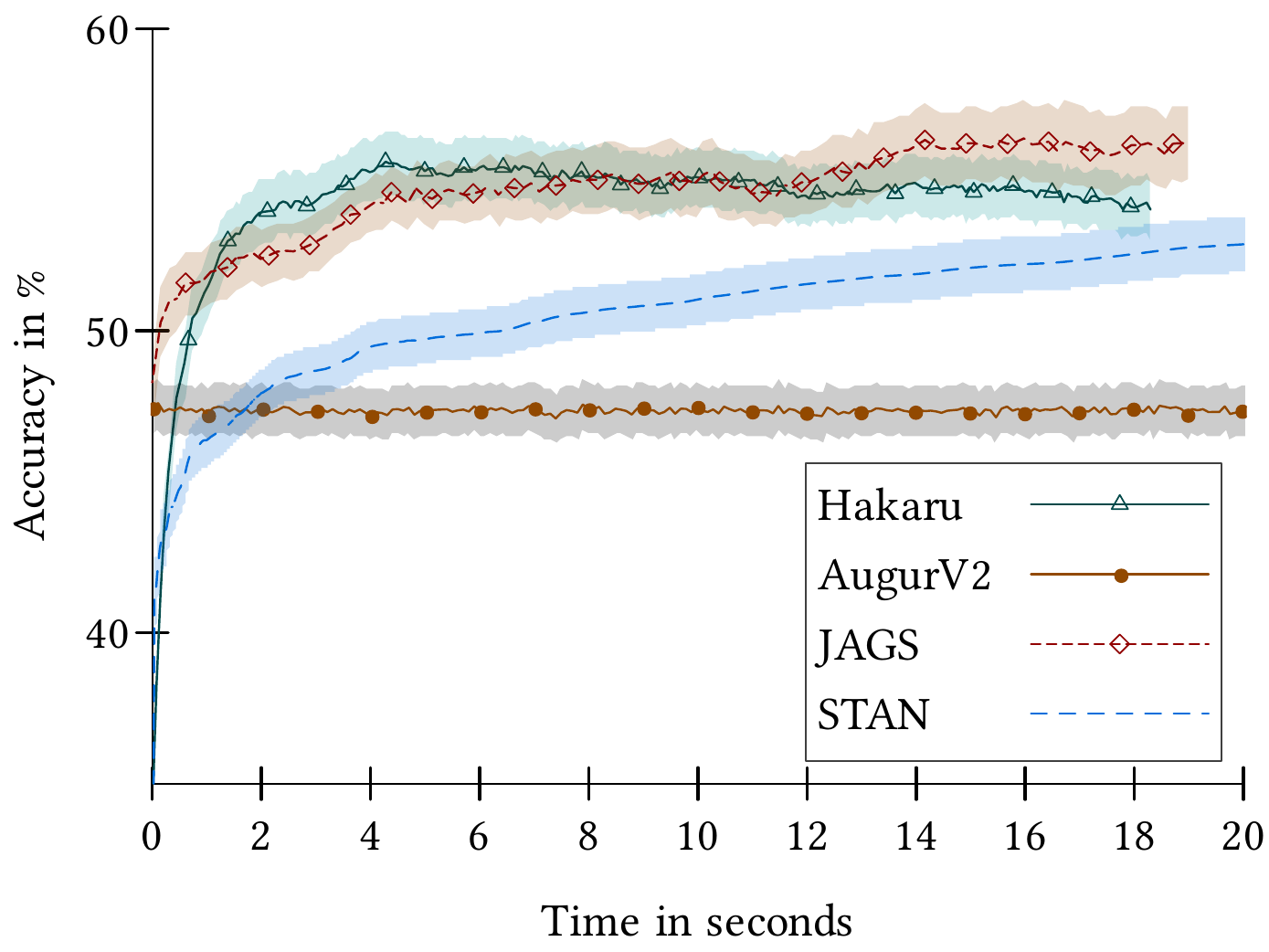}
  \caption{Comparison of samplers for the Gaussian mixture model
    with $n=10000$, $m=50$ and with $n=5000$, $m=25$.  Startup time is
    removed to \cref{fig:startup}.  Curves represent mean accuracy over time; shaded area is
    standard error of $50$ trials with different input data.  Each mark
    on a curve represents $10$ sweeps by Hakaru or JAGS or $100$ sweeps
    by AugurV2\@.  The mixture weights are drawn from the flat Dirichlet
    distribution, so clustering all points together would achieve
    accuracy $\approx 9\%$ for $m=50$ and $\approx 15\%$ for $m=25$.}
  \label{fig:gmm}
\end{figure}

We can compare inference accuracy on this benchmark, because we know the
true labels of our synthetic data.\footnote{For this clustering
task, symmetry (unidentifiability) demands we define accuracy as the
proportion of data points classified correctly under the most favorable
one-to-one correspondence between true and inferred labels.
Hence computing accuracy requires solving the \emph{assignment
problem}.  For STAN\@, which samples \smash{$\vec\theta$}
and~$\vec{x}$ rather than~$\vec{y}$, we plot \emph{expected} accuracy.}\diff{D}{Changed ``unsupervised classification'' to ``clustering''}
\Cref{fig:gmm}\diff{DE}{Describe ``the accuracy obtained by the higher of a random or constant predictor''.  Use markers to make the curves easier to distinguish.} plots the accuracy achieved by each sampler
against wall-clock time.  \hakaru's generated code
achieves higher accuracy compared to STAN's very different algorithm,
and compared to JAGS and AugurV2 after a few seconds.
This is the case even though, as marks on the curves show, AugurV2 is an
order of magnitude faster at performing a sweep than \hakaru and JAGS\@.
We credit our greater accuracy to simplification
eliminating the latent variables \smash{$\vec\theta$} and~$\vec{x}$
(\cref{s:not-enough}).
STAN works well on other models for which simplification has nothing
to~do.\diff{D}{``say that STAN works well on a very different class of models''}




\paragraph{Naive Bayes topic model}

The second benchmark uses the 20 Newsgroups corpus, which consists of
19997 articles classified into 20 newsgroups~\citep{newsgroups}.  We hold out 10\% of the
classifications and use Gibbs sampling to infer them, following
a Dirichlet\hyp multinomial Naive Bayes
model~\citep{mccallum1998comparison,resnik-gibbs}.\footnote{In this
model and the LDA model, to encode that different documents have
different numbers of words, we use two integer arrays of
equal length, one containing word identifiers and one containing
document identifiers. We could as well have used a single ragged array
of integer arrays, where each inner array contains the word identifiers
that make up a document.}\diff{D}{Specify how we encode that different documents have different numbers of words}

\begin{figure}
  \includegraphics[width=.5\linewidth]{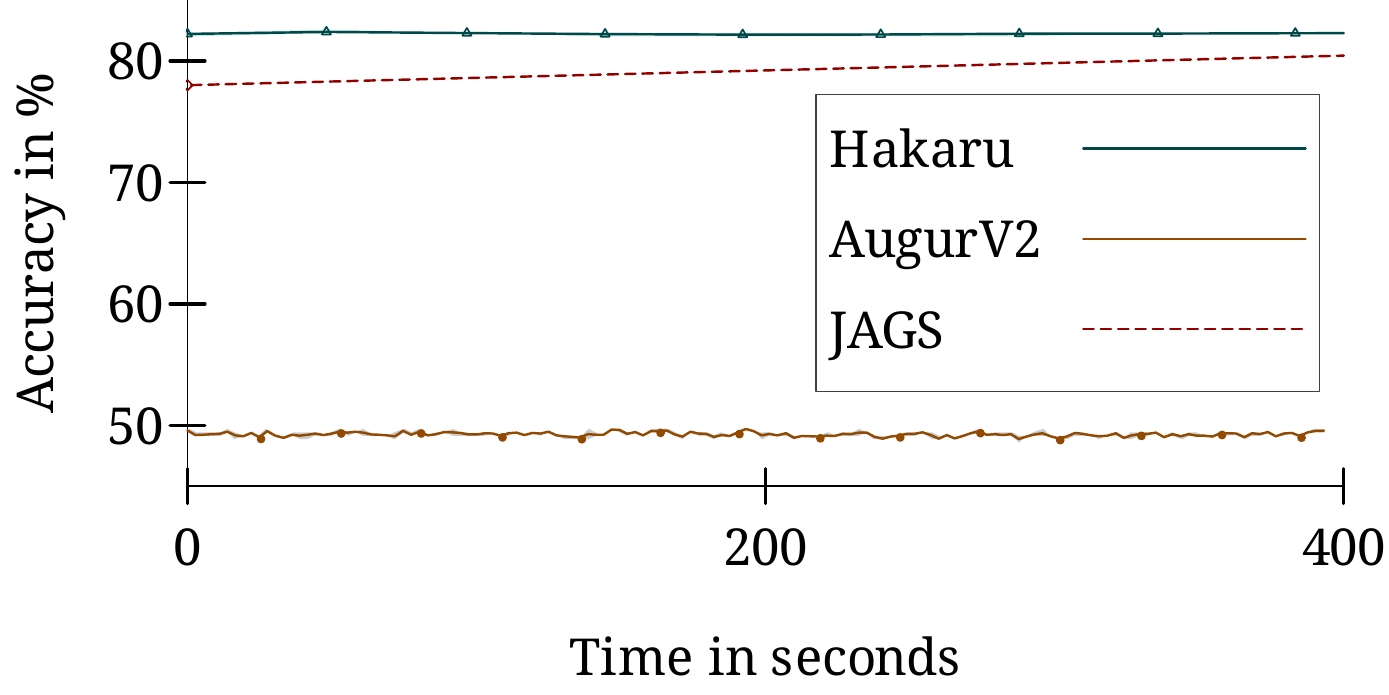}%
  \includegraphics[width=.5\linewidth]{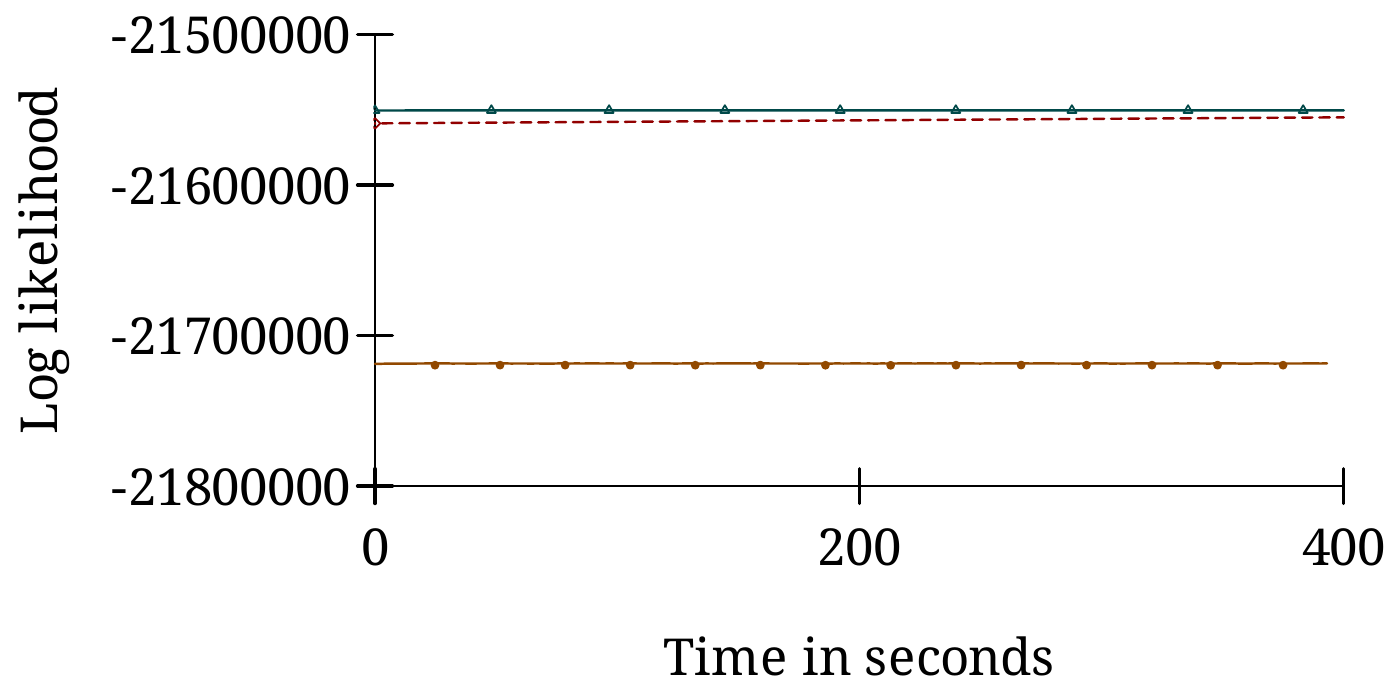}
  \caption{Comparison of Gibbs samplers for Naive Bayes document
    classification.  Curves
    represent mean accuracy or log likelihood over time; shaded area is standard error.
Each mark
    on a curve represents $1$ sweep by Hakaru or $100$ sweeps
    by AugurV2\@; a sweep by JAGS takes more than $500$ seconds.
    The documents are evenly distributed among $20$ newsgroups, so~a
    random or constant classifier would achieve $5\%$ accuracy.
    Log likelihood consists of supervised and inferred labels,
    which are correlated due to eliminating latent variables.}
  \label{fig:nb}
\end{figure}

Again we can compare inference accuracy, because we know the
true labels we hold out.  We also compare the log likelihood of the
samples.  \Cref{fig:nb}\diff{D}{Describe ``the accuracy obtained by the higher of a random or constant predictor''} plots these two metrics against wall-clock time.
As the curves show, \hakaru's generated code achieves higher accuracy
and likelihood right from the first sweep onward.  This is the case even
though, as marks on the curves show, AugurV2 is two orders of magnitude
faster at performing a sweep.  We again credit our simplification
transformation eliminating the latent variables and generating code that
samples no continuous variables.
That is, a~sweep by our generated code is not the same
mathematical operation as a sweep by AugurV2 or JAGS\@.
However, we do not know why JAGS
produces higher-quality samples than AugurV2\@.

For a speed comparison against inference code that has been specialized
and tuned by hand for the same mathematical operation as our generated
code, we also configure MALLET to compute our Gibbs updates, by calling
them 19997-fold cross-validation.  Our generated code is 9\X\ as fast as
MALLET\@, performing an update in $21.32 \pm{} 0.04$ ms while MALLET
takes $189.95 \pm 4.87$ ms.

\begin{figure}
  \includegraphics[width=.5\linewidth]{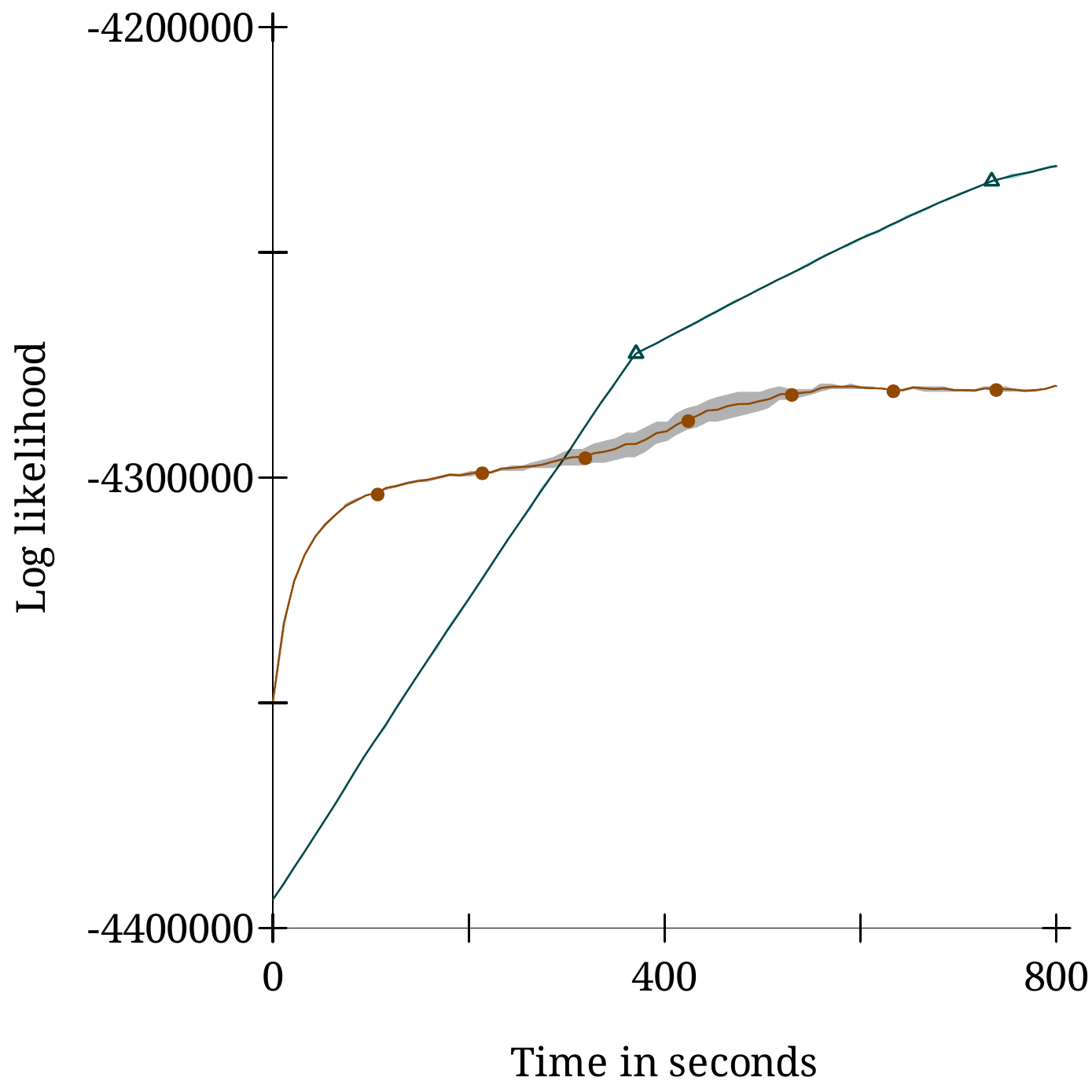}%
  \includegraphics[width=.5\linewidth]{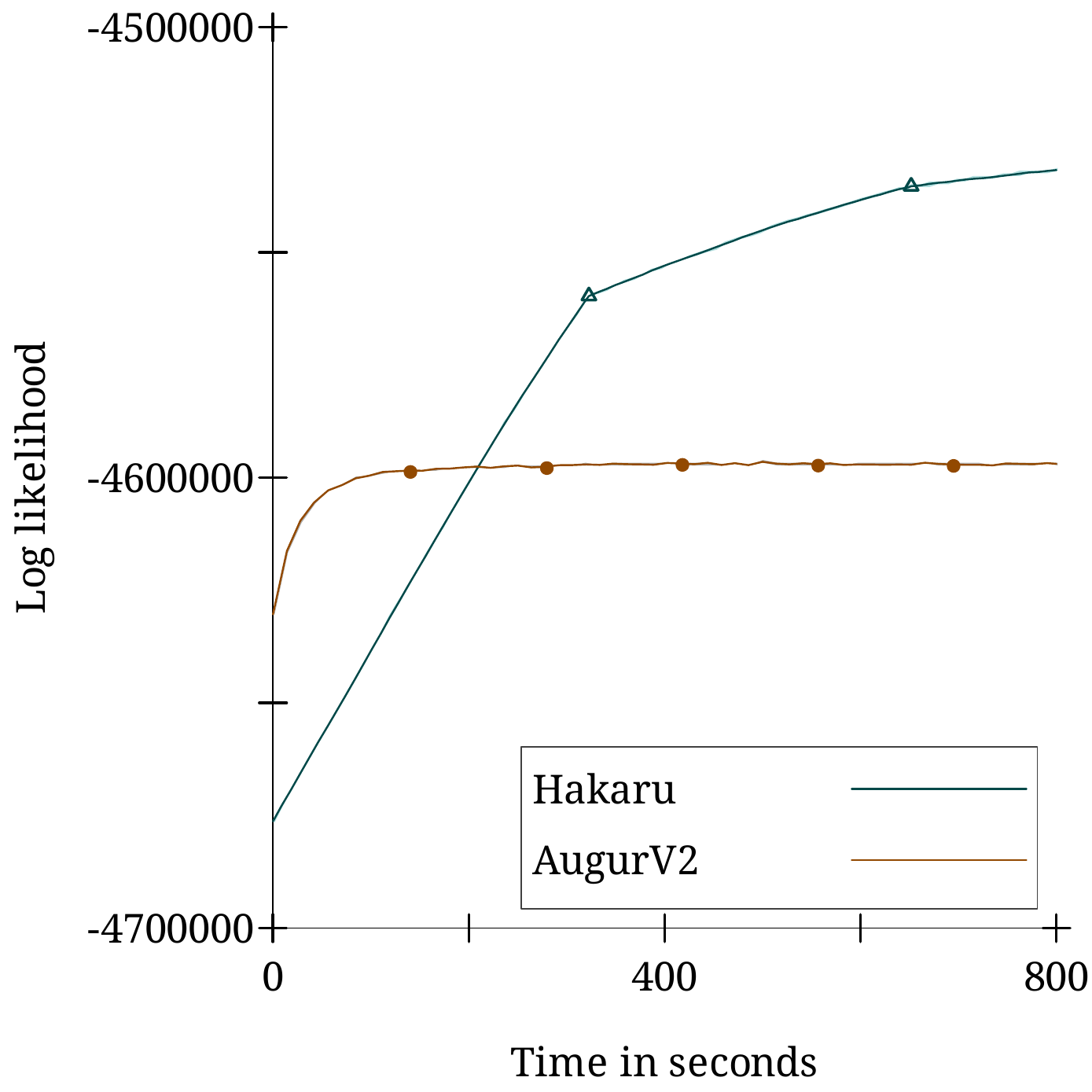}
  \caption{Comparison of Gibbs samplers for the LDA
    model, with $50$ and $100$ topics.  Curves represent mean
    log likelihood over time; shaded area is standard error.  Each mark
    on a curve represents $1$ sweep by Hakaru or $10$ sweeps by
    AugurV2\@.}
  \label{fig:lda}
\end{figure}

\paragraph{Latent Dirichlet Allocation topic model}
The third benchmark applies the LDA model
\citep{blei-latent} to infer topics from the KOS data set
\citep{Dua:2017}, which contains 467714 words drawn from a vocabulary
of 6906\@.
We do not hold out any data.\diff{D}{Explain that \cref{fig:lda} plots log-likelihood, not log-predictive likelihood}

\Cref{fig:lda} plots log likelihood against wall-clock time, for 50
topics and 100 topics, using Hakaru and AugurV2\@.
Here, AugurV2 is more accurate in the first few minutes.
Within 1 sweep, Hakaru's sample likelihood surpasses AugurV2's,
and continues to increase past the bounds of the plot.
We conclude that integrating out latent variables produces a
slower but likelier result on each update.

\paragraph{Compilation and startup time}

Time in the prior figures does not include startup: the time it takes
to initialize a system or generate machine code
for the given model or the given input data.
\Cref{fig:startup} quantifies this startup time separately.  On one
hand, \hakaru has significant ahead-of-time compile time, because the
simplification transformation can take minutes.  We also incur
moderate per-data startup time, for run-time specialization and
machine-code generation.  On the other hand, JAGS incurs negligible
per-model startup time but substantial per-data startup time, because
it unrolls arrays into a graph in memory before sampling.  Moreover,
we have observed the per-data startup time incurred by JAGS to rise
faster than linearly with respect to the input data size. AugurV2\@,
like JAGS\@, does not eliminate latent variables and has negligible
per-model startup time, and like \hakaru has no size-dependent
initialization.
STAN incurs moderate compile and startup times, \diffopen{D}{Discuss STAN burn-in}but its automatic tuning
(\emph{burn-in}) takes tens of minutes, so instead of accounting for
burn-in in \cref{fig:startup}, we show STAN's decent performance and
quick startup by plotting the beginning of burn-in in \cref{fig:gmm}.
We were unable to improve the overall picture by reducing or disabling burn-in.\diffclose{D}

\subsection{Benefits of Each Optimization}
\label{s:ablation}

We perform an ablation study to show how much our optimizations benefit speed.
\Cref{fig:opt-cmp} shows the run time of one sweep of Gibbs sampling with
the larger data size used in~\cref{fig:gmm}. We compare the time with
different optimizations disabled.  We  disable one
optimization at a time, except LICM and loop fusion because loop fusion
requires LICM (\cref{s:licm}).
We never disable simplification (\cref{s:simplify}) because it is
necessary to compile the new class of algorithms at all.
Although these optimizations have a combined
effect, these times give us a general idea of how individual
optimizations affect overall performance.

\input{startup}

\input{ablation}


\diff{AD}{Improved histogram discussion}The measurements show that our performance is made competitive by no
single optimization, but rather by the conjunction of the histogram
transformation and LICM\@: the two optimizations deliver $<2$\X\ speedup
separately but 100\X\ speedup together!
Also, run-time specialization and loop fusion yield 10\X\ and 2\X\
speedups respectively.
We reiterate that it is in the domain of array inference algorithms that
our optimizations can be aggressive and profitable.

\subsection{Exact Inference}


To benchmark exact inference,
we use the \verb|ClinicalTrial| and \verb|LinearRegression| examples from the
R2 system~\citep{nori-r2}.
The \verb|ClinicalTrial| example infers whether a treatment is effective
from the Boolean symptoms of a control group and a treated group of
patients.  The \verb|LinearRegression| example fits a line to
a collection of data points.
In both benchmarks, Bayesian inference efficiently preserves and tracks
the uncertainty of the quantities inferred.  This information can be
useful for making decisions under risk, and is not available through
maximum-likelihood and maximum-a-posteriori estimation (such as ordinary
regression).

For both benchmarks, we compare the code generated by our compilation
pipeline against the code generated by the same pipeline except
replacing the Sham backend (\cref{s:codegen}) by a previous backend that
emits Haskell code.  The latter code is representative of the specialized program
that a practitioner would write by hand in a GPL\@, because
array simplification (\cref{s:simplify}) already delivers that code as
a closed-form formula in both pipelines.
\begin{itemize}
\item
For the \verb|ClinicalTrial| benchmark, the exact solution on $10000$ data
points takes $115.9\,\mu$s to compute (standard deviation $0.1\,\mu$s over
$2000$ trials).
In contrast, the Haskell pipeline takes an average of $409.8\,\mu$s,
which is 3\X\ slower.
\item
For the \verb|LinearRegression| benchmark, the exact solution on $10000$ data
points takes $33\,\mu$s to compute (standard deviation 4\,ns over $2000$ trials).
In contrast, the Haskell pipeline takes an average of $363.4\,\mu$s,
which is 11\X\ slower.
\end{itemize}
These times are orders of magnitude less than even just the startup
times of any approximate inference procedure.

\input{psi}

We also compare the performance of PSI \cite{gehr-psi}, a system for exact
inference that supports arrays, on the two
benchmarks. \diffopen{E}{Updated PSI benchmark results using \texttt{build-release.sh} and the \texttt{--nocheck} flag}\Cref{fig:psi} plots PSI's run times, which increase
with the data size and quickly become prohibitive, because PSI
does not perform compilation and
unrolls all random choices in arrays before reasoning about them.
In both benchmarks, \hakaru is over
5000000\X\ faster while handling over 10\X\ more data.\diffclose{E} Again,
the key to this efficiency is \hakaru's combination of
array transformations and loop optimizations.
\diffopen{E}{Clarified our and PSI's arithmetic}However, also contributing to the speed difference is PSI's use of exact
rational arithmetic throughout.  In contrast, although \hakaru uses
exact arithmetic, it~generates code that uses floating-point arithmetic.\diffclose{E}

\section{Related Work}
\label{s:related}

To situate our work in probabilistic programming, we consider
which components we \emph{specialize} using a domain-specific language and
which components we \emph{reuse} off the shelf.

The difficulty of inference is exacerbated by the ease of composing
models.  To address this, some systems provide a few general-purpose
inference algorithms
\citep{goodman-church,wingate-lightweight,goodman-design,lunn-winbugs,de-salvo-braz-lifted,milch-blog,wu-swift,kiselyov-probabilistic,nori-r2}
or restrict the language to distributions that are continuous
\citep{carpenter-stan}, discrete
\citep{kiselyov-embedded-dsl,pfeffer-design}, or relatively
low-dimensional \citep{gehr-psi}.  Other systems provide a toolbox or
language of inference techniques, so as to specialize
inference to the given model
\citep{figaro,fischer-autobayes,huang-compiling,tristan-augur,wood-new,mansinghka-venture,tran-deep}.
We follow the latter approach.  In particular, by building on prior work on \hakaru~\citep{narayanan-probabilistic,zinkov-composing}, we support a mix of
exact and approximate inference by reusing program transformations such
as simplification and disintegration on
model and inference alike.

Many sophisticated probabilistic programming systems end up (re)implementing
computer algebra
\citep{fischer-autobayes,gehr-psi,de-salvo-braz-lifted-modulo,de-salvo-braz-modulo,tristan-augur,huang-compiling}.
Reusing an existing computer algebra system and
specializing it to the language of patently linear expressions makes it
possible to eliminate latent variables and recognize primitive
distributions without hard-coding patterns such as conjugacy
relationships \citep{carette-simplifying-padl}.  We extend the latter
approach to arrays, further reusing computer algebra to solve
equations in our key \unproduct\ operation.
Our \summarize\ optimization seems related to transforming loops into
list homomorphisms (map-reduce), but we could not find or reuse any
work that makes this relationship clear.

Most probabilistic programming systems either interpret their programs,
or compile or embed them through a GPL\@.
Generating GPU code has also been shown beneficial
\citep{tristan-augur,huang-compiling}.  In contrast, we generate
optimized code through LLVM\@, but specialize our code generation to
take advantage of pure array programs and map-reduce loops.


\begin{acks}                            
  We thank Allen Riddell for help with STAN\@.
  We also thank our anonymous reviewers at PLDI 2018\@, ICFP 2018\@,
    POPL 2019\@, PLDI 2019\@, \emph{and} ICFP 2019\@.

  This research was supported by
  \grantsponsor{DARPA}{DARPA}{https://www.darpa.mil/}
  contract \grantnum{DARPA}{FA8750-14-2-0007},
  \grantsponsor{GS100000001}{NSF}{http://dx.doi.org/10.13039/100000001}
  grants \grantnum{GS100000001}{CCF-1763922} and \grantnum{GS100000001}{CNS-0723054},
  Lilly Endowment, Inc.\ (through its support for the Indiana
  University Pervasive Technology Institute), and the Indiana METACyt
  Initiative.  The Indiana METACyt Initiative at IU is also supported in
  part by Lilly Endowment, Inc.
\end{acks}

\bibliography{pipeline}
\clearpage

\end{document}

%% file: syntax-arrays.tex
\begin{figure}
\leftline{Types\hfil\( T,U \Coloneqq \Real \altern \rplus \altern \Integer \altern
    \iplus \altern \Measure T \altern \Array T \altern \dotsb \)}
\begin{proofrules}[Some primitive distributions (see \cite{carette-simplifying-padl} for more)]
\[
    a : \Real \quad b : \Real
    \justifies \Uniform(a,b) : \Measure\Real
\]
\[
    \mu : \Real \quad \sigma : \rplus
    \justifies \Gaussian(\mu,\sigma) : \Measure\Real
\]
\[
    \alpha : \rplus \quad \beta : \rplus
    \justifies \BetaD(\alpha,\beta) : \Measure\rplus
\]
\[
    e : \Array\rplus
    \justifies \Categorical(e) : \Measure\iplus
\]
\end{proofrules}
\begin{proofrules}[Measure combinators\vphantom{Array}\vspace{-3.5ex}]
\[
    e : T
    \justifies \Ret(e) : \Measure T
\]
\[
    e : \rplus \quad m : \Measure T
    \justifies \Weight(e,m) : \Measure T
\]
\[
    m : \Measure T
    \[ \llap{$[x$} : \rlap{$T]$} \proofoverdots \hphantom{\Measure U}\llap{$m'$} : \Measure U \]
    \justifies \Bind(m,x,m') : \Measure U
\]
\end{proofrules}
\begin{proofrules}[Array constructs\vspace{-3.5ex}]
\[
    e_0 : T \enspace \dotso \enspace e_{n-1} : T
    \justifies \literalArray{e_0,\dotsc,e_{n-1}} : \Array T
\]
\[
    e : \Array T \quad i : \iplus
    \justifies e\idx{i} : T
\]
\[
    n : \iplus \enspace \[ \llap{$[i$} : \rlap{$\iplus]$} \proofoverdots \hphantom{T} \llap{$e$} : T \]
    \justifies \ary(n,i,e) : \Array T
\]~%
\[
    n : \iplus \[ \llap{$[i$} : \rlap{$\iplus]$} \proofoverdots \hphantom{\Measure T} \llap{$m$} : \Measure T \]
    \justifies \Plate(n,i,m) : \Measure(\Array T)
\]
\[
    e : \Array T
    \justifies \size e : \iplus
\]
\end{proofrules}
\caption{Informal term typing rules for distributions and
    (new) for arrays. The bracketed judgments
    indicate the scope of bound variables; for example,
    in~$\Bind(m,x,m')$, the variable~$x$ takes scope over~$m'$ but
    not~$m$.}
\label{fig:syntax}
\end{figure}

%% file: patently-linear.tex
\begin{figure}
$
g \Coloneqq \xh(e)
    \altern  e\cdot g
    \altern  g_1+\dotsb+g_n
    \altern  \IfThenElse(e,g,g)
    \altern  \int_a^b g\,dx
    \altern  \int_\aspace g\,d\vecdotsvec{x}
$\hfil$
\aspace \Coloneqq (a,b) \altern \prod_{i=c}^d \aspace
$
\caption{The grammar of expressions patently linear
    in~$\xh$.  The denotation of~$g$ and the range of~$\xh$
    lie in~$\rplus$.  Metavariables
    $a,b,c,d,e$ stand for expressions, whereas
    $\xh,x,i$ stand for variables.  New is the last
    $g$-production, for integrals over high- and arbitrary-dimensional
    spaces~$\aspace$.
    We omit $g \Coloneqq \sum_{i=a}^{\smash{b}} g$ as we treat distributions
    over~$\Integer$ by analogy to those over~$\Real$.}
\label{fig:patently-linear}
\end{figure}

%% file: forth-arrays.tex
\begin{figure*}
\begin{array}[C]{@{}l<{{}}@{}l@{}}
    \integrate\bigl(\Gaussian(\mu,\sigma),&B,\anintegrand\bigr) = \displaystyle \int_{\prod\nolimits_B (-\infty,\infty)} \Bigl(\prod\nolimits_B \frac{\Exp{-\Frac{(\vecdots{x}\idx{B}-\mu)^2}{2\sigma^2}}}{\sqrt{2\pi}\sigma}\Bigr) \mathinner{\anintegrand(\vecdots{x})} d\vecdots{x} \\
    \integrate\bigl(\Ret(e)              ,&B,\anintegrand\bigr) = \anintegrand\bigl(\ary(B,e)\bigr) \\
    \integrate\bigl(\Weight(e,m)         ,&B,\anintegrand\bigr) = \displaystyle \Bigl(\prod\nolimits_B e\Bigr) \cdot \integrate(m,B,\anintegrand) \\
    \integrate\bigl(\Bind(m,x,m')        ,&B,\anintegrand\bigr) = \integrate\bigl(m,B,\fun{\vecdots{x}} \integrate(\subst{{\vecdots{x}\idx{B}}}{x}{m'},B,\anintegrand)\bigr) \\
    \integrate\bigl(\Plate(e,j,m)        ,&B,\anintegrand\bigr) = \integrate(m,[B,\Loop{j}{0}{e-1}],\anintegrand)
\end{array}

\caption{Converting programs with arrays to patently linear expressions}
\label{fig:forth-arrays}
\end{figure*}

%% file: unproduct.tex
\begin{figure*}
\begin{array}[C]{@{}l<{{}}@{}l@{}>{{}}l@{\quad}Tl@{}}
    \unproduct\bigl(e,&\vec{x}, H\bigr)
    &= \bigl(H[e], \fun{(i,\XI)}1\bigr)
    & if $e$ does not contain $\vec{x}$ free
\\
    \unproduct\bigl(e(\vec{x}\idx{a}),&\vec{x}, H\bigr)
    &= \bigl(1, \fun{(i,\XI)} H[(i=a)\guard e(\XI)] \bigr)
    & if $e$ only uses $\vec{x}$ at index~$a$
\\
    \unproduct\bigl(c^e,&\vec{x}, H^\times\bigr)
    &= \unproduct\bigl(e, \vec{x}, H^\times\bigl[c^{[~]}\bigr]\bigr)
    & where $c$ does not contain $\vec{x}$ free
\\
    \unproduct\bigl(e^c,&\vec{x}, H^\times\bigr)
    &= \unproduct\bigl(e, \vec{x}, H^\times\bigl[[~]^c\bigr]\bigr)
    & where $c$ does not contain $\vec{x}$ free
\\
    \unproduct\bigl(c\cdot e,&\vec{x}, H^+\bigr)
    &= \unproduct\bigl(e, \vec{x}, H^+\bigl[c\cdot[~]\bigr]\bigr)
    & where $c$ does not contain $\vec{x}$ free
\\
    \unproduct\bigl(\prod_{i=a}^b e,&\vec{x}, H^\times\bigr)
    &= \unproduct\bigl(e, \vec{x}, H^\times\bigl[\prod_{i=a}^b[~]\bigr]\bigr)
    & where $a,b$ do not contain $\vec{x}$ free
\\
    \unproduct\bigl(\sum_{i=a}^b e,&\vec{x}, H^+\bigr)
    &= \unproduct\bigl(e, \vec{x}, H^+\bigl[\sum_{i=a}^b[~]\bigr]\bigr)
    & where $a,b$ do not contain $\vec{x}$ free
\\
    \delimiterfactor=850
    \unproduct\bigl(\pw{e_1}{d_1}{e_2}{d_2},&\vec{x}, H\bigr)
    &\multicolumn{2}{@{}l@{}}{{}= \bigl(e'_1\cdot e'_2, g_1\odot g_2\bigr)
    \quad\text{where }(e'_k,g_k) = \unproduct\bigl(e_k,\vec{x},H\bigl[d_k\guard [~]\bigr]\bigr)}
\\
    \unproduct\bigl(e_1 \cdot e_2,&\vec{x}, H^\times\bigr)
    &\multicolumn{2}{@{}l@{}}{{}= \bigl(e'_1\cdot e'_2, g_1\odot g_2\bigr)
    \quad\text{where }(e'_k,g_k) = \unproduct(e_k,\vec{x},H^\times)}
\\
    \unproduct\bigl(e_1 + e_2,&\vec{x}, H^+\bigr)
    &\multicolumn{2}{@{}l@{}}{{}= \bigl(e'_1\cdot e'_2, g_1\odot g_2\bigr)
    \quad\text{where }(e'_k,g_k) = \unproduct(e_k,\vec{x},H^+)}
\\
    \unproduct\bigl(e,&\vec{x}, H\bigr)
    &= \bigl(H[e], \fun{(i,\XI)}1\bigr)
    \makebox[0pt][l]{ as the last resort}
\end{array}
\caption{Rewriting an expression as a product:
    if $\unproduct(e,\vec{x},H) = (e',g)$, then $g$ does not
    contain~$\vec{x}$ free, yet $H[e] = e'\cdot\prod_i g(i,\vec{x}\idx{i})$.
    These rules are applied top-down.
    The first two cases and the last case are the base cases; see the
    text for algebraic reductions that take place in the second case.
    The rest are
    the recursive cases, which simply traverse the
    structure of the input term~$e$ while accumulating the heap~$H$ using
    distributivity.
    In the last three recursive cases, $k$~is $1$ or~$2$,
    and $g_1 \odot g_2$ is short for the pointwise product
    $\fun{(i,\XI)} g_1(i,\XI)\cdot g_2(i,\XI)$.}
\label{fig:unproduct}
\end{figure*}

%% file: reducer.tex
\begin{figure}
\begin{proofrules}[Reducers]
\[
    \[ [j : \iplus] \proofoverdots e : \Real \]
    \justifies \Add(e) \triangleright_j \Real
\]
\[
    b : \iplus
    \[ [j : \iplus] \proofoverdots e : \iplus \]
    \!\!
    \[ [i : \iplus] \proofoverdots r \triangleright_j T \]
    \justifies \Index_i^b(e,r) \triangleright_j \Array T
\]
\[
    \[ [j : \iplus] \proofoverdots e : \Bool \]
    r_1 \triangleright_j T_1 \enspace r_2 \triangleright_j T_2
    \justifies \Split(e,r_1,r_2) \triangleright_j T_1 \times T_2
\]
\[
    r_1 \triangleright_j T_1 \quad r_2 \triangleright_j T_2
    \justifies \Fanout(r_1,r_2) \triangleright_j T_1 \times T_2
\]
\[
    \justifies \Nop \triangleright_j \Unit
\]
\end{proofrules}
\begin{proofrules}[Histograms\vspace{-\proofruleslineskip}]
\[
    a : \iplus \quad b : \iplus \quad
    r \triangleright_j T
    \justifies \Bucket_{j=a}^b(r) : T
\]
\end{proofrules}
\caption{Typing rules for reducer expressions and the histogram
    expressions they constitute}
\label{fig:reducer}
\end{figure}

%% file: summarize.tex
\begin{figure}
\begin{array}[C]{@{}l<{{}}@{}l@{}}
\summarize\bigl(C\bigl[\pwow{e_1}{e}{e_2}\bigr], & j\bigr) =
  \left(\Fanout(m_1,m_2), \fun{(s_1,s_2)} \pwow{f_1(s_1)}{e}{f_2(s_2)}\right)
\\
    \multicolumn{2}{@{\hspace{1pc}}Tl@{}}{%
      where $(m_k,f_k) = \summarize(C[e_k],j)$
      and $e$ does not depend on $j$
    }
\\
\summarize\bigl(C\bigl[\pwow{e_1}{e}{e_2}\bigr],& j\bigr) =
  \bigl(\Split(e,m_1,m_2), \fun{(s_1,s_2)} f_1(s_1) + f_2(s_2)\bigr)
\\
    \multicolumn{2}{@{\hspace{1pc}}Tl@{}}{%
      where $(m_k,f_k) = \summarize(C[e_k],j)$
    }
\\
\summarize\bigl(\pwow{a}{i=e}{0},& j\bigr) =
  \left(\Index_i^m(e, r),
  \fun s \pwow{f(s\idx{i})}{i\in\{0,\dotsc,m-1\}}{0}
  \right)
\\
    \multicolumn{2}{@{\hspace{1pc}}Tl@{}}{where \begin{tabular}[t]{@{}Tl@{}}
      $(r,f) = \summarize(a,j)$, $i$ is a loop-bound variable 
      that does not depend on~$j$, \\ and
      the context entails that $i\in\{0,\dotsc,m-1\}$
                            or $e\in\{0,\dotsc,m-1\}$
    \end{tabular}}
\\
\summarize\bigl(0,& j\bigr) = \bigl(\Nop, \fun s 0\bigr)
\\
\summarize\bigl(e,& j\bigr) = \bigl(\Add(e), \fun s s\bigr)
\\
\end{array}
\caption{Rewriting a summation as a histogram:
    if $\summarize(e,j)=(r,f)$ then
    $\sum_{j=0}^{n-1} e = f\bigl(\Bucket_{j=0}^{n-1}( r )\bigr)$.
The metavariable $C$ denotes a context.
These rules are applied top-down, except the second and third rules
are prioritized by choosing the rule for which the innermost scope of
the free variables $\mathit{FV}(e)\setminus\{j\}$ is outermost.}
\label{fig:summarize}
\end{figure}

%% file: startup.tex
\begin{table}
  \caption{Startup time (mean and standard error in seconds) for different benchmarks and systems before sampling begins}
  \label{fig:startup}
  \begin{tabular}{llr@{${}\pm{}$}lr@{}l@{${}\pm{}$}r@{}l}
    \toprule
    Benchmark & System  & \multicolumn{2}{c}{Compile} & \multicolumn{4}{c}{Startup} \\
    \midrule
      GMM         & \hakaru & 545\hphantom{.0} & 7   &     0&.192 &   0&.002\\
      GMM         & JAGS    & \multicolumn{2}{c}{--} &   223&     &   3&    \\
      GMM         & AugurV2 & \multicolumn{2}{c}{--} &     0&.068 &   0&.001\\
      GMM         & STAN    &  34.3            & 0.1 &     0&.641 &   0&.006\\
      Naive Bayes & \hakaru & 134\hphantom{.0} & 6   &    17&.61  &   0&.09 \\
      Naive Bayes & JAGS    & \multicolumn{2}{c}{--} & 22400&     & 400&    \\
      Naive Bayes & AugurV2 & \multicolumn{2}{c}{--} &     0&.43  &   0&.06 \\
      LDA         & \hakaru & 136\hphantom{.0} & 6   &     2&.904 &   0&.006\\
      LDA         & AugurV2 & \multicolumn{2}{c}{--} &    13&.00  &   0&.08 \\
    \bottomrule
  \end{tabular}
\end{table}

%% file: ablation.tex
\begin{table}
  \caption{Run time in seconds (mean over 1000 trials and standard error) of one sweep of Gibbs sampling with $m=50$ and $n=10000$.
  Slowdown is compared to full optimization.}
\label{fig:opt-cmp}
    \begin{tabular}{lr@{.}l@{${}\pm{}$}lr}
    \toprule
        Optimizations & \multicolumn{3}{c}{Time in seconds} & Slowdown \\
    \midrule
        No optimizations           & 471&4   & 0.6   & 1848\hphantom{.0}\X\\
        No histogram               & 460&6   & 0.2   & 1805\hphantom{.0}\X\\
        No LICM and loop fusion    & 328&7   & 0.1   & 1289\hphantom{.0}\X\\
        No loop fusion             &   0&471 & 0.003 & 1.8\X\\
        No run-time specialization &   2&422 & 0.005 & 9.5\X\\
        Full optimization          &   0&255 & 0.001 & ---\hspace*{1.25em}\\
   \bottomrule\\
   \end{tabular}
\end{table}

%% file: psi.tex
\begin{figure}
\centering
\begin{tikzpicture}[font=\small]
    \begin{axis}[title=\texttt{ClinicalTrial},
                 xmin=0, xmax=90, ymin=0, ymax=700, width=.5\linewidth,
                 axis lines=center,
                 xlabel={Data size}, ylabel={Time in seconds},
                 xlabel near ticks, ylabel near ticks,
                 every axis legend/.append style={anchor=south east, at={(0.98,0.02)}}]
        \addplot+[black, mark size=1pt] coordinates {
(10,0.800)
(20,2.724)
(30,7.372)
(40,22.217)
(50,46.737)
(60,1*60+39.04)
(70,3*60+20.62)
(80,6*60+14.27)
(90,11*60+25.39)
        };
    \end{axis}
\end{tikzpicture}\hfil
\begin{tikzpicture}[font=\small]
    \begin{axis}[title=\texttt{LinearRegression},
                 xmin=0, xmax=900, ymin=0, ymax=250, width=.5\linewidth,
                 axis lines=center,
                 xlabel={Data size}, ylabel={Time in seconds},
                 xlabel near ticks, ylabel near ticks,
                 every axis legend/.append style={anchor=south east, at={(0.98,0.02)}}]
        \addplot+[black, mark size=1pt] coordinates {
(10,0.149)
(20,0.236)
(30,0.325)
(40,0.432)
(50,0.543)
(60,0.693)
(70,0.830)
(80,1.018)
(90,1.150)
(100,1.227)
(200,3.814)
(300,9.437)
(400,18.939)
(500,37.049)
(600,1*60+02.82)
(700,1*60+46.04)
(800,2*60+43.89)
(900,3*60+55.41)
        };
        \addlegendentry{PSI}
    \end{axis}
\end{tikzpicture}%
\caption{PSI performance on exact-inference benchmarks, using \texttt{build-release.sh} and the \texttt{--nocheck} flag}
\label{fig:psi}
\end{figure}
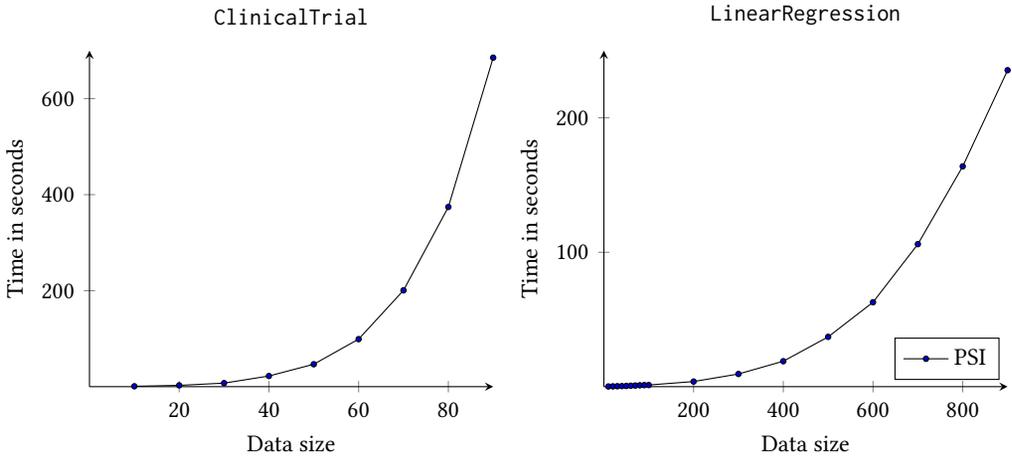

%% file: pipeline.bbl

\begin{thebibliography}{76}


\ifx \showCODEN    \undefined \def \showCODEN     #1{\unskip}     \fi
\ifx \showDOI      \undefined \def \showDOI       #1{#1}\fi
\ifx \showISBNx    \undefined \def \showISBNx     #1{\unskip}     \fi
\ifx \showISBNxiii \undefined \def \showISBNxiii  #1{\unskip}     \fi
\ifx \showISSN     \undefined \def \showISSN      #1{\unskip}     \fi
\ifx \showLCCN     \undefined \def \showLCCN      #1{\unskip}     \fi
\ifx \shownote     \undefined \def \shownote      #1{#1}          \fi
\ifx \showarticletitle \undefined \def \showarticletitle #1{#1}   \fi
\ifx \showURL      \undefined \def \showURL       {\relax}        \fi
\providecommand\bibfield[2]{#2}
\providecommand\bibinfo[2]{#2}
\providecommand\natexlab[1]{#1}
\providecommand\showeprint[2][]{arXiv:#2}

\bibitem[\protect\citeauthoryear{Aho, Sethi, and Ullman}{Aho
  et~al\mbox{.}}{1986}]%
        {aho-dragon}
\bibfield{author}{\bibinfo{person}{Alfred~V. Aho}, \bibinfo{person}{Ravi
  Sethi}, {and} \bibinfo{person}{Jeffrey~D. Ullman}.}
  \bibinfo{year}{1986}\natexlab{}.
\newblock \bibinfo{booktitle}{\emph{Compilers: Principles, Techniques, and
  Tools}}.
\newblock \bibinfo{publisher}{Addison-Wesley Longman Publishing Co., Inc.},
  \bibinfo{address}{Boston, MA, USA}.
\newblock
\showISBNx{0-201-10088-6}


\bibitem[\protect\citeauthoryear{Bayes}{Bayes}{1763}]%
        {bayes-essay}
\bibfield{author}{\bibinfo{person}{Thomas Bayes}.}
  \bibinfo{year}{1763}\natexlab{}.
\newblock \showarticletitle{An Essay towards Solving a Problem in the Doctrine
  of Chances}.
\newblock \bibinfo{journal}{\emph{Philosophical Transactions of the Royal
  Society of London}}  \bibinfo{volume}{53} (\bibinfo{year}{1763}),
  \bibinfo{pages}{370--418}.
\newblock


\bibitem[\protect\citeauthoryear{Betancourt}{Betancourt}{2017}]%
        {betancourt-conceptual}
\bibfield{author}{\bibinfo{person}{Michael Betancourt}.}
  \bibinfo{year}{2017}\natexlab{}.
\newblock \bibinfo{booktitle}{\emph{A Conceptual Introduction to {H}amiltonian
  {M}onte {C}arlo}}.
\newblock \bibinfo{type}{e-Print} 1701.02434.
  \bibinfo{institution}{ar{X}iv.org}.
\newblock
\urldef\tempurl%
\url{https://arxiv.org/abs/1701.02434}
\showURL{%
\tempurl}


\bibitem[\protect\citeauthoryear{Blackwell}{Blackwell}{1947}]%
        {blackwell-conditional}
\bibfield{author}{\bibinfo{person}{David Blackwell}.}
  \bibinfo{year}{1947}\natexlab{}.
\newblock \showarticletitle{Conditional Expectation and Unbiased Sequential
  Estimation}.
\newblock \bibinfo{journal}{\emph{The Annals of Mathematical Statistics}}
  \bibinfo{volume}{18}, \bibinfo{number}{1} (\bibinfo{date}{March}
  \bibinfo{year}{1947}), \bibinfo{pages}{105--110}.
\newblock


\bibitem[\protect\citeauthoryear{Blei, Ng, and Jordan}{Blei
  et~al\mbox{.}}{2003}]%
        {blei-latent}
\bibfield{author}{\bibinfo{person}{David~M. Blei}, \bibinfo{person}{Andrew~Y.
  Ng}, {and} \bibinfo{person}{Michael~I. Jordan}.}
  \bibinfo{year}{2003}\natexlab{}.
\newblock \showarticletitle{Latent {D}irichlet Allocation}.
\newblock \bibinfo{journal}{\emph{Journal of Machine Learning Research}}
  \bibinfo{volume}{3}, \bibinfo{number}{Jan.} (\bibinfo{date}{Jan.}
  \bibinfo{year}{2003}), \bibinfo{pages}{993--1022}.
\newblock


\bibitem[\protect\citeauthoryear{Borgstr{\"o}m, Gordon, Ouyang, Russo, Scibior,
  and Szymczak}{Borgstr{\"o}m et~al\mbox{.}}{2016}]%
        {borgstrom-fabular}
\bibfield{author}{\bibinfo{person}{Johannes Borgstr{\"o}m},
  \bibinfo{person}{Andrew~D. Gordon}, \bibinfo{person}{Long Ouyang},
  \bibinfo{person}{Claudio~V. Russo}, \bibinfo{person}{Adam Scibior}, {and}
  \bibinfo{person}{Marcin Szymczak}.} \bibinfo{year}{2016}\natexlab{}.
\newblock \showarticletitle{{F}abular: Regression Formulas as Probabilistic
  Programming}. In \bibinfo{booktitle}{\emph{Proceedings of the 43th Symposium
  on Principles of Programming Languages ({POPL})}}. \bibinfo{publisher}{ACM
  Press}, \bibinfo{pages}{271--283}.
\newblock


\bibitem[\protect\citeauthoryear{Buntine}{Buntine}{1994}]%
        {buntine-operations}
\bibfield{author}{\bibinfo{person}{Wray~L. Buntine}.}
  \bibinfo{year}{1994}\natexlab{}.
\newblock \showarticletitle{Operations for Learning with Graphical Models}.
\newblock \bibinfo{journal}{\emph{Journal of Artificial Intelligence Research}}
   \bibinfo{volume}{2} (\bibinfo{year}{1994}), \bibinfo{pages}{159--225}.
\newblock


\bibitem[\protect\citeauthoryear{Carette and Shan}{Carette and Shan}{2016}]%
        {carette-simplifying-padl}
\bibfield{author}{\bibinfo{person}{Jacques Carette} {and}
  \bibinfo{person}{Chung-chieh Shan}.} \bibinfo{year}{2016}\natexlab{}.
\newblock \showarticletitle{Simplifying Probabilistic Programs Using Computer
  Algebra}. In \bibinfo{booktitle}{\emph{Practical Aspects of Declarative
  Languages: 18th International Symposium, {PADL} 2016}}
  \emph{(\bibinfo{series}{{L}ecture {N}otes in {C}omputer {S}cience})},
  \bibfield{editor}{\bibinfo{person}{Marco Gavanelli} {and}
  \bibinfo{person}{John~H. Reppy}} (Eds.). \bibinfo{pages}{135--152}.
\newblock


\bibitem[\protect\citeauthoryear{Carpenter, Gelman, Hoffman, Lee, Goodrich,
  Betancourt, Brubaker, Guo, Li, and Riddell}{Carpenter et~al\mbox{.}}{2017}]%
        {carpenter-stan}
\bibfield{author}{\bibinfo{person}{Bob Carpenter}, \bibinfo{person}{Andrew
  Gelman}, \bibinfo{person}{Matthew Hoffman}, \bibinfo{person}{Daniel Lee},
  \bibinfo{person}{Ben Goodrich}, \bibinfo{person}{Michael Betancourt},
  \bibinfo{person}{Marcus Brubaker}, \bibinfo{person}{Jiqiang Guo},
  \bibinfo{person}{Peter Li}, {and} \bibinfo{person}{Allen Riddell}.}
  \bibinfo{year}{2017}\natexlab{}.
\newblock \showarticletitle{{S}tan: A Probabilistic Programming Language}.
\newblock \bibinfo{journal}{\emph{Journal of Statistical Software}}
  \bibinfo{volume}{76}, \bibinfo{number}{1} (\bibinfo{year}{2017}),
  \bibinfo{pages}{1--32}.
\newblock


\bibitem[\protect\citeauthoryear{Casella and Robert}{Casella and
  Robert}{1996}]%
        {casella1996rao}
\bibfield{author}{\bibinfo{person}{George Casella} {and}
  \bibinfo{person}{Christian~P. Robert}.} \bibinfo{year}{1996}\natexlab{}.
\newblock \showarticletitle{{R}ao-{B}lackwellisation of Sampling Schemes}.
\newblock \bibinfo{journal}{\emph{Biometrika}} \bibinfo{volume}{83},
  \bibinfo{number}{1} (\bibinfo{year}{1996}), \bibinfo{pages}{81--94}.
\newblock


\bibitem[\protect\citeauthoryear{Chyzak and Salvy}{Chyzak and Salvy}{1998}]%
        {ChSa98}
\bibfield{author}{\bibinfo{person}{Fr{\'e}d{\'e}ric Chyzak} {and}
  \bibinfo{person}{Bruno Salvy}.} \bibinfo{year}{1998}\natexlab{}.
\newblock \showarticletitle{Non-commutative Elimination in {O}re Algebras
  Proves Multivariate Holonomic Identities}.
\newblock \bibinfo{journal}{\emph{Journal of Symbolic Computation}}
  \bibinfo{volume}{26}, \bibinfo{number}{2} (\bibinfo{year}{1998}),
  \bibinfo{pages}{187--227}.
\newblock


\bibitem[\protect\citeauthoryear{Cook, Gelman, and Rubin}{Cook
  et~al\mbox{.}}{2006}]%
        {cook-validation}
\bibfield{author}{\bibinfo{person}{Samantha~R. Cook}, \bibinfo{person}{Andrew
  Gelman}, {and} \bibinfo{person}{Donald~B. Rubin}.}
  \bibinfo{year}{2006}\natexlab{}.
\newblock \showarticletitle{Validation of Software for {B}ayesian Models Using
  Posterior Quantiles}.
\newblock \bibinfo{journal}{\emph{Journal of Computational and Graphical
  Statistics}} \bibinfo{volume}{15}, \bibinfo{number}{3}
  (\bibinfo{year}{2006}), \bibinfo{pages}{675--692}.
\newblock


\bibitem[\protect\citeauthoryear{De~Raedt, Kimmig, and Toivonen}{De~Raedt
  et~al\mbox{.}}{2007}]%
        {de-raedt-problog}
\bibfield{author}{\bibinfo{person}{Luc De~Raedt}, \bibinfo{person}{Angelika
  Kimmig}, {and} \bibinfo{person}{Hannu Toivonen}.}
  \bibinfo{year}{2007}\natexlab{}.
\newblock \showarticletitle{{P}rob{L}og: A Probabilistic {P}rolog and its
  Application in Link Discovery}. In \bibinfo{booktitle}{\emph{Proceedings of
  the 20th International Joint Conference on Artificial Intelligence}},
  \bibfield{editor}{\bibinfo{person}{Manuela~M. Veloso}} (Ed.).
  \bibinfo{pages}{2462--2467}.
\newblock


\bibitem[\protect\citeauthoryear{de~Salvo~Braz, Amir, and Roth}{de~Salvo~Braz
  et~al\mbox{.}}{2007}]%
        {de-salvo-braz-lifted}
\bibfield{author}{\bibinfo{person}{Rodrigo de Salvo~Braz},
  \bibinfo{person}{Eyal Amir}, {and} \bibinfo{person}{Dan Roth}.}
  \bibinfo{year}{2007}\natexlab{}.
\newblock \showarticletitle{Lifted First-Order Probabilistic Inference}.
\newblock In \bibinfo{booktitle}{\emph{Introduction to Statistical Relational
  Learning}}, \bibfield{editor}{\bibinfo{person}{Lise Getoor} {and}
  \bibinfo{person}{Ben Taskar}} (Eds.). \bibinfo{publisher}{{MIT} Press},
  \bibinfo{pages}{433--451}.
\newblock


\bibitem[\protect\citeauthoryear{de~Salvo~Braz and O'Reilly}{de~Salvo~Braz and
  O'Reilly}{2017}]%
        {de-salvo-braz-lifted-modulo}
\bibfield{author}{\bibinfo{person}{Rodrigo de Salvo~Braz} {and}
  \bibinfo{person}{Ciaran O'Reilly}.} \bibinfo{year}{2017}\natexlab{}.
\newblock \showarticletitle{Exact Inference for Relational Graphical Models
  with Interpreted Functions: Lifted Probabilistic Inference Modulo Theories},
  \bibfield{editor}{\bibinfo{person}{Gal Elidan}, \bibinfo{person}{Kristian
  Kersting}, {and} \bibinfo{person}{Alexander~T. Ihler}} (Eds.).
  \bibinfo{publisher}{{AUAI} Press}.
\newblock


\bibitem[\protect\citeauthoryear{de~Salvo~Braz, O'Reilly, Gogate, and
  Dechter}{de~Salvo~Braz et~al\mbox{.}}{2016}]%
        {de-salvo-braz-modulo}
\bibfield{author}{\bibinfo{person}{Rodrigo de Salvo~Braz},
  \bibinfo{person}{Ciaran O'Reilly}, \bibinfo{person}{Vibhav Gogate}, {and}
  \bibinfo{person}{Rina Dechter}.} \bibinfo{year}{2016}\natexlab{}.
\newblock \showarticletitle{Probabilistic Inference Modulo Theories}. In
  \bibinfo{booktitle}{\emph{Proceedings of the 25th International Joint
  Conference on Artificial Intelligence}},
  \bibfield{editor}{\bibinfo{person}{Subbarao Kambhampati}} (Ed.).
  \bibinfo{publisher}{{AAAI} Press}, \bibinfo{pages}{3591--3599}.
\newblock
\showISBNx{978-1-57735-770-4}
\urldef\tempurl%
\url{http://www.ijcai.org/Abstract/16/506}
\showURL{%
\tempurl}


\bibitem[\protect\citeauthoryear{Dechter}{Dechter}{1998}]%
        {dechter-bucket}
\bibfield{author}{\bibinfo{person}{Rina Dechter}.}
  \bibinfo{year}{1998}\natexlab{}.
\newblock \showarticletitle{Bucket Elimination: A Unifying Framework for
  Probabilistic Inference}.
\newblock In \bibinfo{booktitle}{\emph{Learning and Inference in Graphical
  Models}}, \bibfield{editor}{\bibinfo{person}{Michael~I. Jordan}} (Ed.).
  \bibinfo{publisher}{Kluwer}, \bibinfo{address}{Dordrecht}.
\newblock
\newblock
\shownote{Paperback: {\it Learning in Graphical Models\/}, {MIT} Press.}


\bibitem[\protect\citeauthoryear{Dheeru and Karra~Taniskidou}{Dheeru and
  Karra~Taniskidou}{2017}]%
        {Dua:2017}
\bibfield{author}{\bibinfo{person}{Dua Dheeru} {and} \bibinfo{person}{Efi
  Karra~Taniskidou}.} \bibinfo{year}{2017}\natexlab{}.
\newblock \bibinfo{title}{{UCI} Machine Learning Repository}.
\newblock
\newblock
\urldef\tempurl%
\url{http://archive.ics.uci.edu/ml}
\showURL{%
\tempurl}


\bibitem[\protect\citeauthoryear{Fischer and Schumann}{Fischer and
  Schumann}{2003}]%
        {fischer-autobayes}
\bibfield{author}{\bibinfo{person}{Bernd Fischer} {and} \bibinfo{person}{Johann
  Schumann}.} \bibinfo{year}{2003}\natexlab{}.
\newblock \showarticletitle{{A}uto{B}ayes: A System for Generating Data
  Analysis Programs from Statistical Models}.
\newblock \bibinfo{journal}{\emph{Journal of Functional Programming}}
  \bibinfo{volume}{13}, \bibinfo{number}{3} (\bibinfo{year}{2003}),
  \bibinfo{pages}{483--508}.
\newblock


\bibitem[\protect\citeauthoryear{Flanagan, Sabry, Duba, and Felleisen}{Flanagan
  et~al\mbox{.}}{1993}]%
        {flanagan-anf}
\bibfield{author}{\bibinfo{person}{Cormac Flanagan}, \bibinfo{person}{Amr
  Sabry}, \bibinfo{person}{Bruce~F. Duba}, {and} \bibinfo{person}{Matthias
  Felleisen}.} \bibinfo{year}{1993}\natexlab{}.
\newblock \showarticletitle{The Essence of Compiling with Continuations}. In
  \bibinfo{booktitle}{\emph{Proceedings of the ACM SIGPLAN 1993 Conference on
  Programming Language Design and Implementation}} \emph{(\bibinfo{series}{PLDI
  '93})}. \bibinfo{publisher}{ACM}, \bibinfo{address}{New York, NY, USA},
  \bibinfo{pages}{237--247}.
\newblock
\showISBNx{0-89791-598-4}
\urldef\tempurl%
\url{https://doi.org/10.1145/155090.155113}
\showDOI{\tempurl}


\bibitem[\protect\citeauthoryear{Gehr, Misailovic, and Vechev}{Gehr
  et~al\mbox{.}}{2016}]%
        {gehr-psi}
\bibfield{author}{\bibinfo{person}{Timon Gehr}, \bibinfo{person}{Sasa
  Misailovic}, {and} \bibinfo{person}{Martin~T. Vechev}.}
  \bibinfo{year}{2016}\natexlab{}.
\newblock \showarticletitle{{PSI}: Exact Symbolic Inference for Probabilistic
  Programs}. In \bibinfo{booktitle}{\emph{Proceedings of the 28th International
  Conference on Computer Aided Verification, Part {I}}}
  \emph{(\bibinfo{series}{{L}ecture {N}otes in {C}omputer {S}cience})},
  \bibfield{editor}{\bibinfo{person}{Swarat Chaudhuri} {and}
  \bibinfo{person}{Azadeh Farzan}} (Eds.). \bibinfo{publisher}{Springer},
  \bibinfo{pages}{62--83}.
\newblock
\showISBNx{978-3-319-41527-7}


\bibitem[\protect\citeauthoryear{Gelfand and Smith}{Gelfand and Smith}{1990}]%
        {gelfand-sampling-based}
\bibfield{author}{\bibinfo{person}{Alan~E. Gelfand} {and}
  \bibinfo{person}{Adrian F.~M. Smith}.} \bibinfo{year}{1990}\natexlab{}.
\newblock \showarticletitle{Sampling-Based Approaches to Calculating Marginal
  Densities}.
\newblock \bibinfo{journal}{\emph{J. Amer. Statist. Assoc.}}
  \bibinfo{volume}{85}, \bibinfo{number}{410} (\bibinfo{year}{1990}),
  \bibinfo{pages}{398--409}.
\newblock


\bibitem[\protect\citeauthoryear{Gelman, Carlin, Stern, Dunson, Vehtari, and
  Rubin}{Gelman et~al\mbox{.}}{2014}]%
        {gelman-bayesian}
\bibfield{author}{\bibinfo{person}{Andrew Gelman}, \bibinfo{person}{John~B.
  Carlin}, \bibinfo{person}{Hal~S. Stern}, \bibinfo{person}{David~B. Dunson},
  \bibinfo{person}{Aki Vehtari}, {and} \bibinfo{person}{Donald~B. Rubin}.}
  \bibinfo{year}{2014}\natexlab{}.
\newblock \bibinfo{booktitle}{\emph{{B}ayesian Data Analysis}
  (\bibinfo{edition}{third} ed.)}.
\newblock \bibinfo{publisher}{CRC Press}.
\newblock


\bibitem[\protect\citeauthoryear{Geweke}{Geweke}{2004}]%
        {geweke-getting}
\bibfield{author}{\bibinfo{person}{John Geweke}.}
  \bibinfo{year}{2004}\natexlab{}.
\newblock \showarticletitle{Getting It Right}.
\newblock \bibinfo{journal}{\emph{J. Amer. Statist. Assoc.}}
  \bibinfo{volume}{99}, \bibinfo{number}{467} (\bibinfo{year}{2004}),
  \bibinfo{pages}{799--804}.
\newblock


\bibitem[\protect\citeauthoryear{Giry}{Giry}{1982}]%
        {giry-categorical}
\bibfield{author}{\bibinfo{person}{Mich{\`e}le Giry}.}
  \bibinfo{year}{1982}\natexlab{}.
\newblock \showarticletitle{A Categorical Approach to Probability Theory}. In
  \bibinfo{booktitle}{\emph{Categorical Aspects of Topology and Analysis:
  Proceedings of an International Conference Held at {C}arleton {U}niversity,
  {O}ttawa, {A}ugust 11--15, 1981}},
  \bibfield{editor}{\bibinfo{person}{Bernhard Banaschewski}} (Ed.).
  \bibinfo{publisher}{Springer}, \bibinfo{pages}{68--85}.
\newblock


\bibitem[\protect\citeauthoryear{Goodman, Mansinghka, Roy, Bonawitz, and
  Tenenbaum}{Goodman et~al\mbox{.}}{2008}]%
        {goodman-church}
\bibfield{author}{\bibinfo{person}{Noah~D. Goodman}, \bibinfo{person}{Vikash~K.
  Mansinghka}, \bibinfo{person}{Daniel Roy}, \bibinfo{person}{Keith Bonawitz},
  {and} \bibinfo{person}{Joshua~B. Tenenbaum}.}
  \bibinfo{year}{2008}\natexlab{}.
\newblock \showarticletitle{{C}hurch: A Language for Generative Models}. In
  \bibinfo{booktitle}{\emph{Proceedings of the 24th Conference on Uncertainty
  in Artificial Intelligence}}, \bibfield{editor}{\bibinfo{person}{David~Allen
  McAllester} {and} \bibinfo{person}{Petri Myllym{\"a}ki}} (Eds.).
  \bibinfo{pages}{220--229}.
\newblock


\bibitem[\protect\citeauthoryear{Goodman and Stuhlm{\"u}ller}{Goodman and
  Stuhlm{\"u}ller}{2014}]%
        {goodman-design}
\bibfield{author}{\bibinfo{person}{Noah~D. Goodman} {and}
  \bibinfo{person}{Andreas Stuhlm{\"u}ller}.} \bibinfo{year}{2014}\natexlab{}.
\newblock \bibinfo{title}{The Design and Implementation of Probabilistic
  Programming Languages}.
\newblock \bibinfo{howpublished}{\url{http://dippl.org}}.
\newblock


\bibitem[\protect\citeauthoryear{Griffiths and Steyvers}{Griffiths and
  Steyvers}{2004}]%
        {griffiths-finding}
\bibfield{author}{\bibinfo{person}{Thomas~L. Griffiths} {and}
  \bibinfo{person}{Mark Steyvers}.} \bibinfo{year}{2004}\natexlab{}.
\newblock \showarticletitle{Finding Scientific Topics}.
\newblock \bibinfo{journal}{\emph{Proceedings of the National Academy of
  Sciences}} \bibinfo{volume}{101}, \bibinfo{number}{suppl 1}
  (\bibinfo{year}{2004}), \bibinfo{pages}{5228--5235}.
\newblock
\urldef\tempurl%
\url{https://www.pnas.org/content/101/suppl_1/5228}
\showURL{%
\tempurl}


\bibitem[\protect\citeauthoryear{Hoffman and Gelman}{Hoffman and
  Gelman}{2014}]%
        {hoffman-nuts}
\bibfield{author}{\bibinfo{person}{Matthew~D. Hoffman} {and}
  \bibinfo{person}{Andrew Gelman}.} \bibinfo{year}{2014}\natexlab{}.
\newblock \showarticletitle{The {N}o-{U}-{T}urn {S}ampler: Adaptively Setting
  Path Lengths in {H}amiltonian {M}onte {C}arlo}.
\newblock \bibinfo{journal}{\emph{Journal of Machine Learning Research}}
  \bibinfo{volume}{15}, \bibinfo{number}{1} (\bibinfo{year}{2014}),
  \bibinfo{pages}{1593--1623}.
\newblock


\bibitem[\protect\citeauthoryear{Hoffman, Johnson, and Tran}{Hoffman
  et~al\mbox{.}}{2018}]%
        {hoffman-autoconj}
\bibfield{author}{\bibinfo{person}{Matthew~D. Hoffman},
  \bibinfo{person}{Matthew~J. Johnson}, {and} \bibinfo{person}{Dustin Tran}.}
  \bibinfo{year}{2018}\natexlab{}.
\newblock \showarticletitle{{A}utoconj: Recognizing and Exploiting Conjugacy
  Without a Domain-Specific Language}. In \bibinfo{booktitle}{\emph{Advances in
  Neural Information Processing Systems}},
  \bibfield{editor}{\bibinfo{person}{Samy Bengio}, \bibinfo{person}{Hanna~M.
  Wallach}, \bibinfo{person}{Hugo Larochelle}, \bibinfo{person}{Kristen
  Grauman}, \bibinfo{person}{Nicol\`{o} Cesa-Bianchi}, {and}
  \bibinfo{person}{Roman Garnett}} (Eds.). \bibinfo{pages}{10739--10749}.
\newblock
\urldef\tempurl%
\url{http://papers.nips.cc/paper/8270-autoconj-recognizing-and-exploiting-conjugacy-without-a-domain-specific-language.pdf}
\showURL{%
\tempurl}


\bibitem[\protect\citeauthoryear{Huang, Tristan, and Morrisett}{Huang
  et~al\mbox{.}}{2017}]%
        {huang-compiling}
\bibfield{author}{\bibinfo{person}{Daniel Huang},
  \bibinfo{person}{Jean-Baptiste Tristan}, {and} \bibinfo{person}{Greg
  Morrisett}.} \bibinfo{year}{2017}\natexlab{}.
\newblock \showarticletitle{Compiling {M}arkov Chain {M}onte {C}arlo Algorithms
  for Probabilistic Modeling}. In \bibinfo{booktitle}{\emph{{PLDI} '17:
  Proceedings of the {ACM} Conference on Programming Language Design and
  Implementation}}, \bibfield{editor}{\bibinfo{person}{Albert Cohen} {and}
  \bibinfo{person}{Martin~T. Vechev}} (Eds.). \bibinfo{publisher}{ACM Press},
  \bibinfo{pages}{111--125}.
\newblock
\showISBNx{978-1-4503-4988-8}


\bibitem[\protect\citeauthoryear{Joachims}{Joachims}{1997}]%
        {newsgroups}
\bibfield{author}{\bibinfo{person}{Thorsten Joachims}.}
  \bibinfo{year}{1997}\natexlab{}.
\newblock \showarticletitle{A Probabilistic Analysis of the {R}occhio Algorithm
  with {TFIDF} for Text Categorization}. In
  \bibinfo{booktitle}{\emph{Proceedings of the Fourteenth International
  Conference on Machine Learning}} \emph{(\bibinfo{series}{ICML '97})}.
  \bibinfo{publisher}{Morgan Kaufmann Publishers Inc.}, \bibinfo{address}{San
  Francisco, CA, USA}, \bibinfo{pages}{143--151}.
\newblock
\showISBNx{1-55860-486-3}
\urldef\tempurl%
\url{http://dl.acm.org/citation.cfm?id=645526.657278}
\showURL{%
\tempurl}


\bibitem[\protect\citeauthoryear{Kauers}{Kauers}{2013}]%
        {Kauers2013}
\bibfield{author}{\bibinfo{person}{Manuel Kauers}.}
  \bibinfo{year}{2013}\natexlab{}.
\newblock \showarticletitle{The Holonomic Toolkit}.
\newblock In \bibinfo{booktitle}{\emph{Computer Algebra in Quantum Field
  Theory}}, \bibfield{editor}{\bibinfo{person}{Carsten Schneider} {and}
  \bibinfo{person}{Johannes Bl{\"u}mlein}} (Eds.).
  \bibinfo{publisher}{Springer}, \bibinfo{pages}{119--144}.
\newblock


\bibitem[\protect\citeauthoryear{Kiselyov}{Kiselyov}{2016}]%
        {kiselyov-probabilistic}
\bibfield{author}{\bibinfo{person}{Oleg Kiselyov}.}
  \bibinfo{year}{2016}\natexlab{}.
\newblock \showarticletitle{Probabilistic Programming Language and its
  Incremental Evaluation}. In \bibinfo{booktitle}{\emph{Proceedings of {APLAS}
  2016: 14th {A}sian Symposium on Programming Languages and Systems}}
  \emph{(\bibinfo{series}{{L}ecture {N}otes in {C}omputer {S}cience})},
  \bibfield{editor}{\bibinfo{person}{Atsushi Igarashi}} (Ed.).
  \bibinfo{publisher}{Springer}, \bibinfo{pages}{357--376}.
\newblock


\bibitem[\protect\citeauthoryear{Kiselyov and Shan}{Kiselyov and Shan}{2009}]%
        {kiselyov-embedded-dsl}
\bibfield{author}{\bibinfo{person}{Oleg Kiselyov} {and}
  \bibinfo{person}{Chung-chieh Shan}.} \bibinfo{year}{2009}\natexlab{}.
\newblock \showarticletitle{Embedded Probabilistic Programming}. In
  \bibinfo{booktitle}{\emph{Proceedings of the Working Conference on
  Domain-Specific Languages}} \emph{(\bibinfo{series}{{L}ecture {N}otes in
  {C}omputer {S}cience})}, \bibfield{editor}{\bibinfo{person}{Walid~Mohamed
  Taha}} (Ed.). \bibinfo{publisher}{Springer}, \bibinfo{pages}{360--384}.
\newblock


\bibitem[\protect\citeauthoryear{Koller and Friedman}{Koller and
  Friedman}{2009}]%
        {koller-pgm}
\bibfield{author}{\bibinfo{person}{Daphne Koller} {and} \bibinfo{person}{Nir
  Friedman}.} \bibinfo{year}{2009}\natexlab{}.
\newblock \bibinfo{booktitle}{\emph{Probabilistic Graphical Models: Principles
  and Techniques}}.
\newblock \bibinfo{publisher}{{MIT} Press}.
\newblock


\bibitem[\protect\citeauthoryear{Kolmogorov}{Kolmogorov}{1950}]%
        {kolmogorov-unbiased}
\bibfield{author}{\bibinfo{person}{Andrey~N. Kolmogorov}.}
  \bibinfo{year}{1950}\natexlab{}.
\newblock \showarticletitle{Unbiased Estimates}.
\newblock \bibinfo{journal}{\emph{Izvestiya Akademii Nauk SSSR Seriya
  Matematicheskaya}} \bibinfo{volume}{14}, \bibinfo{number}{4}
  (\bibinfo{year}{1950}), \bibinfo{pages}{303--326}.
\newblock


\bibitem[\protect\citeauthoryear{Liu}{Liu}{1994}]%
        {liu-collapsed}
\bibfield{author}{\bibinfo{person}{Jun~S. Liu}.}
  \bibinfo{year}{1994}\natexlab{}.
\newblock \showarticletitle{The Collapsed {G}ibbs Sampler in {B}ayesian
  Computations with Applications to a Gene Regulation Problem}.
\newblock \bibinfo{journal}{\emph{J. Amer. Statist. Assoc.}}
  \bibinfo{volume}{89}, \bibinfo{number}{427} (\bibinfo{year}{1994}),
  \bibinfo{pages}{958--966}.
\newblock


\bibitem[\protect\citeauthoryear{Liu, Wong, and Kong}{Liu
  et~al\mbox{.}}{1994}]%
        {liu-covariance}
\bibfield{author}{\bibinfo{person}{Jun~S. Liu}, \bibinfo{person}{Wing~Hung
  Wong}, {and} \bibinfo{person}{Augustine Kong}.}
  \bibinfo{year}{1994}\natexlab{}.
\newblock \showarticletitle{Covariance Structure of the {G}ibbs Sampler with
  Applications to the Comparisons of Estimators and Augmentation Schemes}.
\newblock \bibinfo{journal}{\emph{Biometrika}} \bibinfo{volume}{81},
  \bibinfo{number}{1} (\bibinfo{year}{1994}), \bibinfo{pages}{27--40}.
\newblock


\bibitem[\protect\citeauthoryear{Lunn, Thomas, Best, and Spiegelhalter}{Lunn
  et~al\mbox{.}}{2000}]%
        {lunn-winbugs}
\bibfield{author}{\bibinfo{person}{David~J. Lunn}, \bibinfo{person}{Andrew
  Thomas}, \bibinfo{person}{Nicky Best}, {and} \bibinfo{person}{David
  Spiegelhalter}.} \bibinfo{year}{2000}\natexlab{}.
\newblock \showarticletitle{{WinBUGS}---A {B}ayesian Modelling Framework:
  Concepts, Structure, and Extensibility}.
\newblock \bibinfo{journal}{\emph{Statistics and Computing}}
  \bibinfo{volume}{10}, \bibinfo{number}{4} (\bibinfo{year}{2000}),
  \bibinfo{pages}{325--337}.
\newblock


\bibitem[\protect\citeauthoryear{MacKay}{MacKay}{1998}]%
        {mackay-monte}
\bibfield{author}{\bibinfo{person}{David J.~C. MacKay}.}
  \bibinfo{year}{1998}\natexlab{}.
\newblock \showarticletitle{Introduction to {M}onte {C}arlo Methods}.
\newblock In \bibinfo{booktitle}{\emph{Learning and Inference in Graphical
  Models}}, \bibfield{editor}{\bibinfo{person}{Michael~I. Jordan}} (Ed.).
  \bibinfo{publisher}{Kluwer}, \bibinfo{address}{Dordrecht}.
\newblock
\newblock
\shownote{Paperback: {\it Learning in Graphical Models\/}, {MIT} Press.}


\bibitem[\protect\citeauthoryear{Mansinghka, Selsam, and Perov}{Mansinghka
  et~al\mbox{.}}{2014}]%
        {mansinghka-venture}
\bibfield{author}{\bibinfo{person}{Vikash Mansinghka}, \bibinfo{person}{Daniel
  Selsam}, {and} \bibinfo{person}{Yura Perov}.}
  \bibinfo{year}{2014}\natexlab{}.
\newblock \bibinfo{booktitle}{\emph{{V}enture: a Higher-Order Probabilistic
  Programming Platform with Programmable Inference}}.
\newblock \bibinfo{type}{e-Print} 1404.0099.
  \bibinfo{institution}{ar{X}iv.org}.
\newblock


\bibitem[\protect\citeauthoryear{McCallum and Nigam}{McCallum and
  Nigam}{1998}]%
        {mccallum1998comparison}
\bibfield{author}{\bibinfo{person}{Andrew McCallum} {and}
  \bibinfo{person}{Kamal Nigam}.} \bibinfo{year}{1998}\natexlab{}.
\newblock \showarticletitle{A Comparison of Event Models for Naive {B}ayes Text
  Classification}. In \bibinfo{booktitle}{\emph{AAAI-98 workshop on learning
  for text categorization}}, Vol.~\bibinfo{volume}{752}.
  \bibinfo{pages}{41--48}.
\newblock


\bibitem[\protect\citeauthoryear{McCallum}{McCallum}{2002}]%
        {mallet}
\bibfield{author}{\bibinfo{person}{Andrew~Kachites McCallum}.}
  \bibinfo{year}{2002}\natexlab{}.
\newblock \bibinfo{title}{{MALLET: A Machine Learning for Language Toolkit}}.
\newblock
\newblock
\urldef\tempurl%
\url{http://mallet.cs.umass.edu}
\showURL{%
\tempurl}


\bibitem[\protect\citeauthoryear{Meng and {van Dyk}}{Meng and {van
  Dyk}}{1999}]%
        {meng-seeking}
\bibfield{author}{\bibinfo{person}{Xiao-Li Meng} {and}
  \bibinfo{person}{David~A. {van Dyk}}.} \bibinfo{year}{1999}\natexlab{}.
\newblock \showarticletitle{Seeking Efficient Data Augmentation Schemes via
  Conditional and Marginal Augmentation}.
\newblock \bibinfo{journal}{\emph{Biometrika}} \bibinfo{volume}{86},
  \bibinfo{number}{2} (\bibinfo{year}{1999}), \bibinfo{pages}{301--320}.
\newblock


\bibitem[\protect\citeauthoryear{Milch, Marthi, Russell, Sontag, Ong, and
  Kolobov}{Milch et~al\mbox{.}}{2007}]%
        {milch-blog}
\bibfield{author}{\bibinfo{person}{Brian Milch}, \bibinfo{person}{Bhaskara
  Marthi}, \bibinfo{person}{Stuart Russell}, \bibinfo{person}{David Sontag},
  \bibinfo{person}{Daniel~L. Ong}, {and} \bibinfo{person}{Andrey Kolobov}.}
  \bibinfo{year}{2007}\natexlab{}.
\newblock \showarticletitle{{BLOG}: Probabilistic Models with Unknown Objects}.
\newblock In \bibinfo{booktitle}{\emph{Introduction to Statistical Relational
  Learning}}, \bibfield{editor}{\bibinfo{person}{Lise Getoor} {and}
  \bibinfo{person}{Ben Taskar}} (Eds.). \bibinfo{publisher}{{MIT} Press},
  Chapter~13, \bibinfo{pages}{373--398}.
\newblock


\bibitem[\protect\citeauthoryear{Murray, Lund{\'e}n, Kudlicka, Broman, and
  Sch{\"o}n}{Murray et~al\mbox{.}}{2018}]%
        {murray-delayed}
\bibfield{author}{\bibinfo{person}{Lawrence~M. Murray}, \bibinfo{person}{Daniel
  Lund{\'e}n}, \bibinfo{person}{Jan Kudlicka}, \bibinfo{person}{David Broman},
  {and} \bibinfo{person}{Thomas~B. Sch{\"o}n}.}
  \bibinfo{year}{2018}\natexlab{}.
\newblock \showarticletitle{Delayed Sampling and Automatic
  {R}ao-{B}lackwellization of Probabilistic Programs}. In
  \bibinfo{booktitle}{\emph{Proceedings of {AISTATS} 2018: 21st International
  Conference on Artificial Intelligence and Statistics}}
  \emph{(\bibinfo{series}{Proceedings of Machine Learning Research})},
  \bibfield{editor}{\bibinfo{person}{Amos Storkey} {and}
  \bibinfo{person}{Fernando Perez-Cruz}} (Eds.). \bibinfo{pages}{1037--1046}.
\newblock


\bibitem[\protect\citeauthoryear{Narayanan, Carette, Romano, Shan, and
  Zinkov}{Narayanan et~al\mbox{.}}{2016}]%
        {narayanan-probabilistic}
\bibfield{author}{\bibinfo{person}{Praveen Narayanan}, \bibinfo{person}{Jacques
  Carette}, \bibinfo{person}{Wren Romano}, \bibinfo{person}{Chung-chieh Shan},
  {and} \bibinfo{person}{Robert Zinkov}.} \bibinfo{year}{2016}\natexlab{}.
\newblock \showarticletitle{Probabilistic Inference by Program Transformation
  in {H}akaru (System Description)}. In \bibinfo{booktitle}{\emph{Proceedings
  of {FLOPS} 2016: 13th International Symposium on Functional and Logic
  Programming}} \emph{(\bibinfo{series}{{L}ecture {N}otes in {C}omputer
  {S}cience})}, \bibfield{editor}{\bibinfo{person}{Oleg Kiselyov} {and}
  \bibinfo{person}{Andy King}} (Eds.). \bibinfo{publisher}{Springer},
  \bibinfo{pages}{62--79}.
\newblock


\bibitem[\protect\citeauthoryear{Narayanan and Shan}{Narayanan and
  Shan}{2017}]%
        {narayanan-symbolic}
\bibfield{author}{\bibinfo{person}{Praveen Narayanan} {and}
  \bibinfo{person}{Chung-chieh Shan}.} \bibinfo{year}{2017}\natexlab{}.
\newblock \showarticletitle{Symbolic Conditioning of Arrays in Probabilistic
  Programs}.
\newblock \bibinfo{journal}{\emph{Proceedings of the {ACM} on Programming
  Languages}} \bibinfo{volume}{1}, \bibinfo{number}{ICFP}
  (\bibinfo{year}{2017}), \bibinfo{pages}{11:1--11:25}.
\newblock


\bibitem[\protect\citeauthoryear{Neal}{Neal}{2011}]%
        {neal-hamiltonian}
\bibfield{author}{\bibinfo{person}{Radford~M. Neal}.}
  \bibinfo{year}{2011}\natexlab{}.
\newblock \showarticletitle{{MCMC} Using {H}amiltonian Dynamics}.
\newblock In \bibinfo{booktitle}{\emph{Handbook of {M}arkov {C}hain {M}onte
  {C}arlo}}, \bibfield{editor}{\bibinfo{person}{Steve Brooks},
  \bibinfo{person}{Andrew Gelman}, \bibinfo{person}{Galin Jones}, {and}
  \bibinfo{person}{Xiao-Li Meng}} (Eds.). \bibinfo{publisher}{CRC Press},
  Chapter~5.
\newblock


\bibitem[\protect\citeauthoryear{Nori, Hur, Rajamani, and Samuel}{Nori
  et~al\mbox{.}}{2014}]%
        {nori-r2}
\bibfield{author}{\bibinfo{person}{Aditya~V. Nori}, \bibinfo{person}{Chung-Kil
  Hur}, \bibinfo{person}{Sriram~K. Rajamani}, {and} \bibinfo{person}{Selva
  Samuel}.} \bibinfo{year}{2014}\natexlab{}.
\newblock \showarticletitle{{R2}: An Efficient {MCMC} Sampler for Probabilistic
  Programs}. In \bibinfo{booktitle}{\emph{Proceedings of the 28th {AAAI}
  Conference on Artificial Intelligence}},
  \bibfield{editor}{\bibinfo{person}{Carla~E. Brodley} {and}
  \bibinfo{person}{Peter Stone}} (Eds.). \bibinfo{publisher}{{AAAI} Press},
  \bibinfo{pages}{2476--2482}.
\newblock


\bibitem[\protect\citeauthoryear{Obermeyer, Bingham, Jankowiak, Pradhan, and
  Goodman}{Obermeyer et~al\mbox{.}}{2018}]%
        {obermeyer-automated}
\bibfield{author}{\bibinfo{person}{Fritz~H. Obermeyer}, \bibinfo{person}{Eli
  Bingham}, \bibinfo{person}{Martin Jankowiak}, \bibinfo{person}{Neeraj
  Pradhan}, {and} \bibinfo{person}{Noah Goodman}.}
  \bibinfo{year}{2018}\natexlab{}.
\newblock \bibinfo{title}{Automated Enumeration of Discrete Latent Variables}.
  (\bibinfo{year}{2018}).
\newblock
\newblock
\shownote{Poster at PROBPROG 2018.}


\bibitem[\protect\citeauthoryear{Patil, Huard, and Fonnesbeck}{Patil
  et~al\mbox{.}}{2010}]%
        {patil-pymc}
\bibfield{author}{\bibinfo{person}{Anand Patil}, \bibinfo{person}{David Huard},
  {and} \bibinfo{person}{Christopher~J. Fonnesbeck}.}
  \bibinfo{year}{2010}\natexlab{}.
\newblock \showarticletitle{{PyMC}: {B}ayesian Stochastic Modelling in
  {P}ython}.
\newblock \bibinfo{journal}{\emph{Journal of Statistical Software}}
  \bibinfo{volume}{35}, \bibinfo{number}{4} (\bibinfo{date}{July}
  \bibinfo{year}{2010}), \bibinfo{pages}{1--81}.
\newblock


\bibitem[\protect\citeauthoryear{Pearson}{Pearson}{1894}]%
        {Pearson71}
\bibfield{author}{\bibinfo{person}{Karl Pearson}.}
  \bibinfo{year}{1894}\natexlab{}.
\newblock \showarticletitle{{III}. Contributions to the Mathematical Theory of
  Evolution}.
\newblock \bibinfo{journal}{\emph{Philosophical Transactions of the Royal
  Society of London A: Mathematical, Physical and Engineering Sciences}}
  \bibinfo{volume}{185} (\bibinfo{year}{1894}), \bibinfo{pages}{71--110}.
\newblock
\showISSN{0264-3820}
\urldef\tempurl%
\url{https://doi.org/10.1098/rsta.1894.0003}
\showDOI{\tempurl}
\showeprint{http://rsta.royalsocietypublishing.org/content/185/71.full.pdf}


\bibitem[\protect\citeauthoryear{Pfeffer}{Pfeffer}{2007}]%
        {pfeffer-design}
\bibfield{author}{\bibinfo{person}{Avi Pfeffer}.}
  \bibinfo{year}{2007}\natexlab{}.
\newblock \showarticletitle{The Design and Implementation of {IBAL}: A
  General-Purpose Probabilistic Language}.
\newblock In \bibinfo{booktitle}{\emph{Introduction to Statistical Relational
  Learning}}, \bibfield{editor}{\bibinfo{person}{Lise Getoor} {and}
  \bibinfo{person}{Ben Taskar}} (Eds.). \bibinfo{publisher}{{MIT} Press},
  Chapter~14, \bibinfo{pages}{399--432}.
\newblock


\bibitem[\protect\citeauthoryear{Pfeffer}{Pfeffer}{2016}]%
        {figaro}
\bibfield{author}{\bibinfo{person}{Avi Pfeffer}.}
  \bibinfo{year}{2016}\natexlab{}.
\newblock \bibinfo{booktitle}{\emph{Practical Probabilistic Programming}}.
\newblock \bibinfo{publisher}{Manning Publications}.
\newblock


\bibitem[\protect\citeauthoryear{Plummer}{Plummer}{2003}]%
        {Plummer2003}
\bibfield{author}{\bibinfo{person}{Martyn Plummer}.}
  \bibinfo{year}{2003}\natexlab{}.
\newblock \showarticletitle{{JAGS}: A program for analysis of {Bayesian}
  graphical models using {Gibbs} sampling}. In
  \bibinfo{booktitle}{\emph{Proceedings of the 3rd International Workshop on
  Distributed Statistical Computing}}.
\newblock


\bibitem[\protect\citeauthoryear{Pollard}{Pollard}{2001}]%
        {pollard-measure}
\bibfield{author}{\bibinfo{person}{David Pollard}.}
  \bibinfo{year}{2001}\natexlab{}.
\newblock \bibinfo{booktitle}{\emph{A User's Guide to Measure Theoretic
  Probability}}.
\newblock \bibinfo{publisher}{Cambridge University Press}.
\newblock


\bibitem[\protect\citeauthoryear{Poole and Zhang}{Poole and Zhang}{2003}]%
        {poole-exploiting}
\bibfield{author}{\bibinfo{person}{David Poole} {and}
  \bibinfo{person}{Nevin~Lianwen Zhang}.} \bibinfo{year}{2003}\natexlab{}.
\newblock \showarticletitle{Exploiting Contextual Independence In Probabilistic
  Inference}.
\newblock \bibinfo{journal}{\emph{Journal of Artificial Intelligence Research}}
   \bibinfo{volume}{18} (\bibinfo{year}{2003}), \bibinfo{pages}{263--313}.
\newblock


\bibitem[\protect\citeauthoryear{Rabiner}{Rabiner}{1989}]%
        {rabiner-speech}
\bibfield{author}{\bibinfo{person}{Lawrence~R. Rabiner}.}
  \bibinfo{year}{1989}\natexlab{}.
\newblock \showarticletitle{A Tutorial on Hidden {M}arkov Models and Selected
  Applications in Speech Recognition}.
\newblock \bibinfo{journal}{\emph{Proc. IEEE}} \bibinfo{volume}{77},
  \bibinfo{number}{2} (\bibinfo{date}{Feb.} \bibinfo{year}{1989}),
  \bibinfo{pages}{257--286}.
\newblock


\bibitem[\protect\citeauthoryear{Ramsey and Pfeffer}{Ramsey and
  Pfeffer}{2002}]%
        {ramsey-stochastic}
\bibfield{author}{\bibinfo{person}{Norman Ramsey} {and} \bibinfo{person}{Avi
  Pfeffer}.} \bibinfo{year}{2002}\natexlab{}.
\newblock \showarticletitle{Stochastic Lambda Calculus and Monads of
  Probability Distributions}. In \bibinfo{booktitle}{\emph{Proceedings of the
  29th Symposium on Principles of Programming Languages ({POPL})}}.
  \bibinfo{publisher}{ACM Press}, \bibinfo{pages}{154--165}.
\newblock


\bibitem[\protect\citeauthoryear{Rao}{Rao}{1945}]%
        {rao-information}
\bibfield{author}{\bibinfo{person}{C.~Radhakrishna Rao}.}
  \bibinfo{year}{1945}\natexlab{}.
\newblock \showarticletitle{Information and Accuracy Attainable in the
  Estimation of Statistical Parameters}.
\newblock \bibinfo{journal}{\emph{Bulletin of the Calcutta Mathematical
  Society}} \bibinfo{volume}{37}, \bibinfo{number}{3} (\bibinfo{year}{1945}),
  \bibinfo{pages}{81--91}.
\newblock


\bibitem[\protect\citeauthoryear{Resnik and Hardisty}{Resnik and
  Hardisty}{2010}]%
        {resnik-gibbs}
\bibfield{author}{\bibinfo{person}{Philip Resnik} {and} \bibinfo{person}{Eric
  Hardisty}.} \bibinfo{year}{2010}\natexlab{}.
\newblock \bibinfo{booktitle}{\emph{{G}ibbs Sampling for the Uninitiated}}.
\newblock \bibinfo{type}{{T}echnical {R}eport} CS-TR-4956 UMIACS-TR-2010-04
  LAMP-TR-153. \bibinfo{institution}{University of Maryland}.
\newblock


\bibitem[\protect\citeauthoryear{Sanner and Abbasnejad}{Sanner and
  Abbasnejad}{2012}]%
        {sanner-symbolic}
\bibfield{author}{\bibinfo{person}{Scott Sanner} {and} \bibinfo{person}{Ehsan
  Abbasnejad}.} \bibinfo{year}{2012}\natexlab{}.
\newblock \showarticletitle{Symbolic Variable Elimination for Discrete and
  Continuous Graphical Models}. In \bibinfo{booktitle}{\emph{Proceedings of the
  26th {AAAI} Conference on Artificial Intelligence}},
  \bibfield{editor}{\bibinfo{person}{J{\"o}rg Hoffmann} {and}
  \bibinfo{person}{Bart Selman}} (Eds.). \bibinfo{publisher}{{AAAI} Press},
  \bibinfo{pages}{1954--1960}.
\newblock


\bibitem[\protect\citeauthoryear{Shan and Ramsey}{Shan and Ramsey}{2017}]%
        {shan-exact}
\bibfield{author}{\bibinfo{person}{Chung-chieh Shan} {and}
  \bibinfo{person}{Norman Ramsey}.} \bibinfo{year}{2017}\natexlab{}.
\newblock \showarticletitle{Exact {B}ayesian Inference by Symbolic
  Disintegration}. In \bibinfo{booktitle}{\emph{Proceedings of the 44th
  Symposium on Principles of Programming Languages ({POPL})}}.
  \bibinfo{publisher}{ACM Press}, \bibinfo{pages}{130--144}.
\newblock


\bibitem[\protect\citeauthoryear{Staton}{Staton}{2017}]%
        {staton-commutative}
\bibfield{author}{\bibinfo{person}{Sam Staton}.}
  \bibinfo{year}{2017}\natexlab{}.
\newblock \showarticletitle{Commutative Semantics for Probabilistic
  Programming}. In \bibinfo{booktitle}{\emph{Programming Languages and Systems:
  Proceedings of {ESOP} 2017, 26th {E}uropean Symposium on Programming}}
  \emph{(\bibinfo{series}{{L}ecture {N}otes in {C}omputer {S}cience})},
  \bibfield{editor}{\bibinfo{person}{Yang Hongseok}} (Ed.).
  \bibinfo{publisher}{Springer}, \bibinfo{pages}{855--879}.
\newblock


\bibitem[\protect\citeauthoryear{Tran, Hoffman, Saurous, Brevdo, Murphy, and
  Blei}{Tran et~al\mbox{.}}{2017}]%
        {tran-deep}
\bibfield{author}{\bibinfo{person}{Dustin Tran}, \bibinfo{person}{Matthew~D.
  Hoffman}, \bibinfo{person}{Rif~A. Saurous}, \bibinfo{person}{Eugene Brevdo},
  \bibinfo{person}{Kevin Murphy}, {and} \bibinfo{person}{David~M. Blei}.}
  \bibinfo{year}{2017}\natexlab{}.
\newblock \bibinfo{booktitle}{\emph{Deep Probabilistic Programming}}.
\newblock \bibinfo{type}{e-Print} 1701.03757.
  \bibinfo{institution}{ar{X}iv.org}.
\newblock
\newblock
\shownote{5th International Conference on Learning Representations.}


\bibitem[\protect\citeauthoryear{Tristan, Huang, Tassarotti, Pocock, Green, and
  Steele}{Tristan et~al\mbox{.}}{2014}]%
        {tristan-augur}
\bibfield{author}{\bibinfo{person}{Jean-Baptiste Tristan},
  \bibinfo{person}{Daniel Huang}, \bibinfo{person}{Joseph Tassarotti},
  \bibinfo{person}{Adam~C. Pocock}, \bibinfo{person}{Stephen~J. Green}, {and}
  \bibinfo{person}{Guy~Lewis Steele, Jr.}} \bibinfo{year}{2014}\natexlab{}.
\newblock \bibinfo{booktitle}{\emph{{A}ugur: a Modeling Language for
  Data-Parallel Probabilistic Inference}}.
\newblock \bibinfo{type}{e-Print} 1312.3613.
  \bibinfo{institution}{ar{X}iv.org}.
\newblock
\urldef\tempurl%
\url{http://arxiv.org/abs/1312.3613}
\showURL{%
\tempurl}


\bibitem[\protect\citeauthoryear{Venugopal and Gogate}{Venugopal and
  Gogate}{2013}]%
        {venugopal-dynamic}
\bibfield{author}{\bibinfo{person}{Deepak Venugopal} {and}
  \bibinfo{person}{Vibhav Gogate}.} \bibinfo{year}{2013}\natexlab{}.
\newblock \showarticletitle{Dynamic Blocking and Collapsing for {G}ibbs
  Sampling}. In \bibinfo{booktitle}{\emph{Proceedings of the 29th Conference on
  Uncertainty in Artificial Intelligence}},
  \bibfield{editor}{\bibinfo{person}{Ann Nicholson} {and}
  \bibinfo{person}{Padhraic Smyth}} (Eds.). \bibinfo{pages}{664--673}.
\newblock


\bibitem[\protect\citeauthoryear{Wilf and Zeilberger}{Wilf and
  Zeilberger}{1992}]%
        {WilfZeil1992}
\bibfield{author}{\bibinfo{person}{Herbert~S. Wilf} {and}
  \bibinfo{person}{Doron Zeilberger}.} \bibinfo{year}{1992}\natexlab{}.
\newblock \showarticletitle{An Algorithmic Proof Theory for Hypergeometric
  (Ordinary and ``q'') Multisum/Integral Identities}.
\newblock \bibinfo{journal}{\emph{Inventiones mathematicae}}
  \bibinfo{volume}{108} (\bibinfo{year}{1992}), \bibinfo{pages}{557--633}.
\newblock


\bibitem[\protect\citeauthoryear{Wingate, Stuhlm{\"u}ller, and Goodman}{Wingate
  et~al\mbox{.}}{2011}]%
        {wingate-lightweight}
\bibfield{author}{\bibinfo{person}{David Wingate}, \bibinfo{person}{Andreas
  Stuhlm{\"u}ller}, {and} \bibinfo{person}{Noah~D. Goodman}.}
  \bibinfo{year}{2011}\natexlab{}.
\newblock \showarticletitle{Lightweight Implementations of Probabilistic
  Programming Languages Via Transformational Compilation}. In
  \bibinfo{booktitle}{\emph{Proceedings of {AISTATS} 2011: 14th International
  Conference on Artificial Intelligence and Statistics}}
  \emph{(\bibinfo{series}{JMLR Workshop and Conference Proceedings})},
  \bibfield{editor}{\bibinfo{person}{Geoffrey Gordon}, \bibinfo{person}{David
  Dunson}, {and} \bibinfo{person}{Miroslav Dud{\'\i}k}} (Eds.).
  \bibinfo{publisher}{{MIT} Press}, \bibinfo{pages}{770--778}.
\newblock


\bibitem[\protect\citeauthoryear{Wood, van~de Meent, and Mansinghka}{Wood
  et~al\mbox{.}}{2014}]%
        {wood-new}
\bibfield{author}{\bibinfo{person}{Frank Wood}, \bibinfo{person}{Jan~Willem
  van~de Meent}, {and} \bibinfo{person}{Vikash Mansinghka}.}
  \bibinfo{year}{2014}\natexlab{}.
\newblock \showarticletitle{A New Approach to Probabilistic Programming
  Inference}. In \bibinfo{booktitle}{\emph{Proceedings of {AISTATS} 2014: 17th
  International Conference on Artificial Intelligence and Statistics}}
  \emph{(\bibinfo{series}{JMLR Workshop and Conference Proceedings})}.
  \bibinfo{pages}{1024--1032}.
\newblock


\bibitem[\protect\citeauthoryear{Wu, Li, Russell, and Bod{\'\i}k}{Wu
  et~al\mbox{.}}{2016}]%
        {wu-swift}
\bibfield{author}{\bibinfo{person}{Yi Wu}, \bibinfo{person}{Lei Li},
  \bibinfo{person}{Stuart~J. Russell}, {and} \bibinfo{person}{Rastislav
  Bod{\'\i}k}.} \bibinfo{year}{2016}\natexlab{}.
\newblock \showarticletitle{{S}wift: Compiled Inference for Probabilistic
  Programming Languages}. In \bibinfo{booktitle}{\emph{Proceedings of the 25th
  International Joint Conference on Artificial Intelligence}},
  \bibfield{editor}{\bibinfo{person}{Subbarao Kambhampati}} (Ed.).
  \bibinfo{publisher}{{AAAI} Press}, \bibinfo{pages}{3637--3645}.
\newblock
\showISBNx{978-1-57735-770-4}
\urldef\tempurl%
\url{http://www.ijcai.org/Abstract/16/512}
\showURL{%
\tempurl}


\bibitem[\protect\citeauthoryear{Zhang and Poole}{Zhang and Poole}{1994}]%
        {zhang-simple}
\bibfield{author}{\bibinfo{person}{Nevin~Lianwen Zhang} {and}
  \bibinfo{person}{David~L. Poole}.} \bibinfo{year}{1994}\natexlab{}.
\newblock \showarticletitle{A Simple Approach to {B}ayesian Network
  Computations}. In \bibinfo{booktitle}{\emph{Proceedings of the 10th
  {C}anadian Conference on Artificial Intelligence}}.
  \bibinfo{pages}{171--178}.
\newblock


\bibitem[\protect\citeauthoryear{Zhang and Poole}{Zhang and Poole}{1996}]%
        {zhang-exploiting}
\bibfield{author}{\bibinfo{person}{Nevin~Lianwen Zhang} {and}
  \bibinfo{person}{David~L. Poole}.} \bibinfo{year}{1996}\natexlab{}.
\newblock \showarticletitle{Exploiting Causal Independence in {B}ayesian
  Network Inference}.
\newblock \bibinfo{journal}{\emph{Journal of Artificial Intelligence Research}}
   \bibinfo{volume}{5} (\bibinfo{year}{1996}), \bibinfo{pages}{301--328}.
\newblock


\bibitem[\protect\citeauthoryear{Zinkov and Shan}{Zinkov and Shan}{2017}]%
        {zinkov-composing}
\bibfield{author}{\bibinfo{person}{Robert Zinkov} {and}
  \bibinfo{person}{Chung-chieh Shan}.} \bibinfo{year}{2017}\natexlab{}.
\newblock \showarticletitle{Composing Inference Algorithms as Program
  Transformations}, \bibfield{editor}{\bibinfo{person}{Gal Elidan},
  \bibinfo{person}{Kristian Kersting}, {and} \bibinfo{person}{Alexander~T.
  Ihler}} (Eds.). \bibinfo{publisher}{{AUAI} Press}.
\newblock


\end{thebibliography}
